\documentclass[apjl]{emulateapj}
\usepackage{comment}
\usepackage{ifthen}
\usepackage{amsmath}
\usepackage{amsfonts}
\usepackage{amssymb}
\usepackage{wasysym}
\usepackage{subfigure}
\usepackage{epsfig}

\newcommand{\luna}{{\tt LUNA}}
\newcommand{\multi}{{\sc MultiNest}}
\newcommand{\cofiam}{{\tt CoFiAM}}

\newboolean{emulateapj}
\setboolean{emulateapj}{true}

\newboolean{astroph}
\setboolean{astroph}{true}
\shortauthors{Kipping et al.}
\shorttitle{II. The Hunt for Exomoons}
\ifthenelse{\boolean{emulateapj}}{
    \newcommand{\titledag}{$\dagger$}
}{
    \newcommand{\titledag}{\dagger}
}
\def\mathbi#1{\textbf{\em #1}}

\begin{document}

%% Titlepage
\title {The Hunt for Exomoons with Kepler (HEK):\\
 II. Analysis of Seven Viable Satellite-Hosting Planet Candidates
\altaffilmark{\titledag}}

%% Authors
\author{
	{\bf D.~M.~Kipping\altaffilmark{1,2},
             J.~Hartman\altaffilmark{3},
             L.~A.~Buchhave\altaffilmark{4,5},\\
             A.~R.~Schmitt\altaffilmark{6},
             G.~\'A.~Bakos\altaffilmark{3,7,8},
             D.~Nesvorn\'y\altaffilmark{9}
	}
}
\altaffiltext{1}{Harvard-Smithsonian Center for Astrophysics,
		Cambridge, MA 02138, USA; email: dkipping@cfa.harvard.edu}

\altaffiltext{2}{NASA Carl Sagan Fellow}

\altaffiltext{3}{Dept. of Astrophysical Sciences, Princeton University,
		Princeton, NJ 05844, USA}

\altaffiltext{4}{Niels Bohr Institute, University of Copenhagen, DK-2100, 
		Copenhagen, Denmark}

\altaffiltext{5}{Centre for Star and Planet Formation, Natural History Museum of 
		Denmark, University of Copenhagen, DK-1350, Copenhagen, Denmark}

\altaffiltext{6}{Citizen Science}

\altaffiltext{7}{Alfred P. Sloan Fellow}

\altaffiltext{8}{Packard Fellow}

\altaffiltext{9}{Dept. of Space Studies, Southwest Research Institute, 
1050 Walnut St., Suite 300, Boulder, CO 80302, USA}

\altaffiltext{$\dagger$}{
Based on archival data of the \emph{Kepler} telescope. 
}

%% EOF authors

% #####################################################################
%% abstract
\begin{abstract}
%++++++++++++++++++++++++++++++++++++++++++++++++++++++++++++++++++++++
\begin{comment}
\end{comment}
%++++++++++++++++++++++++++++++++++++++++++++++++++++++++++++++++++++++

From the list of 2321 transiting planet candidates announced by the
\emph{Kepler Mission}, we select seven targets with favorable properties 
for the capacity to dynamically maintain an exomoon and present a detectable 
signal. These seven candidates were identified through our automatic target 
selection (TSA) algorithm and target selection prioritization (TSP) filtering,
whereby we excluded systems exhibiting significant time-correlated noise and
focussed on those with a single transiting planet candidate of radius less than 
6$\,R_{\oplus}$. We find no compelling evidence for an exomoon around any of the 
seven KOIs but constrain the satellite-to-planet mass ratios for each. For four 
of the seven KOIs, we estimate a 95\% upper quantile of $M_S/M_P<0.04$, which 
given the radii of the candidates, likely probes down to sub-Earth masses. We 
also derive precise transit times and durations for each candidate and find no 
evidence for dynamical variations in any of the KOIs. With just a few systems
analyzed thus far in the on-going HEK project, projections on $\eta_{\leftmoon}$
would be premature, but a high frequency of large moons around 
Super-Earths/Mini-Neptunes would appear to be incommensurable with our results
so far.

\end{abstract}

% #####################################################################
%% keywords

%% EOF keywords
%% EOF titlepage

% #####################################################################
%% Introduction
\section{INTRODUCTION}
\label{sec:intro}

%% INTRODUCTION
%%

The ``Hunt for Exomoons with Kepler'' (HEK) project is the first systematic
survey for moons around planets outside of our solar system \citep{hek:2012}. 
Whilst many planets around our Sun host one or more satellites, there is no 
empirical evidence for moons around the hundreds of extrasolar planets detected
in recent years. At best, one can interpret the possible detection of a
circumplanetary disc by \citet{mamajek:2012} as a putative moon-forming region. 
The \emph{Kepler Mission} \citep{borucki:2009} is the most suitable instrument
available for detecting exomoons thanks to the large number of target stars, long 
temporal baselines, nearly continuous monitoring and very precise photometry. By 
monitoring the timing of exoplanet transits, \citet{kipping:2009} have estimated 
that \emph{Kepler} should be sensitive to $\mathcal{O}[M_{\oplus}]$ mass 
exomoons. In addition, \emph{Kepler} is designed to detect 
$\mathcal{O}[R_{\oplus}]$ radius transiting bodies and moons may be found in a 
similar way \citep{luna:2011}. It is therefore argued that Earth-mass/radius 
moons should be detectable. Although there are no moons this large or massive in 
our solar system, HEK seeks to answer whether this is true for all exoplanetary 
systems or not.

A detailed description of the goals and methods of the HEK project are discussed
in \citet{hek:2012}. To date, the analysis of only one system for exomoons has
been published by the HEK project in \citet{nesvorny:2012}. In this case, the
target planetary candidate, KOI-872.01, was identified as being a target of
opportunity (TSO) due to the presence of very large transit timing variations
(TTV), enabling us to detect a second non-transiting planet in the system (and 
confirm the planetary nature of KOI-872.01). The non-moon origin of these TTVs
demonstrated the importance of the careful interpretation of dynamical effects.

In this work, we will provide an analysis of seven planetary candidates (KOIs)
identified through our automatic target selection (TSA) method. Each candidate
therefore satisfies the criteria of having the capability to host an Earth-mass
moon plus sufficient signal-to-noise to make such a detection feasible.
Consequently, null-detections have much greater significance for understanding
the frequency of large moons around viable planet hosts, $\eta_{\leftmoon}$.

\section{TARGET SELECTION}
\label{sec:TS}

%% TARGET SELECTION
%%

\subsection{Automatic Target Selection (TSA)}
\subsubsection{Overview}

The HEK project treats the Kepler Objects of Interest (KOIs) as a list of
potential moon-hosting targets in much the same way that Kepler itself treats
the Kepler Input Catalogue (KIC) stars as a list of potential planet-hosting
targets. At the time of writing, 2321 KOIs have been reported by 
\citet{batalha:2012} (B12). Searching individual systems for signs of an 
exomoon is a time expensive task in terms of computational demands and
human manpower \citep{nesvorny:2012}. As a result, the HEK project performs 
a target selection (TS) procedure to select only the most viable candidates 
for detailed analysis. In Paper I \citep{hek:2012}, we discussed the three 
principal TS methods employed by HEK: i) automatic target selection (TSA) 
ii) visual target selection (TSV) and iii) target selection opportunities 
(TSO). Details on all three methods are discussed in \citet{hek:2012}, but 
this work will make use of TSA only. A dedicated TSV survey will be 
presented in a subsequent work.

After the TSA stage, we also apply a target selection prioritization (TSP) 
selection process, which identifies the optimal targets for an exomoon hunt.

\subsubsection{Modifications to the TSA algorithm}

The automatic target selection (TSA) algorithm has been slightly modified since 
Paper I \citep{hek:2012}. The main modification is to accommodate a continuous, 
and thus more realistic, minimum planetary mass estimation function. This
mass function is required to estimate the maximum stable moon mass around each
KOI as described in \citet{hek:2012}. Previously, we considered three regimes: 
i) Super-Earths ($R_P<2.0$\,$R_{\oplus}$) ii) Neptunes 
($2.0$\,$R_{\oplus}<R_P<6.0$\,$R_{\oplus}$) and iii) Jupiters 
($R_P>6.0$\,$R_{\oplus}$). The mass was estimated for Super-Earths using a
terrestrial-scaling law from \citet{valencia:2006}, whereas Neptunes were
assumed to have a constant density of $1.7$\,g\,cm$^{-3}$ for reasons discussed
in \citet{hek:2012}. Jupiters were not considered at all due to the higher
potential for a false-positive \citep{santerne:2012}. These minimum mass 
estimates are required to evaluate the dynamical capacity of each KOI for 
hosting a moon and thus a \emph{minimum} mass provides a conservative lower 
limit.

In this revised TSA algorithm, we make two major changes to the mass function:
1) we ensure a continuous mass-function 2) we allow this mass function to go
into the Jupiter-regime. The first improvement is inspired by the fact that in 
the Super-Earth/Neptune regime, the \citet{valencia:2006} mass function quickly 
exceeds 10\,$M_{\oplus}$, which leads to an abundance of TSA targets at this 
boundary (since higher mass planets have a better chance of hosting a moon). 
The second improvement allows us to consider the Jupiters as well and thus 
expand our search somewhat.

The \citet{valencia:2006} expression of $R\sim M^{0.27}$ is invertible to
$M\sim R^{3.7}$, in units of Earth radii and masses. For $R=1.863$\,$R_{\oplus}$
the mass hits 10\,$M_{\oplus}$, which we consider a sensible upper limit for
Super-Earth masses. Recall that TSA is primarily interested in a conservative
estimate of the planetary mass in order to ensure selected candidates have the 
best chance of being a good target. We therefore consider the Super-Earth
regime to be modified by this new radius limit. To bridge the gap between
Super-Earths and Neptunes in a continuous manner, we fix the mass to be
$10$\,$M_{\oplus}$ until $(4/3)\pi\rho_{\mathrm{Neptune}}
R_P^3=10$\,$M_{\oplus}$, which occurs for $R_P=3.186$\,$R_{\oplus}$ (we assume
spherical planets). The region between is dubbed ``Mini-Neptunes'' and such 
objects are assumed to have a mass of 10\,$M_{\oplus}$.

Our previous boundary between Neptunes and Jupiters is also discontinuous. The 
mass of Jupiter-radius planets varies widely, but a sensible lower limit is
to assume a Jovian bulk density. Much lower-density Jupiters do exist, so
called inflated gas giants, but are thought to be due to their high irradiation
environment since they are typically hot-Jupiters \citep{burrows:2007}. TSA 
automatically excludes hot-Jupiters since they are too close to their star to 
maintain an exomoon \citep{weidner:2010}. We therefore adopt a Jovian density of 
1.326\,g\,cm$^{-3}$. To bridge the density discontinuity between Neptunes and 
Jupiters, we assume the bulk density linearly drops off between 6\,$R_{\oplus}$ 
and 7\,$R_{\oplus}$ to create a continuous mass function. Finally, for a Jovian 
density object, once the radius exceeds $10.963$\,$R_{\oplus}$ the mass will 
exceed a Jupiter mass\footnote{This does not occur at $71,492,000$\,km$=$
1\,$R_J$ since this the \emph{equatorial} radius of Jupiter and the planet is 
non-spherical.}. For such cases, we set the upper limit on the mass to be 
1\,$M_J$. This is again in line with the conservative mass estimate requirements 
of TSA. In summary, we have:

\begin{itemize}
\item[i)] ``Super-Earths''; $0<\frac{R_P}{R_{\oplus}}\leq1.863$
\item[ii)] ``Mini-Neptunes''; $1.863<\frac{R_P}{R_{\oplus}}\leq3.186$
\item[iii)] ``Neptunes''; $3.186<\frac{R_P}{R_{\oplus}}\leq6$
\item[iv)] ``sub-Jupiters''; $6<\frac{R_P}{R_{\oplus}}\leq7$
\item[v)] ``Jupiters''; $7<\frac{R_P}{R_{\oplus}}\leq10.963$
\item[vi)] ``super-Jupiters''; $10.963<\frac{R_P}{R_{\oplus}}\leq\infty$
\end{itemize}

These six regimes are described by the following mass-function:

\begin{equation*}
M_P =
\begin{cases}
M_{\oplus} (\frac{R_P}{R_{\oplus}})^{3.7}  & \text{if } 0<\frac{R_P}{R_{\oplus}}\leq1.863 ,\\
10 M_{\oplus}  & \text{if } 1.863<\frac{R_P}{R_{\oplus}}\leq3.186 ,\\
\frac{4}{3}\pi R_P^3\rho_{\mathrm{Nep}} & \text{if } 3.186<\frac{R_P}{R_{\oplus}}\leq6 ,\\
\frac{4}{3}\pi R_P^3\rho_{\mathrm{subJup}} & \text{if } 6<\frac{R_P}{R_{\oplus}}\leq7 ,\\
\frac{4}{3}\pi R_P^3\rho_{\mathrm{Jup}} & \text{if } 7<\frac{R_P}{R_{\oplus}}\leq10.963 ,\\
M_{\mathrm{Jup}} & \text{if } 10.963<\frac{R_P}{R_{\oplus}}\leq\infty ,
\end{cases}
\label{eqn:massfunction}
\end{equation*}

where

\begin{align}
\rho_{\mathrm{subJup}} &= \rho_{\mathrm{Nep}} \Bigg[1-\Big(\frac{R_P-6\,R_{\oplus}}{R_{\oplus}}\Big)\Big(1-\frac{\rho_{\mathrm{Jup}}}{\rho_{\mathrm{Nep}}}\Big)\Bigg].
\end{align}

%%% NEW MASS FUNCTION
\begin{figure}
\begin{center}
\includegraphics[width=8.4 cm]{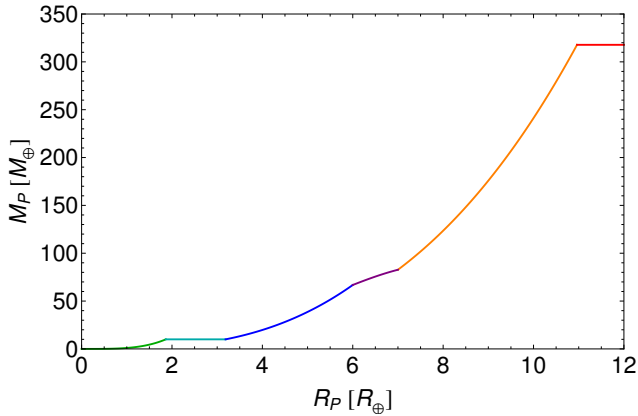}
\caption{\emph{Assumed mass function of Kepler Objects of Interest (KOIs) 
used in the updated Automatic Target Selection (TSA) algorithm of this work. The 
six regimes, from low-to-high radius, are i) Super-Earth ii) Mini-Neptune iii) 
Neptune iv) sub-Jupiter v) Jupiter vi) Super-Jupiter. Expressions for mass 
function are provided in \S~\ref{eqn:massfunction}. The eight triangles represent
the Solar System planets.
}} 
\label{fig:modelerrors}
\end{center}
\end{figure}

\subsubsection{TSA inputs}

The TSA algorithm can be executed for several different inputs. Critically,
one can choose the maximum orbital distance that an exomoon can reside at
(in units of the Hill radius) $f_{\mathrm{max}}$, and the tidal dissipation 
factor of the host planet $Q_P$, both of which strongly affect the maximum 
allowed exomoon mass via the expressions of \citet{barnes:2002}. TSAs are 
defined to satisfy the criterion that the host planet can maintain an Earth-mass
moon for 5\,Gyr \citep{hek:2012} and thus these inputs affect the number of TSA
candidates identified. 

An additional freedom is that one can alter the requirement of the 
signal-to-noise ratio (SNR) for a moon transit. We define SNR as an Earth-radius 
transit depth divided by the combined differential photometric precision (CDPP) 
\citep{christiansen:2012} over 6 hours (see Equation~\ref{eqn:SNR}). Note that 
the CDPP values are taken from the B12 tables, as with all 
other TSA inputs. One can see that enforcing a higher SNR condition will 
naturally reduce the number of TSA candidates found. 

\begin{align}
SNR &= \frac{(R_{\oplus}/R_{\star})^2}{\mathrm{CDPP}_6}.
\label{eqn:SNR}
\end{align}

Finally, one can choose whether we look at host planets in the Neptune-regime 
and smaller or whether we expand our search to include Jupiter-sized objects 
(which tend to have a higher false-positive-rate, \citealt{santerne:2012}). In 
this work, we treat quarters 1 to 9 from Kepler as the survey data used to look 
for exomoons. This leads to another criterion for TSA that $P_P<193.6$\,days, so 
that at least three transits exist in the Q1-9 \emph{Kepler} photometry; a 
minimum requirement for exomoon searches. At the time of writing, Q10-13 have 
also recently become available and we treat these data as follow-up photometry
with application for particularly interesting candidates. Any candidate already 
selected as a target of opportunity (TSO) by the HEK project was not included
in the TSA lists. We summarize the results of making these modifications in 
Table~\ref{tab:TSAsummary}.

\subsubsection{TSA results}

\begin{table}
\caption{\emph{
Summary of the number of candidates identified using TSA from the 2321 KOIs 
available for various inputs to the TSA algorithm. The left column modifies the 
maximum allowed exomoon distance from the planet, in units of the Hill radius, 
$f_{\mathrm{max}}$, for several reasonable guesses. The second column presents
the results for two different assumptions of the tidal dissipation factor, 
$Q_P$. The final column shows a vector-form of the number of candidates found
for the assumption of $\{SNR>1,SNR>2,SNR>3\}$.
}} % title of Table
\centering % used for centering table
\begin{tabular}{c c c} % centered columns (9 columns)
\hline
$f_{\mathrm{max}}$ & $Q_P$ & Candidates found \\ [0.5ex] % inserts table
%heading
\hline
$R_P\leq6$\,$R_{\oplus}$ & & \\
\hline
0.9309 & $10^5$ & $\{274,92,34\}$ \\
0.4805 & $10^5$ & $\{91,35,13\}$ \\
0.3333 & $10^5$ & $\{39,19,8\}$ \\
0.2500 & $10^5$ & $\{16,7,4\}$ \\
0.9309 & $10^4$ & $\{156,59,20\}$ \\
0.4805 & $10^4$ & $\{42,20,8\}$ \\
0.3333 & $10^4$ & $\{9,4,2\}$ \\
0.2500 & $10^4$ & $\{0,0,0\}$ \\
\hline
$R_P>6$\,$R_{\oplus}$ & & \\
\hline
0.9309 & $10^5$ & $\{26,5,2\}$ \\
0.4805 & $10^5$ & $\{13,2,2\}$ \\
0.3333 & $10^5$ & $\{9,2,2\}$ \\
0.2500 & $10^5$ & $\{1,0,0\}$ \\
0.9309 & $10^4$ & $\{20,4,2\}$ \\
0.4805 & $10^4$ & $\{9,2,2\}$ \\
0.3333 & $10^4$ & $\{1,0,0\}$ \\
0.2500 & $10^4$ & $\{0,0,0\}$ \\ [1ex]
\hline\hline %inserts single line
\end{tabular}
\label{tab:TSAsummary} % is used to refer this table in the text
\end{table}

Curiously, Table~\ref{tab:TSAsummary} reveals that Jupiter-sized objects offer
very little improvement in the number of viable candidates for moon hunting.
This is certainly due to a dearth of Jupiter-sized objects in the 
\emph{Kepler}-sample more than anything else (B12). 

For regular satellites, such as the Galilean satellites around Jupiter,
the formation of large moons is enhanced by a massive primary
\citep{canup:2006,sasaki:2010,ogihara:2012}. However, if the proposed
mass scaling law of \citep{canup:2006} holds true, that $M_S/M_P<10^{-4}$,
then such moons will be undetectable using Kepler \citep{kipping:2009}.
The moons we seek are therefore most likely irregular satellites arriving
through capture or impact (e.g. Triton-Neptune \citealt{agnor:2006}).
In such a case, \citet{porter:2011} argue that the mass of the primary has 
a much weaker effect with Neptunes and Jupiters retaining captured
satelites with broadly equivalent efficiencies. It is therefore important
to stress that Jupiter-sized KOIs do not hold a special significance for
target selection over Neptunes.

Due to the relatively small improvement in the overall number of candidates
offered by Jupiters, combined with their higher false-positive rate 
\citep{santerne:2012}, we will not consider Jovian TSA candidates in this work,
but we will return to them later in a future HEK survey.

\subsubsection{Selecting a TSA category}

Table~\ref{tab:TSAsummary} presents twenty-four different viable inputs for the 
TSA algorithm (for Neptunes or smaller). We must now select which input to use.
The first point to bear in mind is that an excellent candidate will appear in
multiple categories. For example, if a candidate satisfies SNR$>3$ then it will
of course also satisfy SNR$>1$. The task is therefore simply to move from
the most conservative estimate to the most optimistic and stop at the point at
which we have the desired number of candidates.

Due to computational constraints, we estimated we could analyze a handful of 
targets in this work. We therefore attempted to select a category which yields 
around a dozen or so candidates and apply the final target selection 
prioritization (TSP) stage to filter out the best of those.

We choose to work with $Q_P=10^5$ in what follows, since this option finds 
dramatically more high SNR signals and thus the best chance for success.
Next, we only consider candidates where the expected SNR$>2$ to balance between 
a high SNR and a significant number of candidates. We only consider KOIs with 
radii below $R_P<6\,R_{\oplus}$ (as reported by B12) for 
reasons discussed earlier. Finally, we opt for $f_{\mathrm{max}}=0.4805$ 
which bounds all prograde satellites yet still returns 35 TSA candidates, which
are listed in Table~\ref{tab:allTSAs}.

\begin{table*}
\caption{\emph{Kepler candidates identified for an exomoon search using 
TSA. SNR is defined in Equation~\ref{eqn:SNR}. The multiplicity denotes the 
number of KOIs in each system. Bold and bold-italic highlighted rows are those 
targets chosen for target selection prioritization (TSP) with the former being
those which we accepted by TSP and the latter being those rejected.
$\dagger$ = KOI dropped by Kepler-team.
* = Durbin-Watson statistic (Equation~\ref{eqn:durbin}) indicated high 
($>3$\,$\sigma$) probability of correlated noise at 30\,minute timescale in both 
PA and PDC data and the candidate was consequently rejected.
}} % title of Table
\centering % used for centering table
\begin{tabular}{r r r c c c} % centered columns (9 columns)
\hline
KIC & KOI & $P_P$ [days] & SNR (B12) & $\bar{\mathrm{SNR}}$ (MAST) & Multiplicity \\ [0.5ex] % inserts table
%heading
\hline
\emph{\textbf{3425851.01}}* & \emph{\textbf{268.01}} & \emph{\textbf{110.4}} & \emph{\textbf{5.76}} & $\mathbi{6.65}\mathbf{\pm}\mathbi{0.36}$ & \emph{\textbf{1}} \\
\textbf{11623629.01} &  \textbf{365.01} &  \textbf{81.7} & \textbf{6.39} & $\mathbf{5.45\pm0.62}$ & \textbf{1} \\
\emph{\textbf{7296438.01}}$\dagger$ &  \emph{\textbf{364.01}} & \emph{\textbf{173.9}} & \emph{\textbf{4.68}} & $\mathbi{4.57}\mathbf{\pm}\mathbi{0.19}$ & \emph{\textbf{1}} \\
\textbf{5966322.01} & \textbf{303.01} & \textbf{60.9} & \textbf{2.62} & $\mathbf{3.18\pm0.35}$ & \textbf{1} \\
8292840.02  &  260.02 & 100.3 & 2.69 & $3.29\pm0.27$ & 2 \\
9451706.01  &  271.01 &  48.6 & 2.32 & $2.96\pm0.38$ & 2 \\
\emph{\textbf{9414417.01}}* & \emph{\textbf{974.01}} & \emph{\textbf{53.5}} & \emph{\textbf{2.28}} & $\mathbi{2.87}\mathbf{\pm}\mathbi{0.25}$ & \emph{\textbf{1}} \\
9002278.03  &  701.03 & 122.4 & 2.18 & $2.77\pm0.11$ & 3 \\
\textbf{9965439.01} & \textbf{722.01} & \textbf{46.4} & \textbf{2.22} & $\mathbf{2.31\pm0.19}$ & \textbf{1} \\
\textbf{11622600.01} & \textbf{1876.01} & \textbf{82.5} & \textbf{3.17} & $\mathbf{2.27\pm0.10}$ & \textbf{1} \\
\emph{\textbf{7199397.01}}* & \emph{\textbf{75.01}} & \emph{\textbf{105.9}} & \emph{\textbf{2.61}} & $\mathbi{2.22}\mathbf{\pm}\mathbi{0.11}$ & \emph{\textbf{1}} \\
\textbf{11297236.01} & \textbf{1857.01} & \textbf{88.6} & \textbf{2.10} & $\mathbf{2.19\pm0.09}$ & \textbf{1} \\
\emph{\textbf{9349482.01}}* & \emph{\textbf{2020.01}} & \emph{\textbf{111.0}} & \emph{\textbf{3.72}} & $\mathbi{2.17}\mathbf{\pm}\mathbi{0.22}$ & \emph{\textbf{1}} \\
\textbf{10810838.01} & \textbf{174.01} & \textbf{56.4} & \textbf{2.04} & $\mathbf{2.15\pm0.16}$ & \textbf{1} \\
10471621.01 & 2554.01 &  39.8 & 3.19 & $2.10\pm0.33$ & 2 \\
\textbf{7761545.01} & \textbf{1472.01} & \textbf{85.4} & \textbf{2.60} & $\mathbf{2.04\pm0.35}$ & \textbf{1} \\
8686097.01  &  374.01 & 172.7 & 2.09 & $1.92\pm0.07$ & 1 \\
12121570.01 & 2290.01 &  91.5 & 2.73 & $1.88\pm0.07$ & 1 \\
9661979.01  & 2132.01 &  69.9 & 2.11 & $1.87\pm0.14$ & 1 \\
2449431.01 & 2009.01 &  86.7 & 3.82 & $1.81\pm0.25$ & 1 \\
2443393.01  & 2603.01 &  73.7 & 2.46 & $1.79\pm0.07$ & 1 \\ 
5526717.01  & 1677.01 &  52.1 & 2.65 & $1.77\pm0.16$ & 2 \\
10027247.01 & 2418.01 &  86.8 & 3.66 & $1.70\pm0.08$ & 1 \\
11037335.01 & 1435.01 &  40.7 & 2.22 & $1.54\pm0.07$ & 2 \\
12400538.01 & 1503.01 &  76.1 & 2.00 & $1.20\pm0.22$ & 1 \\
10015937.01 & 1720.01 &  59.7 & 3.98 & $1.07\pm0.07$ & 1 \\
11656918.01 & 1945.01 &  82.5 & 2.31 & $1.06\pm0.08$ & 2 \\
11176127.03 & 1430.03 &  77.5 & 3.18 & $1.05\pm0.10$ & 3 \\
8611781.01  & 2185.01 &  77.0 & 2.34 & $1.05\pm0.11$ & 1 \\
8892157.02  & 2224.02 &  86.1 & 2.92 & $1.01\pm0.08$ & 2 \\
6765135.01  & 2592.01 & 175.6 & 2.53 & $0.80\pm0.05$ & 1 \\
8758204.01 & 2841.01 & 159.4 & 3.49 & $0.75\pm0.08$ & 1 \\ 
9030537.01 & 1892.01 &  62.6 & 4.18 & $0.71\pm0.10$ & 1 \\
8240904.02  & 1070.02 & 107.7 & 2.66 & $0.40\pm0.09$ & 3 \\
8240904.03  & 1070.03 &  92.8 & 2.66 & $0.40\pm0.09$ & 3 \\ [1ex]
\hline\hline %inserts single line
\end{tabular}
\label{tab:allTSAs} % is used to refer this table in the text
\end{table*}

\subsection{Target Selection Prioritization (TSP)}
\subsubsection{Overview}

The TSA algorithm has identified 35 KOIs as being suitable for an exomoon
analysis. With this more manageable number, we can apply some more time intensive
selection criteria as part of the target selection prioritization (TSP)
process. In this work, we consider three TSP criteria, which sequentially
increase in time requirements to evaluate:

\begin{itemize}
\item[{\tiny$\blacksquare$}] KOI must be in a single-transiting system
\item[{\tiny$\blacksquare$}] SNR should hold-up when queried from the MAST 
archive
%\item[{\tiny$\blacksquare$}] How much data is available (both long and short 
%cadence)
\item[{\tiny$\blacksquare$}] KOI should not exhibit excessive time-correlated
noise
%\item[{\tiny$\blacksquare$}] ... and if not, do the targets exhibit TTVs?
%\item[{\tiny$\blacksquare$}] Any relevant publications on this KOI already?
%\item[{\tiny$\blacksquare$}] Is the target in the HZ?
\end{itemize}

We discuss each of these three criteria in the following subsections.

\subsubsection{Multiplicity}

The first criterion eliminates KOIs which would require a more complicated and
involved analysis due to their multiple nature. Multiples should induce
transit timing variations (TTVs) on one another, which is also a signature of
exomoons. Eliminating these KOIs does not eliminate the possibility of
planet-induced TTVs by any means (as recently demonstrated by the 
counter-example of KOI-872 \citealt{nesvorny:2012}), but it does make our task
simpler. Later HEK surveys may relax this constraint. Of the 35 TSAs, 11 were 
found to reside in multiple transiting systems and were rejected. 

\subsubsection{Cross-referencing CDPPs}

The TSA algorithm works by reading in a list of planet and star parameters
for each KOI. The major source for such parameters comes from 
B12. This work also includes estimates of the CDPP over
6 hours timescale, which TSA uses to estimate the SNR. During our investigation, 
we noticed several cases where the CDPP values reported in the tables of 
B12 did not agree with those reported by MAST when queried.
We therefore decided to cross-reference the SNRs calculated from the 
B12 CDPP to those given by MAST.

In addition to the CDPP values differing between B12 and MAST,
KOIs with a low $T_{\mathrm{eff}}$ host star may yield unreliable $R_{\star}$
estimates for reasons discussed in detail in \citet{muirhead:2012}. These 
authors provide improved $R_{\star}$ estimates for such systems, which we use 
where available to compute a revised SNR. Since MAST provides the CDPP values 
for each quarter, we evaluate the mean and standard deviation of the SNR across 
all LC quarters.

We find that 12 of the remaining 24 KOIs have a mean SNR$<2.0$ when we used the
revised values and these candidates are summarily rejected. This leaves
us with 12 KOIs.

\subsubsection{Removing KOIs with excessive time-correlated noise}

With 12 KOIs remaining, we are now ready to consider the most time-consuming
TSP test. In the limit of a perfectly well-behaved star and instrument, the
noise should be purely due to photon noise and thus behave as a Poisson
distribution. Since the number of photons is large, the noise is very well
described as a Gaussian distribution and has no frequency dependency; so-called
``white noise''.

Time-correlated noise refers to noise which has the property that the 
probability distribution of values for a given measurement is not independent of
previous measurements. This is a problem for the HEK project since 
time-correlated noise can mimic dips, bumps and distortions due to an exomoon. 
Whilst many methods exist to tackle time-correlated noise, they require various 
assumptions about the data's behavior and invariably greater computational 
overhead. Since a significant fraction of \emph{Kepler's} targets have their 
photometry dominated by uncorrelated noise \citep{jenkins:2010}, the simplest 
strategy to deal with time-correlated noise is reject any KOIs exhibiting an 
excess on the timescale of interest. On timescales of days to weeks, one 
invariably finds time-correlated flux modulations, which could be considered a 
form of time-correlated noise, typically due to focus drift or stellar rotation.
However, the timescale of interest for transiting exomoons is of 
order-of-magnitude one hour. Therefore, these long term variations do not affect
our analysis and should be detrended out appropriately 
(see \S\ref{sec:detrending}).

At this timescale of interest, it makes no difference to us whether excessive
time-correlated noise is of instrumental or stellar origin since we have no
intention of attempting to correct for it. Instead, our strategy is simply to
reject all candidates showing excessive time-correlated noise. The question 
then becomes, what do we define as excessive time-correlated 
noise?

There is a dizzying number of metrics at our disposal for this task and we
here seek a simple, computational efficient expression. A classic metric is
the \citet{durbin:1950} statistic, $d$, which uses autocorrelation to test 
whether a time series is positively or negatively autocorrelated. The 
Durbin-Watson statistic is given by

\begin{align}
d &= \frac{ \sum_{i=2}^N (r_i - r_{i-1})^2 }{ \sum_{i=1}^N r_i^2 },
\label{eqn:durbin}
\end{align}

where $r_i$ are the residuals and $N$ is the number of data points.
The value of $d$ always lies between 0 and 4, with 2 representing an absence of
autocorrelation, $d<2$ representing positively-autocorrelated noise (expected
for instrumental/astrophysical sources) and $d>2$ representing 
negatively-autocorrelated noise (anomalous and unphysical; we do not expect to
see a significant excess of this). 

In calculating $d$, there exists a degree of freedom regarding over what 
cadence should we evaluate the statistic i.e. what timescale do we consider
most relevant for an exomoon search? An exomoon transit or distortion could 
occur on a timescale of a few minutes or a few hours and so we selected 
30\,minutes for the simplicity that no binning is required for LC
data and that the timescale is consistent with exomoon features. The next 
question is what value of $d$ should one consider acceptable?

Although test statistics are available via lookup tables for
instance, these statistics assume regularly spaced time series which we often
do not have for \emph{Kepler} data, mostly due to outlier rejections and data 
gaps. Instead of using such statistics, we generate 1000 Monte Carlo simulations 
of Gaussian noise for the exact time sampling of a given data set to reproduce 
the expected posterior distribution of $d$ for data with no time-correlated 
noise. The Gaussian noise is generated assuming each data point is described by 
a Gaussian distribution with a standard deviation given by its associated 
uncertainty and a mean of unity.

The quantity $d$ is only calculated on data locally surrounding transits to 
within twice the timescale of the Hill sphere, $2 T_{\mathrm{Hill}}$ (with the 
planetary transit itself excluded), since only this data is relevant for our 
exomoon hunt. We compute $d$ for each and every transit event and then take the 
mean over all transit epochs, $\bar{d}$. The same process is applied to the 1000 
synthetic time series, which behave as Gaussian noise with the exact same 
cadence and time sampling as the original data. We also apply the same 
final-stage local linear detrending used on the real data to every synthetic 
data set (as is described later in \S\ref{sec:detrending}). The 1000 synthetic 
time series are converted into 1000 $\bar{d}$ metrics in the same as the 
original data and this is used to compute a probability distribution of 
$\bar{d}$ in the case of Gaussian noise.

We then compare the real $\bar{d}$ metric with the simulated distribution to 
evaluate whether our data set is consistent with a lack of autocorrelated noise. 
Any KOIs which show $>3$\,$\sigma$ autocorrelation (as determined by the 
$\bar{d}$ metric) in both the PA (Photometric Analysis) and PDC (Pre-search Data
Conditioning) detrended LC data are rejected (SC data is not used in this 
selection phase but is used later). We do not anticipate that this will remove
genuine moon signals since such events would be temporally localized rather
whereas autocorrelation at a 30\,minute timescale must be present throughout
the entire time series (for the particular transit epoch under analysis).
After applying this test to each target, we find that only 7 of the 12 remaining 
KOIs pass this test, as listed in Table~\ref{tab:durbins}. Note that all 12 KOIs 
were fully detrended in exactly the same way, as is described in 
\S\ref{sec:detrending}. However, only 7 of these 12 are actually fitted with a 
transit light curve model - the most resource intensive stage of the entire 
process.

\begin{table}
\caption{\emph{Durbin-Watson ($d$) statistics of TSAs. Each vector displays two
numbers; the first for the PA data and the second for the PDC-MAP. \emph{Kepler}
data usually yields $d<2$ indicating positive serial correlation, as would be
expected for some hidden systematic error source. Rows in italics are those
rejected for having excessive autocorrelation.
}} % title of Table
\centering % used for centering table
\begin{tabular}{l c c} % centered columns (9 columns)
\hline
KOI & $\bar{d}$ of LC & (1-FAP) of autocorrelation ($\sigma$) \\ [0.5ex] % inserts table
%heading
\hline
\emph{KOI-364.01}	& \emph{-} & \emph{-} \\
KOI-303.01		& $\{1.922,1.893\}$ & $\{2.0,2.6\}$   \\
\emph{KOI-974.01}	& $\emph{\{1.423,1.498\}}$ & $\emph{\{11.9,10.8\}}$ \\
\emph{KOI-268.01}	& $\emph{\{1.446,1.415\}}$ & $\emph{\{9.5,10.5\}}$  \\
KOI-1472.01		& $\{1.951,1.945\}$ & $\{1.4,1.4\}$   \\
KOI-722.01		& $\{1.951,1.944\}$ & $\{1.7,1.9\}$   \\
KOI-365.01		& $\{1.905,1.847\}$ & $\{2.2,3.1\}$   \\
KOI-174.01		& $\{1.947,1.999\}$ & $\{1.5,0.6\}$   \\
\emph{KOI-75.01}	& $\emph{\{1.422,1.471\}}$ & $\emph{\{13.6,13.4\}}$ \\
\emph{KOI-2020.01}	& $\emph{\{1.643,1.633\}}$ & $\emph{\{6.4,6.7\}}$   \\
KOI-1857.01 		& $\{1.901,1.902\}$ & $\{2.3,2.3\}$ \\
KOI-1876.01		& $\{1.909,1.930\}$ & $\{0.5,0.4\}$ \\ [1ex]
\hline\hline %inserts single line
\end{tabular}
\label{tab:durbins} % is used to refer this table in the text
\end{table}

\section{DETRENDING THE DATA WITH \cofiam}
\label{sec:detrending}

%% DETRENDING
%%

Data is detrended using a custom algorithm which we dub Cosine Filtering with
Autocorrelation Minimization (\cofiam), which is described in this section.

\subsection{Pre-Detrending Cleaning}

In all cases, we performed the detrending procedure twice; once for the
PA data and once for the PDC-MAP data. In what follows, each \emph{transit} is 
always analyzed independently of the others i.e. we obtain a detrended light
curve unique to each transit event, not each quarter. The first step is to 
visually inspect each quarter and remove any exponential ramps, flare-like 
behaviours and instrumental discontinuities in the data. We make no attempt to 
correct these artefacts and simply exclude them from the photometry manually. We 
then remove all transits using the B12 ephemerides and clean 
the data of 3\,$\sigma$ outliers from a moving median smoothing curve with a 
20-point window (for both LC and SC data). 

\subsection{Cosine Filtering with Linear Minimization (\cofiam)}

The remaining unevenly spaced data is then regressed using a discrete series of 
harmonic cosine functions, which act as a high-pass, low-cut filter 
\citep{ahmed:1974}. The functional form is given by

\begin{align}
f_k(t_i) = a_0 + \sum_{k=1}^{N_{\mathrm{order}}} \Bigg[ x_k \sin\Big(\frac{2\pi t_i k}{2D} \Big) + y_k \cos\Big(\frac{2\pi t_i k}{2D} \Big) \Bigg],
\label{eqn:linearcosinefilter}
\end{align}

where $D$ is the total baseline of the data under analysis, $t_i$ are the time
stamps of the data, $x_k$ \& $y_k$ are model variables and $N_{\mathrm{order}}$ 
is the highest harmonic order. Equation~\ref{eqn:linearcosinefilter} may be
more compactly expressed as a cosine function with a phase term, but the above
format illustrates how the equation is linear with respect to $x_k$ and $y_k$.
This means that we can employ weighted linear minimization, which is not only 
computationally quicker than non-linear methods, but also guaranteed to reach
the global minimum. In our regression, the data are weighted by the inverse of 
their reported standard photometric errors.

Harmonic filtering has been previously used to correct CoRoT (e.g. 
\citealt{mazeh:2010}) and \emph{Kepler} light curves (e.g. 
\citealt{kipbak:2011a}; \citealt{kipbak:2011b}; \citealt{darkest:2011}) and is 
attractive for its simplicity, computational efficiency and ability to preserve 
the transit shape.

\subsection{Frequency Protection}

There are many possible choices of $N_{\mathrm{order}}$, but above a certain
threshold the harmonics will start to appear at the same timescale as the
transit shape and thus distort the profile, which is undesirable. The transit
light curve of a planet can be considered to be a trapezoid to an excellent 
approximation, for which an analytic Fourier decomposition is available.
\citet{waldmann:2012} showed that a equatorial trapezoidal transit light curve 
is described by the following Fourier series:

\begin{align}
f(t) = 8 \sqrt{2} \delta \Bigg( \sin(1/T_{14}) + \frac{\sin(3/T_{14})}{9} - \frac{\sin(5/T_{14})}{25} - ... \Bigg).
\end{align}

Under the approximation of a trapezoidal light curve then, the lowest frequency
is thus $(1/T_{14})$. Another way of putting this is that the highest 
periodicity is $T_{14}$ (i.e. the transit duration) and so if we protect this
timescale, and all shorter timescales, the transit light curve should be
minimally distorted. Let us therefore choose to protect a timescale $x T_{14}$,
where $x$ is a real number greater than unity and $T_{14}$ is the 
first-to-fourth contact duration reported in B12. The timescale
$x T_{14}$ can be protected by imposing

\begin{align}
N_{\mathrm{order}} &= \frac{2 D}{4 x T_{14}}.
\end{align}

Ideally, one would wish to impose $x\gg1$ in all cases to provide some cushion,
but in reality such a condition means $N_{\mathrm{order}}$ is small and the 
ability of the regression algorithm to obtain a reasonable fit to the data 
becomes poor. In contrast, going to higher values of $N_{\mathrm{order}}$ leads 
to a better regression in the $\chi^2$ sense, but increases the risk of higher 
harmonics distorting the transit profile since strictly speaking we 
require $x\gg1$. Therefore, there exists a trade-off between these two effects 
and one might expect an optimal choice of $x$ to exist for any given data set.
Selecting such an optimum requires a quantitative metric which we aim to optimize.

\subsection{Autocorrelation Minimization}

For any optimization problem, one must first define what it is we wish to 
optimize or minimize. In this work, we identify the primary objective to be that 
the transit light curve contains the lowest possible degree of time-correlated 
noise around each transit, which could lead to false-positive moon signals. We 
therefore require some metric to quantify the amount of autocorrelation and 
optimize against. As discussed earlier, the Durbin-Watson statistic is a useful 
tool to this end and evaluates the degree of first-order autocorrelation in a 
time series. Our objective is therefore to choose a value of $x$ such that the 
Durbin-Watson statistic is consistent with the lowest quantity of 
autocorrelation, when evaluated on the data surrounding the planetary transit. 
Another way of putting this is that we wish to choose the value of 
$N_{\mathrm{order}}$ which minimizes $(d-2)^2$, when evaluated on the data 
within $2 T_{\mathrm{Hill}}$ of the time of transit minimum (where 
$T_{\mathrm{Hill}}$ is the Hill timescale).

Before we can begin our optimization search, one must define the allowed range
of $N_{\mathrm{order}}$ through which we can search. Recall that our expressions
protect a timescale $x T_{14}$ and thus one must choose $x>1$ in order to
not disturb the transit profile. Consequently, we chose the lowest allowed value 
of $x$ to be 3, such the timescale $3T_{14}$ is never perturbed. This factor of
three cushion is to allow for transit duration changes, leakage of the harmonics 
for real transits and longer exomoon transits. Protecting this timescale 
corresponds to a maximum allowed value of $N_{\mathrm{order}}$ of

\begin{align}
N_{\mathrm{order,max}} &= \frac{2 D}{12 T_{14}}.
\end{align}

Our detrending algorithm, \cofiam, regresses the \emph{Kepler} time series
to the harmonic series given by Equation~\ref{eqn:linearcosinefilter} in a least
squares sense and repeats this regression for every possible integer choice of
$N_{\mathrm{order}}$ between 1 and $N_{\mathrm{order,max}}$. In this way, we
explore dozens of different regressions which all satisfy the conditions of
providing a good least squares fit to the light curve and protect a timescale 
$\geq 3 T_{14}$. We then simply scan through the final list of $d$ values and
define the optimal detending function to be the harmonic order which minimized 
$(d-2)^2$ when evaluated on the out-of-transit data within $2 T_{\mathrm{Hill}}$
of the eclipse. If a quarter contains more than one transit, we always repeat 
the entire procedure for each transit to ensure the data associated with the
transit is fully optimized for our exomoon hunt.

As before, the $d$ statistic is computed on a timescale of 30\,minutes which
means that no binning is required for the LC data. For the SC data, we bin the
data up the long-cadence data rate.

Once the optimal detrending function has been found, we divide all data within
$2 T_{\mathrm{Hill}}$ (including the transit) by $f_k(t_k)$ to correct for
the long-term variations. We also apply a second outlier rejection of 
10\,$\sigma$ filtering (to allow for unusual anomalies in the transit) from a
moving 5-point median. For short-cadence data, we instead use a 3\,$\sigma$
filtering on a 20-point moving median. In some cases, we relaxed this outlier 
rejection when we felt the filter was removing potential exomoon signals. We 
clip out the data within $\pm2 T_{\mathrm{Hill}}$ and apply a linear fit 
through the out-of-transit data to remove any residual trend, which acts as a 
final normalization. The process is repeated for all transits and the surviving 
light curves are stitched together to form a single input file for our fitting 
code. Any transit epoch with $>0$ data points in the window 
$\tau\pm2T_{\mathrm{Hill}}$ is accepted into the final file. Additionally, the 
final file always uses SC data over LC data, where such data exists.

\section{LIGHT CURVE FITS}
\label{sec:fits}

%% FITS
%%

\subsection{Overview}
\label{sub:fitsoverview}

Model light curves of a transiting planet are generated using the \luna\
algorithm described in \citet{luna:2011}. \luna\ is an analytic photodynamic 
light curve modeling algorithm, optimized for a planet with a satellite. \luna\ 
accounts for auxiliary transits, mutual events, non-linear limb darkening and 
the dynamical motion of the planet and its satellite with respect to the host 
star. In the case of a zero-radius and zero-mass satellite, the \luna\ 
expressions are equivalent to the familiar \citet{mandel:2002} algorithm.

For any given model, we regress the data to the model parameters using the
\multi\ algorithm \citep{feroz:2008,feroz:2009}. \multi\ is a multimodal
nested sampling (see \citealt{skilling:2004}) algorithm designed to calculate
the Bayesian evidence of each model regressed, along with the parameter 
posteriors. By comparing the Bayesian evidences of different models, one may
conduct Bayesian model selection, which has the advantage of featuring a
built-in Occam's razor. For each KOI in our survey, we always regress
the following models as a minimum requirement:

\begin{itemize}
\item[{\tiny$\blacksquare$}] Planet-only model with variable baselines using 
theoretical quadratic limb darkening coefficients; model 
$\mathcal{V}_{\mathrm{P}}$.
\item[{\tiny$\blacksquare$}] Planet-only model with variable baselines 
and free quadratic limb darkening coefficients; model
$\mathcal{V}_{\mathrm{P,LD}}$.
\item[{\tiny$\blacksquare$}] If $\log\mathcal{Z}(\mathcal{V}_{\mathrm{P,LD}}) >
\log\mathcal{Z}(\mathcal{V}_{\mathrm{P}})$, then we repeat 
$\mathcal{V}_{\mathrm{P}}$ fixing the LD coefficients to the maximum 
a-posteriori LD coefficients from model $\mathcal{V}_{\mathrm{P,LD}}$ (the best
LD coefficients will be adopted throughout from this point on), in a fit which
we dub as $\mathcal{V}_{\mathrm{P,MAP}}$.
\item[{\tiny$\blacksquare$}] Planet-only model with flat baseline over all 
epochs, $\mathcal{F}_{\mathrm{P}}$. If 
$\log\mathcal{Z}(\mathcal{F}_{\mathrm{P}}) > 
\log\mathcal{Z}(\mathcal{V}_{\mathrm{P}})$ then we use a flat baseline in all
following fits (to reduce the number of free parameters), which was found to be
always true thanks to \cofiam\ employing a final-stage normalization.
\item[{\tiny$\blacksquare$}] Planet-only model with variables times of transit 
minimum (i.e. TTVs); model $\mathcal{F}_{\mathrm{TTV}}$.
\item[{\tiny$\blacksquare$}] Planet-only model with each transit possessing 
unique transit parameters to allow for both TTVs and TDVs; model
$\mathcal{V}_{\mathrm{V}}$.
\item[{\tiny$\blacksquare$}] Planet-with-satellite fit; model
$\mathcal{F}_{\mathrm{S}}$.
\item[{\tiny$\blacksquare$}] Planet-with-satellite fit assuming a zero-mass 
moon; model $\mathcal{F}_{\mathrm{S,M0}}$.
\item[{\tiny$\blacksquare$}] Planet-with-satellite fit assuming a zero-radius 
moon; model $\mathcal{F}_{\mathrm{S,R0}}$.
\end{itemize}

The question as to why we switch from local baselines to a global baseline is
discussed in the next subsection, \S\ref{sub:planetonly}.
In all fits, we use the same data set throughout, which is usually the PA data.
This is because the PDC-MAP data is subject to numerous detrending processes
that do not necessarily preserve exomoon signals. However, if the PA data yields
a $\bar{d}$ statistic with more than 3\,$\sigma$ confidence of autocorrelation
but the PDC-MAP is below 3\,$\sigma$, then the PDC-MAP data is used in the fits
instead. However, Table~\ref{tab:durbins} reveals how there is no such instance
in the sample of KOIs studied in this work.

\subsection{Planet-only Fits}
\label{sub:planetonly}

The first stage of our fitting process always begins with planet-only fits. The
purpose of these fits is to i) verify or obtain reliable limb darkening 
parameters ii) serve as a baseline for comparison with the planet-with-moon 
fits. Initially, we employ fixed limb darkening coefficients, calculating 
theoretical values from a \citet{kurucz:2006} style-atmosphere integrated over 
the \emph{Kepler}-bandpass. The computation is performed by a code written by I. 
Ribas and associated details can be found in \citet{kipbak:2011a}. 

In these fits, the five basic transit parameters are $p$, $b$, 
$\rho_{\star}^{2/3}$, $P$ and $\tau$. The choice of these five parameters is 
fairly commonplace in the exoplanet literature, except for perhaps 
$\rho_{\star}^{2/3}$. This parameter is used so that the posteriors of 
$\rho_{\star}^{2/3}$ have a uniform prior and can be utilized with the Multibody 
Asterodensity Profiling (MAP) technique, as discussed in \citet{map:2012}. 
Although it is not the purpose of this work to conduct MAP, the posteriors are 
available upon request so that these studies can be facilitated in the future 
without re-executing the light curve fits.

For all KOIs, an estimate of $P$ and $\tau$ exists from 
B12 which we use to define a uniform prior of $\pm 1$\,day
either side of the estimate for both $P$ and $\tau$. The other three basic
parameters also have uniform priors of $0<p<1$, $0<b<2$ and 
$7.6499$\,kg$^{2/3}$\,m$^{-2}$\,$<\rho_{\star}^{2/3}<
6097.85$\,kg$^{2/3}$\,m$^{-2}$ (covering the main-sequence of stars between 
spectral types of M5 to F0 with a factor of 10 cushion at each boundary 
\citealt{cox:2000}). The transit epoch, $\tau$, can be centered on any one of 
the transits in the Q1-9 time series. We choose an epoch which is nearest to the 
median time stamp of Q1-9 data, in order to minimize degeneracy between $P$ and 
$\tau$ in the fits \citep{pal:2009}.

Whether the fit is using a variable baseline ($\mathcal{V}_{\mathrm{P}}$)
or a flat baseline ($\mathcal{F}_{\mathrm{P}}$), we adopt a uniform prior of
$0.95<\mathrm{OOT}<1.05$. For $\mathcal{F}_{\mathrm{P}}$ model, the total
number of free parameters to be marginalized over (and thus explored by
\multi) is just 5+1=6. The $\mathcal{V}_{\mathrm{P}}$ model has 5+$N$ free 
parameters. Experience with \multi\ shows that fitting models with more than 
$\gtrsim$20 free parameters becomes dramatically more time-consuming and so the 
flat baseline model is useful later for exomoon fits with a greater number of 
basic parameters. It is for this reason that we transition from local 
baselines to a global baseline as the model complexity increases.

When fitting for limb darkening parameters, we use quadratic limb darkening
so that only two degrees of freedom are required, yet the curvature of the
light curve can be modelled effectively \citep{claret:2000}. For 
consistency, we employ quadratic limb darkening when we use the fixed limb 
darkening parameters too. We fit for the terms $u_1$ and $u_1+u_2$ since they
are bounded by the physically motivated lower and upper limits of
$0<u_1<2$ and $0<u_1+u_2<1$ (see \citealt{carter:2009} and \citealt{hek:2012}).

In both the planet-only fits, and all subsequent fits, we account for the
integration time of the long-cadence data using the resampling method 
\citep{binning:2010} with $N_{\mathrm{resam}} = 5$ (as generally recommended in 
\citealt{binning:2010}). All \multi\ fits will also employ 4000 live
points, as recommended by \citet{feroz:2009} for evidence calculations.

\subsection{TTV \& TDV Fits}

A transit timing variation (TTV) fit is performed by assigning each transit
epoch a unique time of transit minimum, $\tau_i$. All other parameters are kept
global as before in $\mathcal{F}_{\mathrm{P}}$. This TTV fit is dubbed
$\mathcal{F}_{\mathrm{TTV}}$. The period prior is changed to a Gaussian prior
assigned from the posterior of $P$ from the model $\mathcal{F}_{\mathrm{P}}$
fits. Without a constraining prior on $P$, allowing every transit epoch to
have a unique transit time would mean that $P$ would be fully degenerate with
the $\tau_i$ parameters.

Transit duration variations (TDVs) may be due to velocity changes or 
impact-parameter changes of the observed planet. These changes induce not only
changes in the transit duration, but changes in the derived $a/R_{\star}$ (and 
thus $\rho_{\star}^{2/3}$) and $b$. We also search for changes in the apparent 
$p$ value due to spot activity or exomoon mutual events, for example. 
Additionally, TDVs are expected to occur in dynamic systems exhibiting TTVs 
too. With so many degrees of freedom required, the fits would certainly involve 
a large number of free parameters making the regression very time consuming with
\multi. To solve this, one may paradoxically increase the number of degrees
of freedom again by allowing for variable baselines (OOT). By doing so, all
six transit parameters are independent for each epoch (i.e. there are no fitted
global parameters) and thus the fits can be conducted on each epoch separately 
and then the sum of the log Bayesian evidences will give the global log Bayesian 
evidence. These individual transit fits are very fast to execute and may be run 
simultaneously. For these reasons, we dub the regression 
$\mathcal{V}_{\mathrm{V}}$ meaning variable transit parameters for each epoch and 
the subscript V denotes variable baselines too.

TDVs may be defined in several ways, unlike TTVs which have less ambiguous
definition. In this paper, we define the TDVs to be the variation of the
parameter $\tilde{T}_{B*}/2$, where $\tilde{T}_{B*}$ is the duration for the
planet's center to enter to stellar disc and subsequently leave. We use
this definition rather than the first-to-fourth contact duration, for example,
since $\tilde{T}_{B*}$ has the lowest relative uncertainty when the limb
darkening coefficients are fixed \citep{carter:2008}. We divide this duration
by two since the theoretical uncertainties on $\tilde{T}_{B*}$ are exactly
twice that of $\tau$ \citep{carter:2008} and thus our derived TTVs and TDVs
should exhibit similar scatter and scale, which makes for useful comparisons.

\subsection{Planet-with-Moon Fits}

In general, we make the following assumption: exomoons may be randomly oriented
but have nearly circular orbits due to tidal dissipation (e.g. see 
\citealt{porter:2011,heller:2013}). Therefore, our survey-mode fits do not 
consider fitting for the orbital eccentricity of the exomoon. We find that this 
dramatically improves the speed and stability of our fits using \multi.

In contrast to exomoons, there is no reason to expect exoplanets to have
zero eccentricity, especially at long periods, due to the much longer
circularization timescales and the possibility for planet-planet forcing or
Kozai migration. Despite this, we choose to assume a circular orbit
in the survey light curve fits in this paper. The advantage of doing so
is firstly to save \multi\ exploring an additional two free parameters 
and secondly to save solving Kepler's equation numerically at every
time stamp. 

We justify our choice on the basis that the maximum stable orbital
separation of a moon decreases rapidly with respect to the host
planet's orbital eccentricity, as shown by \citet{domingos:2006}.
Here, the authors find the maximum separation scales as $\sim(1-e_P)$.
Further, the maximum stable exomoon mass around a host planet
scales as this separation to the index of $13/2$, as shown by
\citet{barnes:2002}. We therefore expect that the maximum moon mass
around a host planet scales as $\sim(1-e_P)^{13/2}$. On this basis,
an eccentricity of even 0.1 halves the maximum stable exomoon mass
and an eccentricity of 0.3 reduces it by an order-of-magnitude. Future
HEK surveys may explore eccentric planet solutions, but for this paper
computational constraints limit our survey to circular systems for the
reasons discussed.

Another assumption we make is that exomoons orbit in a prograde sense.
The gravitational influence of a satellite induces transit timing variations
(TTV) and velocity-induced transit duration variations (TDV-V) 
\citep{kipping:2009a}. Both of these effects are insensitive to the sense of
the moon's orbital motion. Additionally, satellites induce transit impact
parameter induced transit duration variations (TDV-TIP), which typically have
an amplitude of around an order-of-magnitude less than that of TDV-V. However,
TDV-TIP is sensitive to the sense of orbital motion \citep{kipping:2009b}. 
Therefore, by treating all exomoons as prograde in our survey, a retrograde moon 
would have not its TDV-TIP effect modeled correctly and thus an implicit 
assumption is therefore that TDV-V$\gg$TDV-TIP. As with the eccentricity 
assumption, this allows us to halve the parameter volume to be scanned through 
and thus expedite the fitting procedure.

In addition to the standard planet-with-moon model, we try two ``unphysical'' 
fits where we fix the moon's mass and then the radius to be zero. These fits are 
useful in the vetting stage since a moon detection should not yield an improved 
Bayesian evidence with unphysical properties, such as zero-radius. These fits 
are dubbed $\mathcal{F}_{\mathrm{S,M0}}$ and $\mathcal{F}_{\mathrm{S,R0}}$ for 
the zero-mass and radius cases respectively.

A careful choice of the parameter set and priors is crucial for the moon's
parameters, since the expected signal-to-noise is low and thus priors can be
expected to play an increasingly significant role in the derived results.
In \citet{hek:2012}, we suggested $M_S/M_P$, $R_S/R_P$, $P_S$, $\rho_P^{2/3}$, 
$\phi_S$, $i_S$ and $\Omega_S$ with uniform priors for all. In this work,
we have found that these priors were not fully adequate and our greater
experience has led us to propose a modification. Firstly, $P_S$ is now fitted
with a Jeffrey's prior since it spans several orders of magnitude and the
low periods require dense sampling due to the bunching up of harmonics.
Secondly, we have exchanged $i_S$ for $\cos i_S$ to impose an isotropic prior.
Thirdly, we have exchanged $\rho_P^{2/3}$ for $a_{SP}/R_P$ (the separation
between the planet and moon in units of the planetary radius). This last change
is geometrically motivated and means that \multi\ scans for moon transits
evenly in time and space from the primary planet event. Except for $P_S$, all
terms have uniform priors.

The parameter $\cos i_S$ has the intuitively obvious prior of 
$\mathcal{U}\{-1,1\}$. Similarly, we use $\mathcal{U}\{-\pi,\pi\}$ for
$\phi_S$ and $\Omega_S$ with the exception that the boundary conditions are
periodic and thus the parameters are considered ``wrap-around'' parameters in
\multi. The mass and radius ratios have the very simple boundary conditions of
being between zero and unity i.e. $\mathcal{U}\{0,1\}$.

The upper boundary condition on $P_S$ is given by $P_P/\sqrt{3}$, which
represents the edge of the Hill sphere as proved by \citet{kipping:2009a}
(i.e. $f_{\mathrm{max}}\leq1$). The lower boundary is less obvious and in 
\citet{hek:2012} we proposed $\sim2$\,hours as a rough estimate. Since 
$0<R_S/R_P<1$, the maximum size of the moon is $R_P$. If this is the case, 
then the closest separation allowed before contact would occur is $2R_P$. 
On this assumption, one may derive a lower limit for $P_S$. From 
Equation~7 of \citet{weighing:2010}, one can re-arrange to make $P_S$ the 
subject:

\begin{align}
P_S &= \frac{ (3\pi)^{1/2} (a_{SB}/R_{\star})^{3/2} (1+M_S/M_P) (M_S/M_P)^{1/2} }{ G^{1/2} \rho_S^{1/2} s^{3/2} }.
\end{align}

Substituting $a_{SB}$ with $a_{SP}/(1+M_S/M_P)$ and then replacing $a_{SP}$ with
the minimum allowed value of $2R_P$ we have:

\begin{align}
P_{S,\mathrm{min}} &= \sqrt{\frac{24\pi}{G}} \Big(\frac{p}{1+M_S/M_P}\Big)^{3/2} \frac{(1+M_S/M_P) (M_S/M_P)^{1/2}}{\rho_S^{1/2} s^{3/2}}.
\end{align}

Exploiting the fact $\rho_S = \rho_P (M_S/M_P) (R_S/R_P)^{-3}$ and cleaning
up the expression we find:

\begin{align}
P_{S,\mathrm{min}} &= \sqrt{\frac{24\pi}{G}} \sqrt{ \frac{1}{\rho_P (1+M_S/M_P)} }
\end{align}

The maximum allowed value of $(1+M_S/M_P)$ is 2 and the maximum physically
plausible value of $\rho_P$ is 27950\,kg\,m$^{-3}$ \citep{hek:2012}. This
yields $P_{S,\mathrm{min}} = 0.0520311$\,days (1.25\,hours) which we employ as 
our lower boundary condition for $P_S$.

For $a_{SP}/R_P$, the lower boundary condition is simply 2, which for 
$0<R_S/R_P<1$ guarantees no contact (for a circular satellite orbit). The
maximum requires another small derivation. We consider the maximum to be the
Hill sphere, $R_{\mathrm{Hill}} = a_{B*} [M_P/(3M_{\star})]^{1/3}$; therefore we 
have:

\begin{align}
\Big( \frac{a_{SP}}{R_P} \Big)_{\mathrm{max}} &= \frac{(a_{B*}/R_{\star})}{p} \Big(\frac{M_P}{3 M_{\star}}\Big)^{1/3}.
\end{align}

Replacing the mass terms with densities, we obtain:

\begin{align}
\Big( \frac{a_{SP}}{R_P} \Big)_{\mathrm{max}} &= 3^{-1/3} p \rho_P^{1/3} \rho_{\star}^{-1/3} (a_{B*}/R_{\star}).
\end{align}

If we assume $(M_P+M_S)\ll M_{\star}$, then $\rho_{\star} = (3\pi (a_{B*}/R_{\star})^3)/(G P_B^2)$
which can be substituted into the above expression to give:

\begin{align}
\Big( \frac{a_{SP}}{R_P} \Big)_{\mathrm{max}} &= \frac{ G^{1/3}\rho_P^{2/3} }{ 3^{2/3}\pi^{1/3} } P_B^{2/3}.
\end{align}

To estimate this value, we adopt the maximum allowed planetary density of
$\rho_{P,\mathrm{max}} = 27950$\,kg\,m$^{-3}$ \citep{hek:2012} and
use the maximum a-posteriori value of $P_B$ from the planet-only fits
(technically model $\mathcal{F}_{\mathrm{P}}$). $a_{SP}/R_P$ is then fitted 
with a uniform prior between 2 and this maximum value.

Finally, we instruct \multi\ to ignore any trials which yield an unphysical
density for the planet, which we consider to be between $80$\,kg\,m$^{-3}$ 
and $27950$\,kg\,m$^{-3}$ \citep{hek:2012}. We apply the same constraint to
the satellite except the lower allowed limit is 0\,kg\,m$^{-3}$. This is
imposed so that zero-mass moons can be explored meaning that in the case of
a null-detection the posterior of $M_S/M_P$ can still reach zero. For models, 
$\mathcal{F}_{\mathrm{S,M0}}$ and $\mathcal{F}_{\mathrm{S,R0}}$ we deactivate 
the constraint on the satellite density since the satellite is specifically 
defined to be unphysical.

\subsection{Detection Criteria}
\label{sub:criteria}

An exomoon has never been detected and thus one is forced to seriously consider
what constitutes a ``detection'' in such a new area. The signal-to-noise will
inevitably be at, or close to, the limit of \emph{Kepler}, the signal will
vary in phase and time and may manifest simply as a slight distortion to a
planet's transit profile. If one fits a planet-with-moon model to real data, the 
extra degrees of freedom will inevitably lead to an improved $\chi^2$ relative 
to a planet-only fit. Clearly an improved $\chi^2$ is not sufficient to claim
an exomoon has been found. This concern was one of the driving reasons why the 
HEK project adopted Bayesian model selection \citep{hek:2012} available through 
\multi, since such comparisons implicitly penalize models for using extra 
parameters (Occam's razor). However, even the Bayesian evidence is not a tool 
which can be wielded blindly to claim exomoon detections.

Although our detrending process \cofiam\ minimizes the amount of 
autocorrelation, the data will always possess some quantity of time-correlated
noise. The likelihood function employed by \luna\ is a Gaussian likelihood
expression (see Equation~20 of \citealt{hek:2012}) and so this assumption will 
never be strictly true. It is therefore possible, and in fact quite common, 
that the Bayesian evidence of a planet-with-moon fit will be superior to a 
planet-only fit for an isolated planet, even with the built-in Occam's razor 
of Bayesian model selection. A possible remedy would be to employ a more 
sophisticated likelihood function but the computational demands
of \multi\ make this unrealistic as fits typically take weeks to run even on
modern clusters. Instead, we stress that a superior Bayesian evidence is not
tantamount to a detection with exomoon fits since \luna\ can generate distortions
both in- and out-of-transit which can describe certain time-correlated noise
features.

We therefore consider that a superior Bayesian evidence of a planet-with-moon
fit to a planet-only fit (i.e. $\log\mathcal{Z}(\mathcal{F}_{\mathrm{S}}) 
> \log\mathcal{Z}(\mathcal{F}_{\mathrm{P}})$) is a requirement for a 
``detection'', but not a proof in of itself. As discussed in \citet{hek:2012}, 
we set the significance level of this improvement to be in 4\,$\sigma$ or 
greater in order to qualify. This discussion therefore indicates that other 
detection criteria are required.

One of the easiest tests is that the posteriors should be physically plausible.
For example, as shown in \citet{weighing:2010}, exomoons allow us to measure
the ratio $M_P/M_{\star}$ and which for an assumed $M_{\star}$ yields $M_P$. It 
is easy to check whether the derived $M_P$ is consistent with the derived $R_P$ 
from known planet populations. For example, a candidate planet yielding a 
Jupiter-mass and an Earth-radius can be easily dismissed.

Further, a planet-with-moon model must be superior to both unphysical moon
models considered i.e. the zero-mass moon and the zero-radius moon models.
If the zero-mass moon model is superior, one should suspect starspots or
correlated noise to be responsible. If the zero-radius moon is superior, one
should suspect TTVs from a non-moon origin. In both cases, it is also possible
that the signal-to-noise of the moon signal is presently too low to make a
confirmed detection (but more data may change this).

We also require that both the mass and radius of the exomoon can be considered
independently detected. This means that the posteriors of both $M_S/M_P$
and $R_S/R_P$ must not be converged at zero in the planet-with-moon
fits. These battery of tests form our requirements for an exomoon to be 
\emph{considered} plausible and so far we have described four basic detection
criteria, summarized as:

\begin{itemize}
\item[{\textbf{B1}}] Improved evidence of planet-with-moon fits at 
$\geq 4$\,$\sigma$ confidence
\item[{\textbf{B2}}] Planet-with-moon evidences indicate both a mass and radius 
preference
\item[{\textbf{B3}}] Parameter posteriors are physical (e.g. $\rho_P$, 
$\rho_{\star}$)
\item[{\textbf{B4}}] Mass and radius of moon converge away from zero
\end{itemize}

Should these tests be passed or perhaps a candidate only marginally fails some
of these criteria, we may consider further investigation. We discuss here
three quick general follow-up criteria which can be implemented. Since this
paper's survey only uses Q1-9 data, subsequent \emph{Kepler} data may be
treated as follow-up photometry. One simple check then is that all four
basic criteria are satisfied when the new data is included and the model
refitted. Further, the significance of the moon candidate should be enhanced
by the new data and yield broadly the same set of parameters. Without even
fitting the new data, another simple check is to extrapolate the best-fit
light curve model from the moon hypothesis into the times of the new
observations and compare (in a $\chi^2$ sense) the ``predictive power'' of 
the moon model relative to a simple planet-only model. We summarize these
general follow-up detection criteria below:

\begin{itemize}
\item[{\textbf{F1}}] All four basic criteria are still satisfied when new data 
is included
\item[{\textbf{F2}}] The predicitive power of the moon model is superior (or at
least equivalent) to that of a planet-only model
\item[{\textbf{F3}}] A consistent and statistically enhanced signal is recovered
with the inclusion of more data
\end{itemize}

Even after passing all of these tests, the candidate should not be blindly
accepted as a confirmed detection. Candidate specific tests and follow-up may be 
needed too, if for example the star shows rotational modulations and the 
candidate moon exhibits mutual events (which may in fact be star spot 
crossings). Exploration of perturbing planet solutions causing TTVs/TDVs may be 
needed, as in the case of \citet{nesvorny:2012}. Target specific tests will be 
investigated appropriately should the need arise.

%%%%%%%%%%%%%%%%%%%%%%%%%%%%%%%%%%%%%%%%%%%%%%%%%%%%%%%%%%%%%%%%%%%%%%%%%%%%%%%%

\subsection{Excluded Moons}

In cases of null-detections, one of the aims of the HEK project is to provide
limits on what moons can be excluded. There are many possible choices for which
parameters we provide excluded limits. Two terms of particular interest are 
the mass ratio $(M_S/M_P)$ and the radius ratio ($R_S/R_P$). In general, we find 
the radius ratio to be untrustworthy due to the effects of starspots and
time-correlated noise. Further, the parameter is positively-biased since we
impose a likelihood penalty for high-density (i.e. low-radius) moon solutions.
Therefore, we opt to provide upper limits on the mass ratio $M_S/M_P$ for each 
null-detection.

The excluded limits must be understood in terms of the adopted priors. For 
example, when we posit that there is a 95\% probability of there existing no 
moon of $M_S/M_P>x$, the statement is only meaningful when combined with the 
adopted priors e.g. uniform prior in $a_{SP}/R_P$ and Jeffrey's prior in $P_S$.
This subtle point is important when interpreting the 95\% and 3\,$\sigma$
quoted limits. However, we also make available the full posteriors revealing
all relevant cross-correlations for those wishing to investigate the frequency
of moons in more detail.

One caveat with the provided upper limits is that \multi\ may have located a
spurious signal and spurious detections cannot be used to define upper limits
on $M_S/M_P$. Spurious detections occur because the code attempts to locate the 
best modes which explain the data i.e. the best model fits. In many cases, the 
solution can be dismissed using some of the detection criteria already discussed 
but the derived posteriors of $M_S/M_P$ and/or $R_S/R_P$ still converge to 
non-zero values due to perhaps time-correlated noise or starspots. Simply taking 
the 95\% quantile of these posteriors does not technically translate to an 
excluded upper limit estimate. Indeed these upper limits can approach unity if 
the model fit converged to a binary-planet solution, for example. In such 
spurious detection cases, all we can say for certain is that we are unable to 
detect an exomoon but we caution that meaningful upper limits on $M_S/M_P$ 
and $R_S/R_P$ is not guaranteed for each system analyzed.

In addition to providing limits on $M_S/M_P$, we also compute limits for two 
more observable-centric defined terms, the moon-induced TTV and TDV amplitudes.
Specifically, we calculate the quantiles of the distribution of the 
root-mean-square (r.m.s.) amplitudes of the two effects using the expressions
from \citet{thesis:2011}, $\delta_{\mathrm{TTV}}$ and $\delta_{\mathrm{TDV}}$. 
The TDV r.m.s. amplitude is defined as the sum of the TDV-V and TDV-TIP effects,
for reasons discussed in Chapter 6 of \citet{thesis:2011}. It should be stressed
that excluding a certain moon-induced TTV amplitude does not equate to excluding
a TTV amplitude induced by other effects too (and similarly for TDV).

\section{RESULTS}
\label{sec:results}

\subsection{KOI-722.01}
\label{sub:koi722}

%% KOI-722
%%

\subsubsection{Data selection}

After detrending with \cofiam, the PA and PDC-MAP data were found to have a 
1.7\,$\sigma$ and 1.9\,$\sigma$ confidence of autocorrelation on a 30\,minute 
timescale respectively and therefore both were acceptable ($<3$\,$\sigma$). 
In general, we always prefer to use the raw data and so we opted for the PA data 
in all subsequent analysis of this system. Short-cadence data is available for 
quarter 9 and this data displaced the corresponding long-cadence quarter in our 
analysis.

\subsubsection{Planet-only fits}

% Limb darkening
When queried from MAST, the KIC effective temperature and surface gravity
were reported as $T_{\mathrm{eff}} = 6133$\,K and $\log g = 4.628$ 
\citep{brown:2011}. Using these values, we estimated quadratic limb darkening 
coefficients $u_1 = 0.3694$ and $(u_1+u_2) = 0.6564$ (as described in 
\S\ref{sub:planetonly}). The initial two models we regressed were 
$\mathcal{V}_{\mathrm{P}}$ and $\mathcal{V}_{\mathrm{P,LD}}$ where the former 
uses the aforementioned limb darkening coefficients as fixed values and the 
latter allows the two coefficients to be free parameters. We find that
$\log\mathcal{Z}(\mathcal{V}_{\mathrm{P,LD}}) - 
\log\mathcal{Z}(\mathcal{V}_{\mathrm{P}}) = 0.05\pm0.26$ indicating essentially
no preference between the two models. Given that the data is equally
well-described by either theoretical or fitted coefficients, we opt for
the theoretical limb darkening coefficients since they are more physically
motivated.

% V -> P
KOI-722.01 has a period of $P_P = 46.40630 \pm 0.00022$\,days (as 
determined by model $\mathcal{V}_{\mathrm{P,LD}}$) and exhibits 14 complete 
transits in Q1-Q9 from epochs -8 to +8 (epochs +3, +4 and +7 are missing).
As is typical for all cases, $\log\mathcal{Z}(\mathcal{F}_{\mathrm{P}}) > 
\log\mathcal{Z}(\mathcal{V}_{\mathrm{P}})$ indicating that allowing for 14 
independent baseline parameters is unnecessary relative to a single baseline 
term (thanks to the \cofiam\ final stage normalizations).

\begin{figure*}
\subfigure[\textbf{KOI-722.01} ($-7.2$\,$\sigma$)
\label{fig:KOI722_TTVs}]
{\epsfig{file=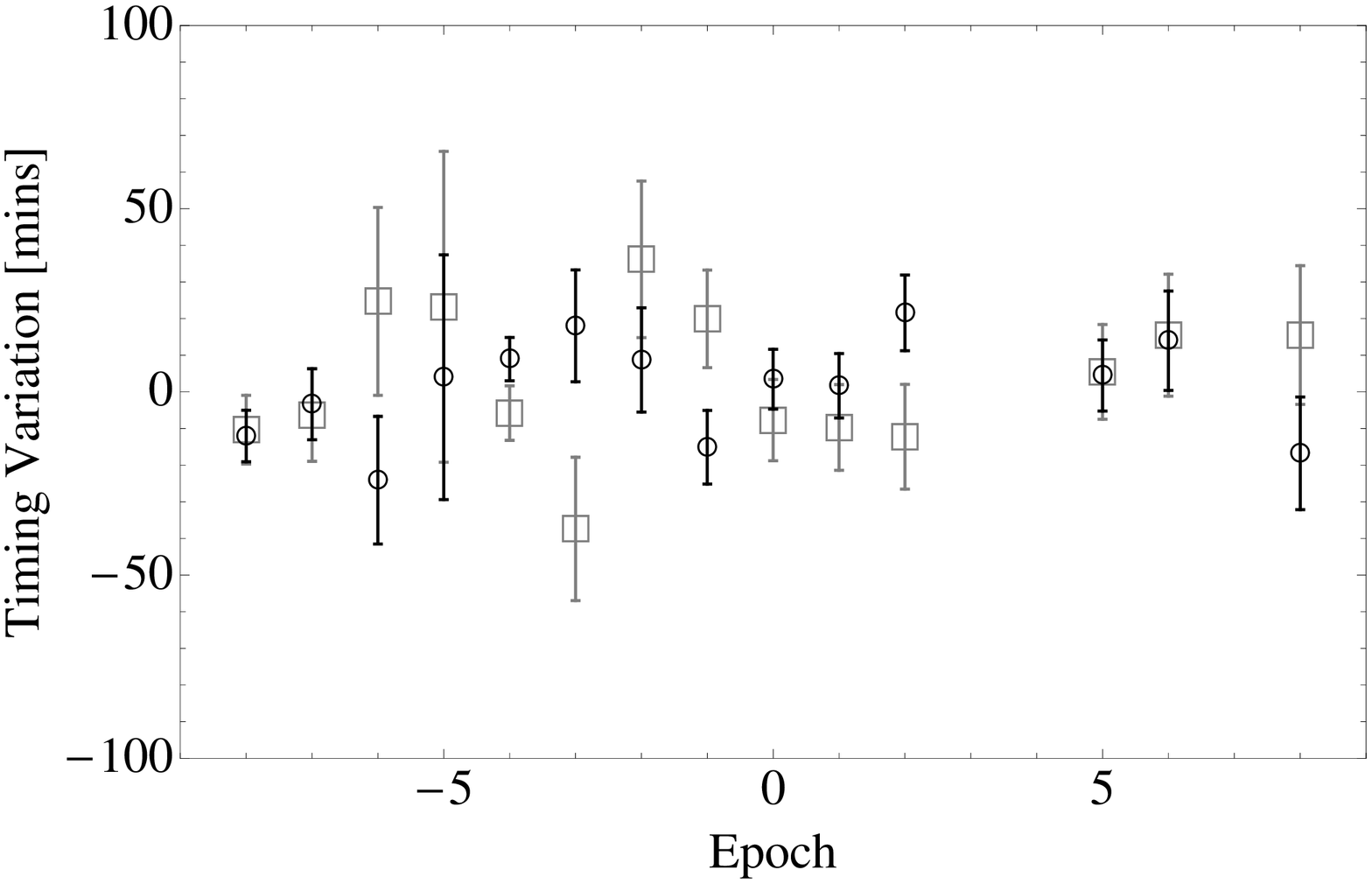,width=8.0 cm}}
\subfigure[\textbf{KOI-365.01} ($-8.3$\,$\sigma$)
\label{fig:KOI365_TTVs}]
{\epsfig{file=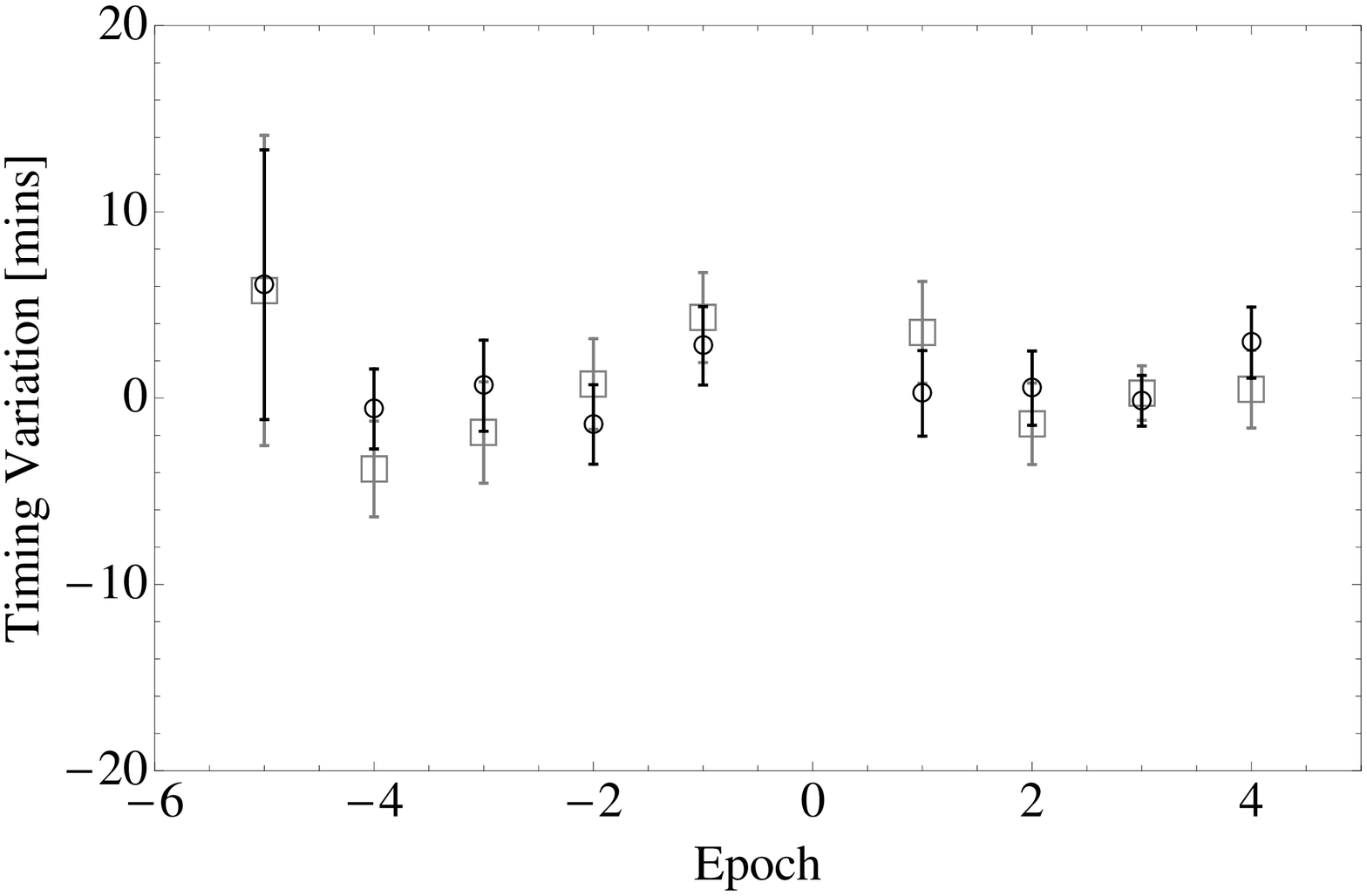,width=8.0 cm}}\\
\subfigure[\textbf{KOI-174.01} ($-6.4$\,$\sigma$)
\label{fig:KOI174_TTVs}]
{\epsfig{file=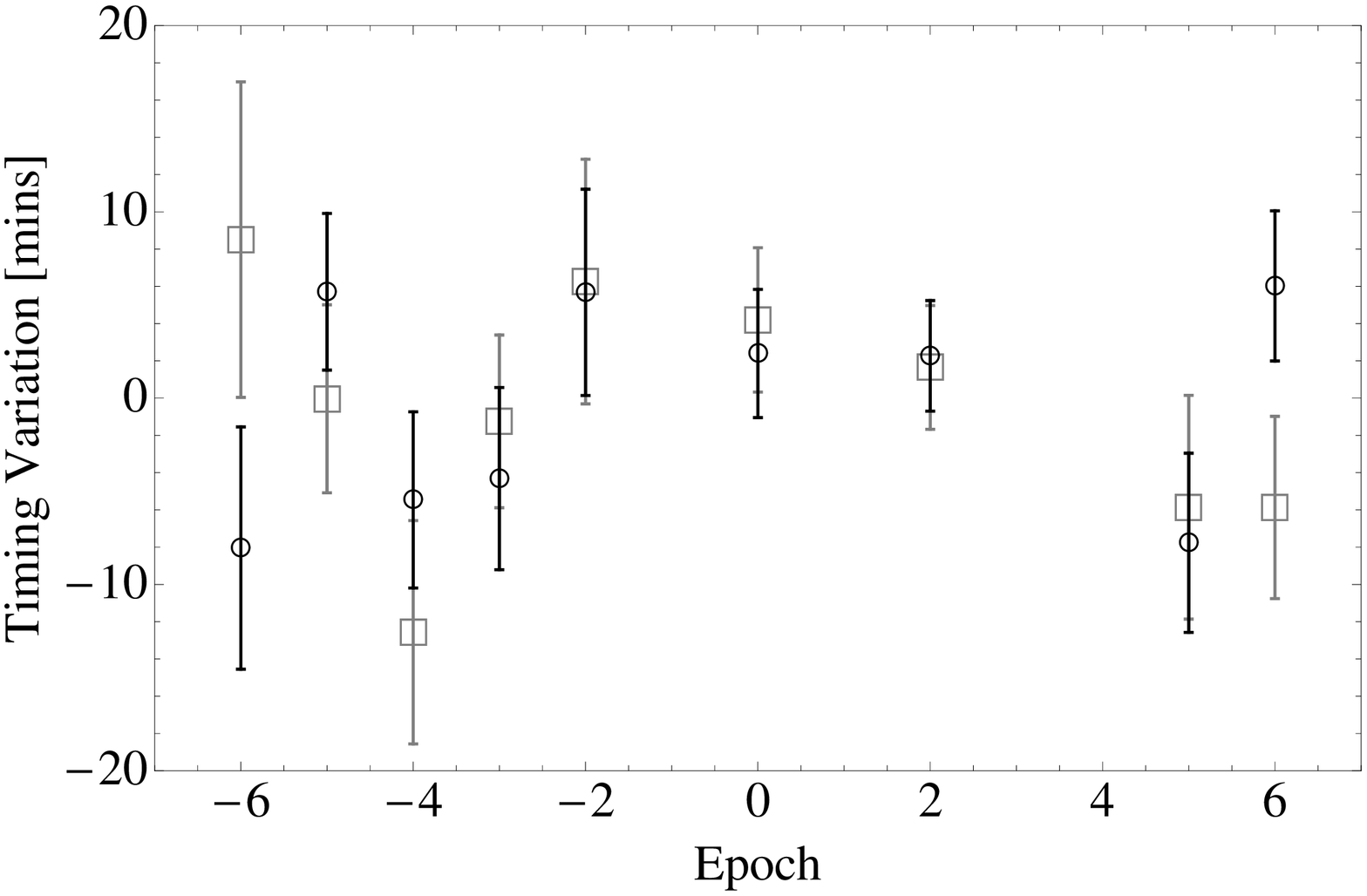,width=8.0 cm}}
\subfigure[\textbf{KOI-1472.01} ($-6.0$\,$\sigma$)
\label{fig:KOI1472_TTVs}]
{\epsfig{file=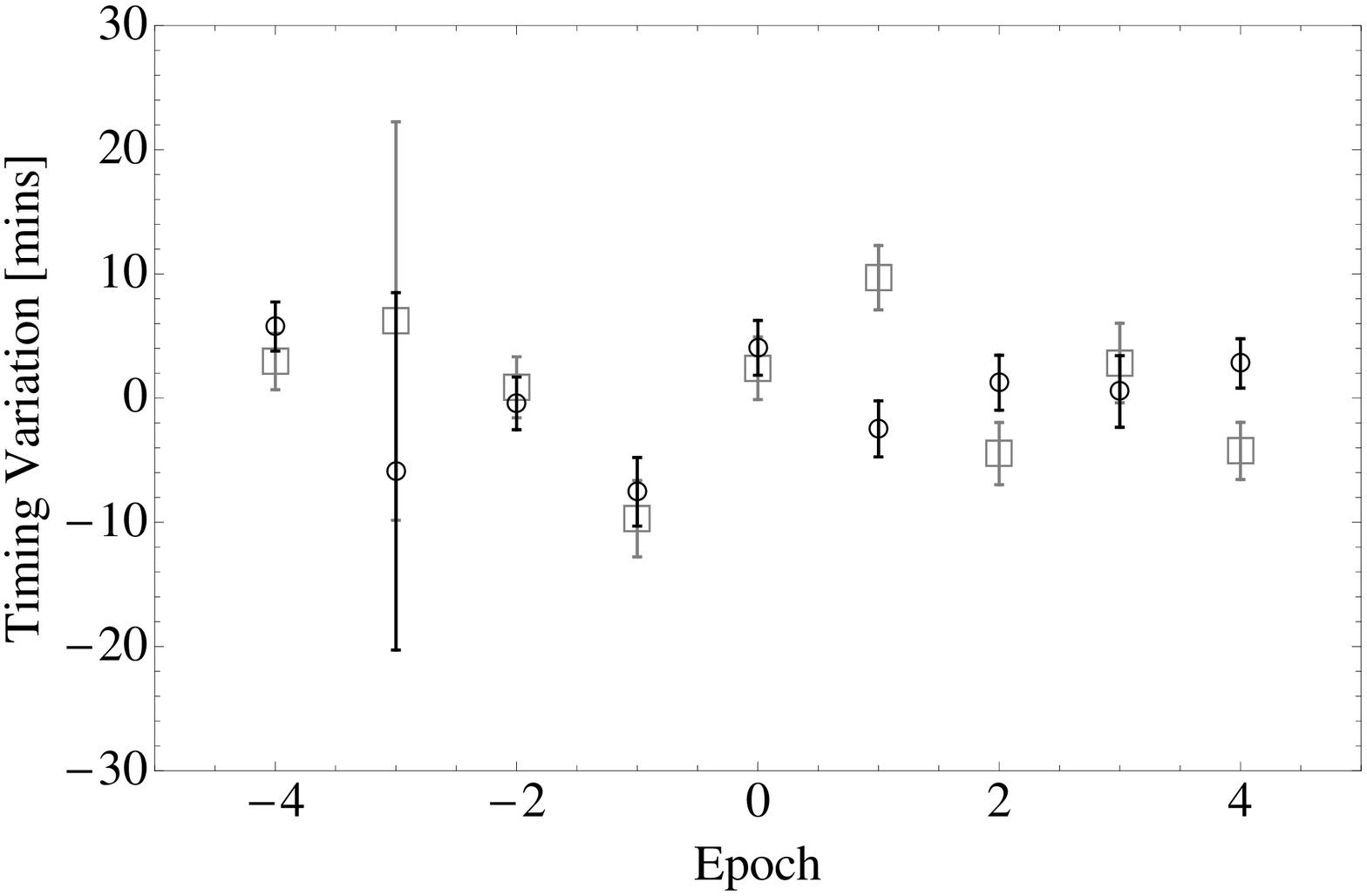,width=8.0 cm}}\\
\subfigure[\textbf{KOI-1857.01} ($-5.8$\,$\sigma$)
\label{fig:KOI1857_TTVs}]
{\epsfig{file=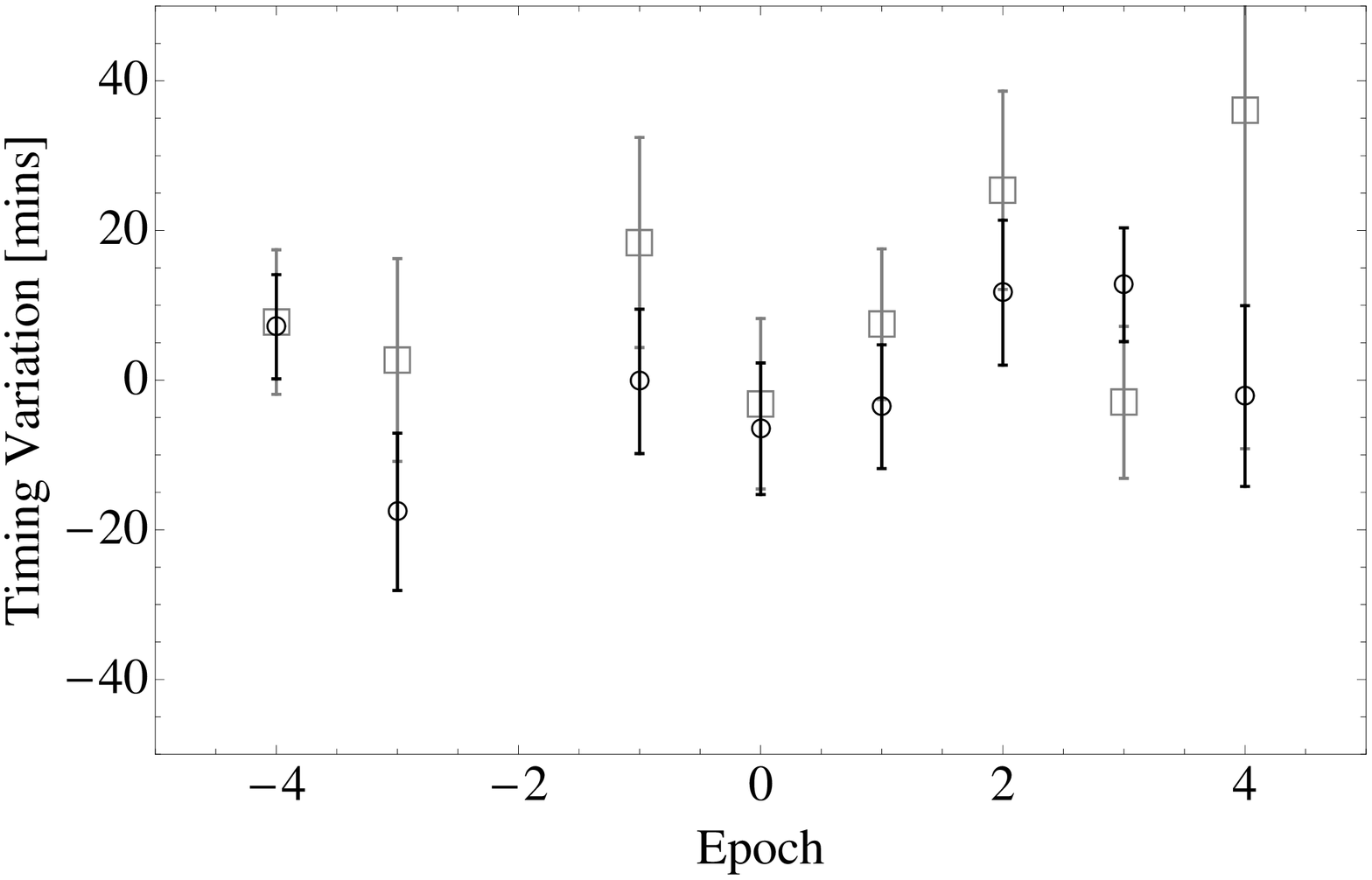,width=8.0 cm}}
\subfigure[\textbf{KOI-303.01} ($-6.6$\,$\sigma$)
\label{fig:KOI303_TTVs}]
{\epsfig{file=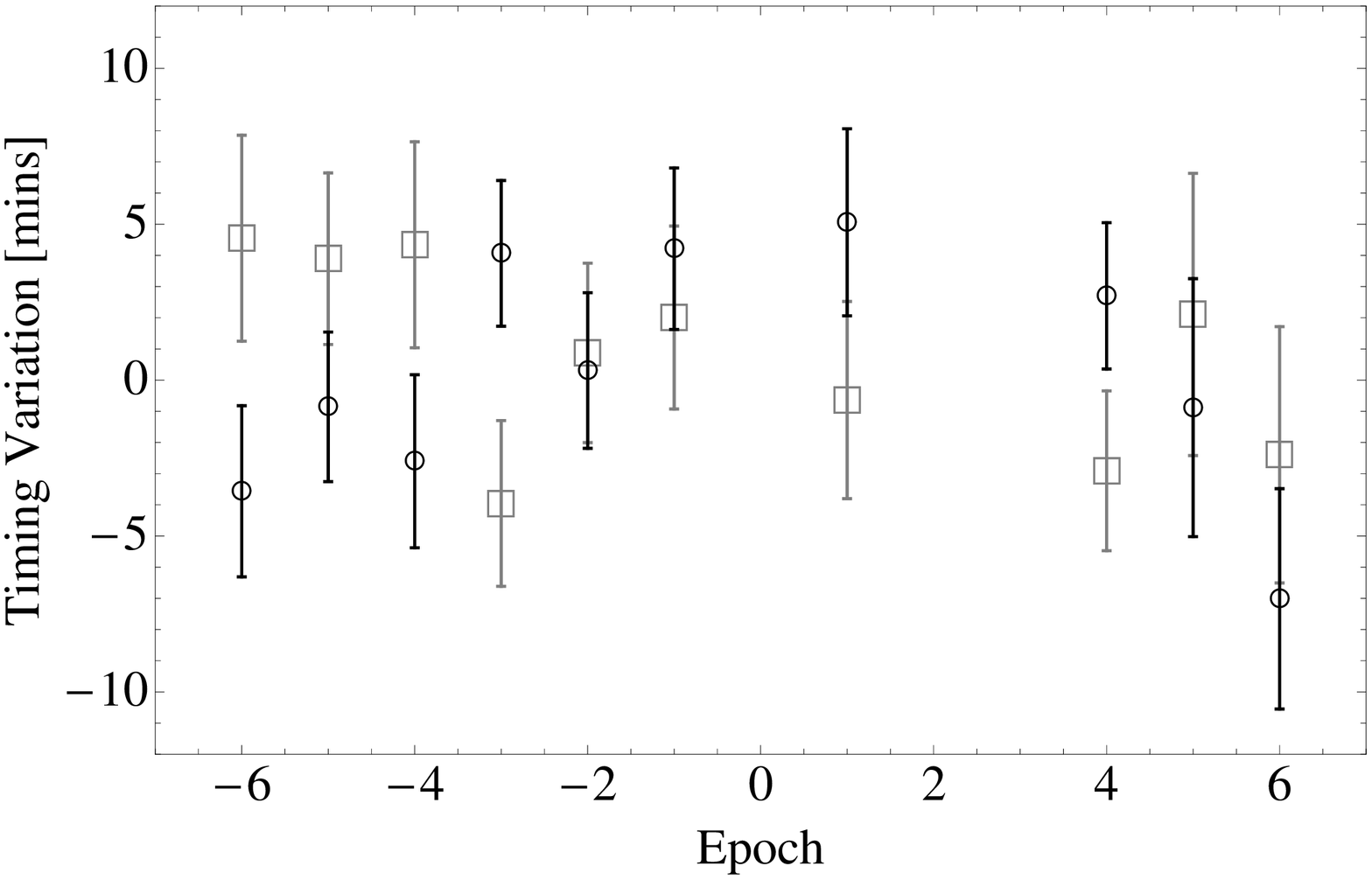,width=8.0 cm}} \\
\subfigure[\textbf{KOI-1876.01} ($-5.5$\,$\sigma$)
\label{fig:KOI1876_TTVs}]
{\epsfig{file=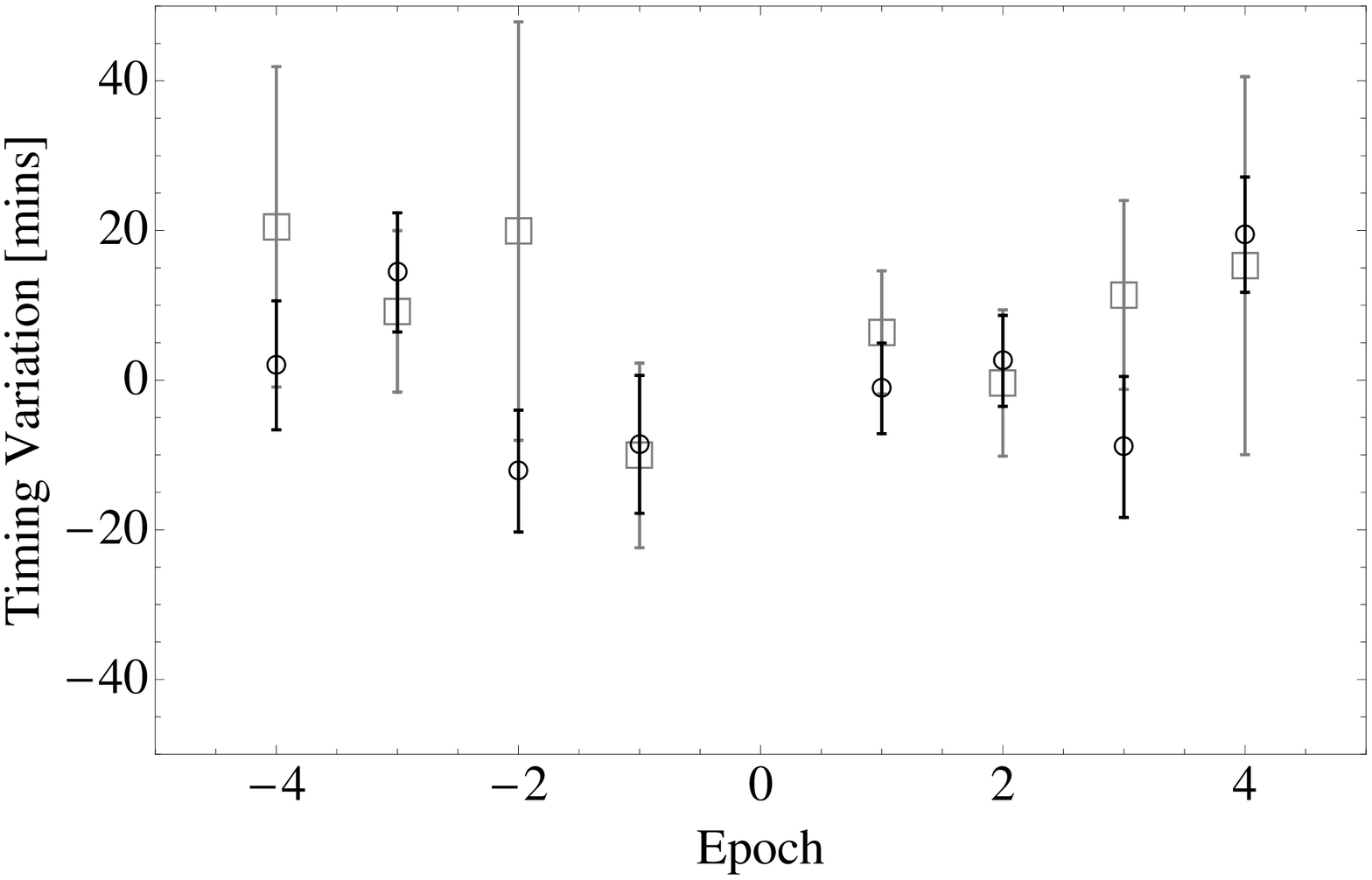,width=8.0 cm}}
\caption{
TTVs (black circles) and TDVs (gray squares) from model fit 
$\mathcal{V}_{\mathrm{V}}$ of the seven KOIs investigated here. TTVs computed 
relative to the maximum a-posteriori ephemeris derived by model 
$\mathcal{F}_{\mathrm{P}}$. TDVs computed relative to the median duration from 
model $\mathcal{F}_{\mathrm{P}}$. After the title of each sub-figure, we provide 
the confidence (in sigmas) of a TTV model over a static model in parentheses for 
each KOI (negative values imply a static model is favored).
\label{fig:TTVs}}
\end{figure*}

% TTV
We find no evidence for TTVs in KOI-722.01, with 
$\log\mathcal{Z}(\mathcal{F}_{\mathrm{TTV}}) 
- \log\mathcal{Z}(\mathcal{F}_{\mathrm{P}}) = -27.82\pm0.18$, which is formally 
an 7.2\,$\sigma$ preference for a static model over a TTV model. The timing 
precision on the 14 transits ranged from 5-17\,minutes (see 
Table~\ref{tab:KOI722_TTVs}). The TTVs, shown in Fig.~\ref{fig:KOI722_TTVs}, 
show no clear pattern and exhibit a standard deviation of 
$\delta_{\mathrm{TTV}} = 12.1$\,minutes and $\chi_{\mathrm{TTV}}^2 = 21.6$ 
for 14-2 degrees of freedom.

% TDV
The TTV+TDV model fit, $\mathcal{V}_{\mathrm{V}}$, finds consistent transit 
times with those derived by model $\mathcal{F}_{\mathrm{TTV}}$. We also find no 
clear pattern or excessive scatter in the TDVs, visible in 
Fig.~\ref{fig:KOI722_TTVs}. The standard deviation of the TDVs is found to be 
$\delta_{\mathrm{TDV}} = 20.1$\,minutes and we determine 
$\chi_{\mathrm{TDV}}^2 = 15.7$ for 14-1 degrees of freedom.

%%% TTV FIGURES

%%% KOI-722 TTV Table
\begin{table*}
\caption{\emph{Transit times and durations for KOI-722.01. The model used to 
calculate the supplied values is provided in parentheses next to each column
heading. BJD$_{\mathrm{UTC}}$ times offset by $2,400,000$ days.
}} % title of Table
\centering % used for centering table
\begin{tabular}{c c c c c c c} % centered columns (3 columns)
\hline
Epoch & $\tau$ [BJD$_{\mathrm{UTC}}$] ($\mathcal{F}_{\mathrm{TTV}}$) & TTV [mins] ($\mathcal{F}_{\mathrm{TTV}}$)
      & $\tau$ [BJD$_{\mathrm{UTC}}$] ($\mathcal{V}_{\mathrm{V}}$) & TTV [mins] ($\mathcal{V}_{\mathrm{V}}$)
      & $\tilde{T}$ [mins] ($\mathcal{V}_{\mathrm{V}}$) & TDV [mins] ($\mathcal{V}_{\mathrm{V}}$) \\ [0.5ex] % inserts table
%heading
\hline
-8 & $54979.5804_{-0.0058}^{+0.0057}$ & $-4.2 \pm 8.3$  
   & $54979.5798_{-0.0050}^{+0.0048}$ & $-12.0 \pm 7.0$
   & $396_{-20}^{+18}$ & $-10.3 \pm 9.4$ \\
-7 & $55025.9889_{-0.0064}^{+0.0073}$ & $-2.2 \pm 9.9$ 
   & $55025.9921_{-0.0074}^{+0.0061}$ & $-3.4 \pm 9.7$
   & $404_{-24}^{+27}$ & $-6 \pm 13$ \\
-6 & $55072.382_{-0.011}^{+0.012}$    & $-22 \pm 17$    
   & $55072.384_{-0.012}^{+0.012}$    & $-24 \pm 17$
   & $466_{-57}^{+46}$ & $+25 \pm 26$ \\
-5 & $55118.8010_{-0.0056}^{+0.0054}$ & $-4.9 \pm 8.0$  
   & $55118.810_{-0.012}^{+0.034}$    & $+4 \pm 33$
   & $464_{-42}^{+127}$ & $+23 \pm 42$ \\
-4 & $55165.2211_{-0.0038}^{+0.0037}$ & $+13.8 \pm 5.4$ 
   & $55165.2198_{-0.0041}^{+0.0041}$ & $+9.0 \pm 5.9$
   & $406_{-16}^{+13}$ & $-5.8 \pm 7.4$ \\
-3 & $55211.6189_{-0.0073}^{+0.0085}$ & $+1 \pm 11$     
   & $55211.6325_{-0.0129}^{+0.0083}$ & $+18 \pm 15$
   & $342_{-32}^{+47}$ & $-37 \pm 20$ \\
-2 & $55258.0267_{-0.0062}^{+0.0056}$ & $+1.7 \pm 8.4$  
   & $55258.0324_{-0.0099}^{+0.0098}$ & $+9 \pm 14$
   & $489_{-47}^{+39}$ & $+36 \pm 21$ \\
-1 & $55304.4253_{-0.0057}^{+0.0058}$ & $-10.4 \pm 8.3$ 
   & $55304.4222_{-0.0067}^{+0.0073}$ & $-15 \pm 10$
   & $457_{-28}^{+26}$ & $+20 \pm 13$ \\
+0 & $55350.8409_{-0.0048}^{+0.0048}$ & $+2.0 \pm 6.9$  
   & $55350.8415_{-0.0054}^{+0.0060}$ & $3.5 \pm 8.2$
   & $402_{-26}^{+18}$ & $-8 \pm 11$ \\
+1 & $55397.2460_{-0.0055}^{+0.0054}$ & $-0.9 \pm 7.8$  
   & $55397.2466_{-0.0057}^{+0.0065}$ & $1.7 \pm 8.8$
   & $398_{-26}^{+20}$ & $-10 \pm 12$ \\
+2 & $55443.6628_{-0.0055}^{+0.0082}$ & $+13.3 \pm 9.9$ 
   & $55443.6668_{-0.0074}^{+0.0069}$ & $+22 \pm 10$
   & $393_{-30}^{+28}$ & $-12 \pm 14$ \\
+5 & $55582.8744_{-0.0057}^{+0.0059}$ & $-0.3 \pm 8.4$  
   & $55582.8741_{-0.0070}^{+0.0064}$ & $+4.5 \pm 9.7$
   & $428_{-25}^{+27}$ & $+5 \pm 13$ \\
+6 & $55629.2800_{-0.0040}^{+0.0042}$ & $-2.4 \pm 5.9$  
   & $55629.2870_{-0.0074}^{+0.0114}$ & $+14 \pm 14$
   & $448_{-26}^{+41}$ & $+15 \pm 17$ \\
+8 & $55722.0752_{-0.0074}^{+0.0065}$ & $-30 \pm 10$    
   & $55722.0784_{-0.0089}^{+0.0124}$ & $-17 \pm 15$
   & $448_{-34}^{+42}$ & $+16 \pm 19$ \\ [1ex]
\hline\hline %inserts single line
\end{tabular}
\label{tab:KOI722_TTVs} % is used to refer this table in the text
\end{table*} % title of Table

\subsubsection{Moon fits}

A planet-with-moon fit, $\mathcal{F}_{\mathrm{S}}$, is preferable to a 
planet-only fit at a formally high significance level (7.5\,$\sigma$), 
passing detection criteriopn B1 (see Table~\ref{tab:KOI722_evidences}). 
KOI-722.01 fails detection criterion B2 though, since a zero-mass moon 
model fit yields a higher Bayesian evidence at 2.8\,$\sigma$ confidence.

Investigating further, one finds the parameter posteriors to be ostensibly
unphysical. Physical parameters of the candidate solution may be estimated
by combining our posteriors with the estimated stellar parameters of
B12 ($M_{\star}=1.08$\,$M_{\odot}$ and 
$R_{\star}=0.83$\,$R_{\odot}$). The moon is found to lie at a highly inclined 
orbit and exhibit physically consistent parameters of 
$M_S=1.13_{-0.78}^{+1.48}$\,$M_{\oplus}$ and $R_S=0.906_{-0.055}^{+0.055}$\,
$R_{\oplus}$. In contrast, the planet has a very high density with parameters 
$M_P=245_{-75}^{+98}$\,$M_{\oplus}$ and 
$R_P = 2.031_{-0.047}^{+0.064}$\,$R_{\oplus}$. We consider these parameters to 
be likely unphysical and thus KOI-722.01 fails detection criterion B3. Finally,
the posterior for $M_S/M_P$ does not converge away from zero and thus no
mass signal can be considered to be detected, failing criterion B4. This is 
also consistent with the fact that a zero-mass moon model gave a higher 
Bayesian evidence. We therefore conclude the model fit of 
$\mathcal{F}_{\mathrm{S}}$ is an exomoon false-positive and no convincing 
evidence for a satellite around KOI-722.01 exists in Q1-9.

\begin{figure*}
\begin{center}
\includegraphics[width=18.0 cm]{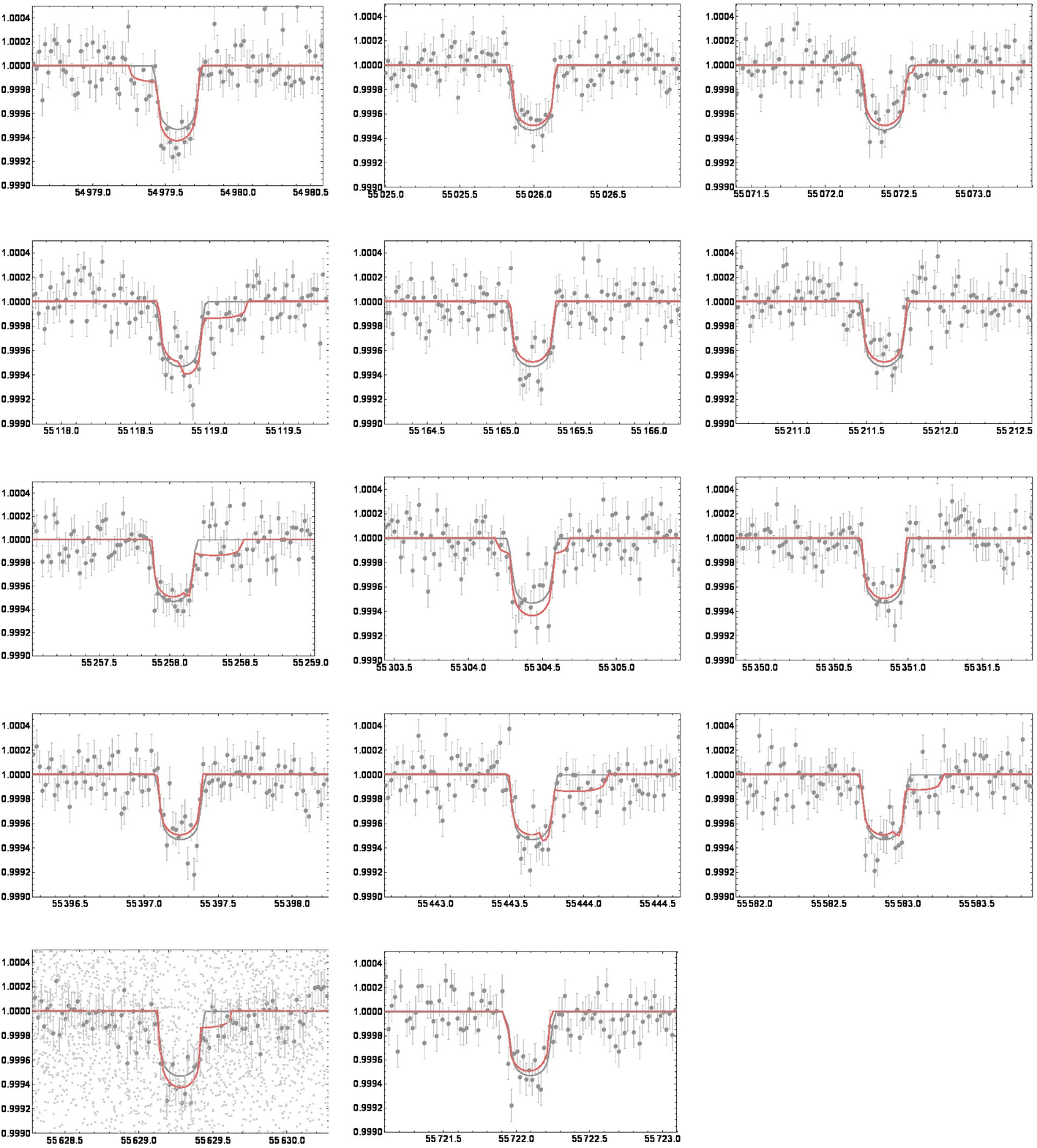}
\caption{\emph{From left-to-right then top-to-bottom we show
the chronological sequence of transits of KOI-722.01. The
first 8 panels show the Q1-9 data and the maximum a-posteriori
light curve fit of a planet-only model (gray line) and a moon model
(red line). Note that the figure temporally zooms-in on the
transits of the planet and the candidate moon.}} 
\label{fig:KOI722_Fs}
\end{center}
\end{figure*}

\begin{figure*}
\subfigure[\textbf{KOI-722.01} ($M_S/M_P<0.016$)
\label{fig:KOI722_Msp}]
{\epsfig{file=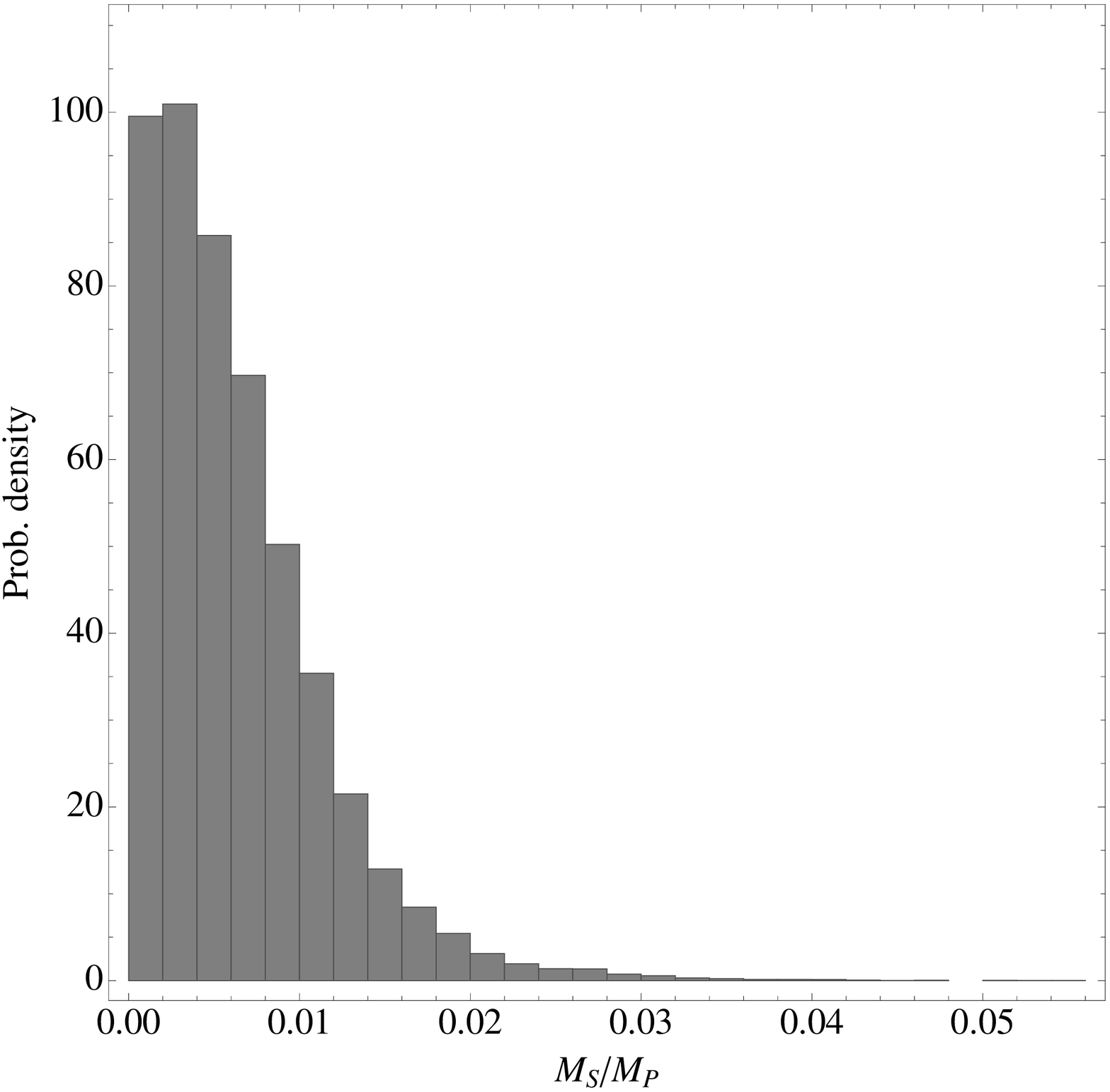,width=5.5 cm}}
\subfigure[\textbf{KOI-365.01} ($M_S/M_P<0.69$)
\label{fig:KOI365_Msp}]
{\epsfig{file=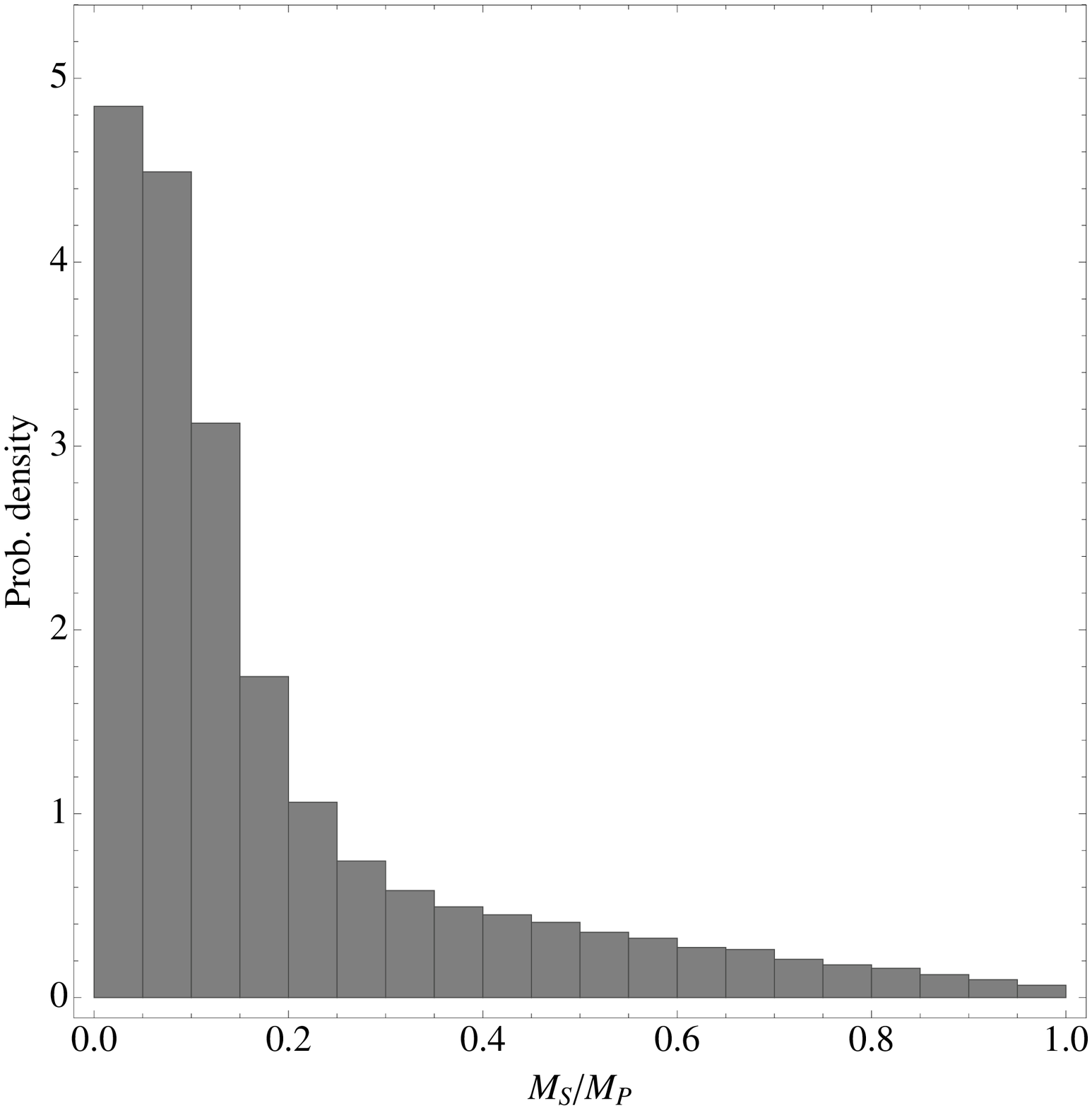,width=5.5 cm}}
\subfigure[\textbf{KOI-174.01} ($M_S/M_P<0.86$)
\label{fig:KOI174_Msp}]
{\epsfig{file=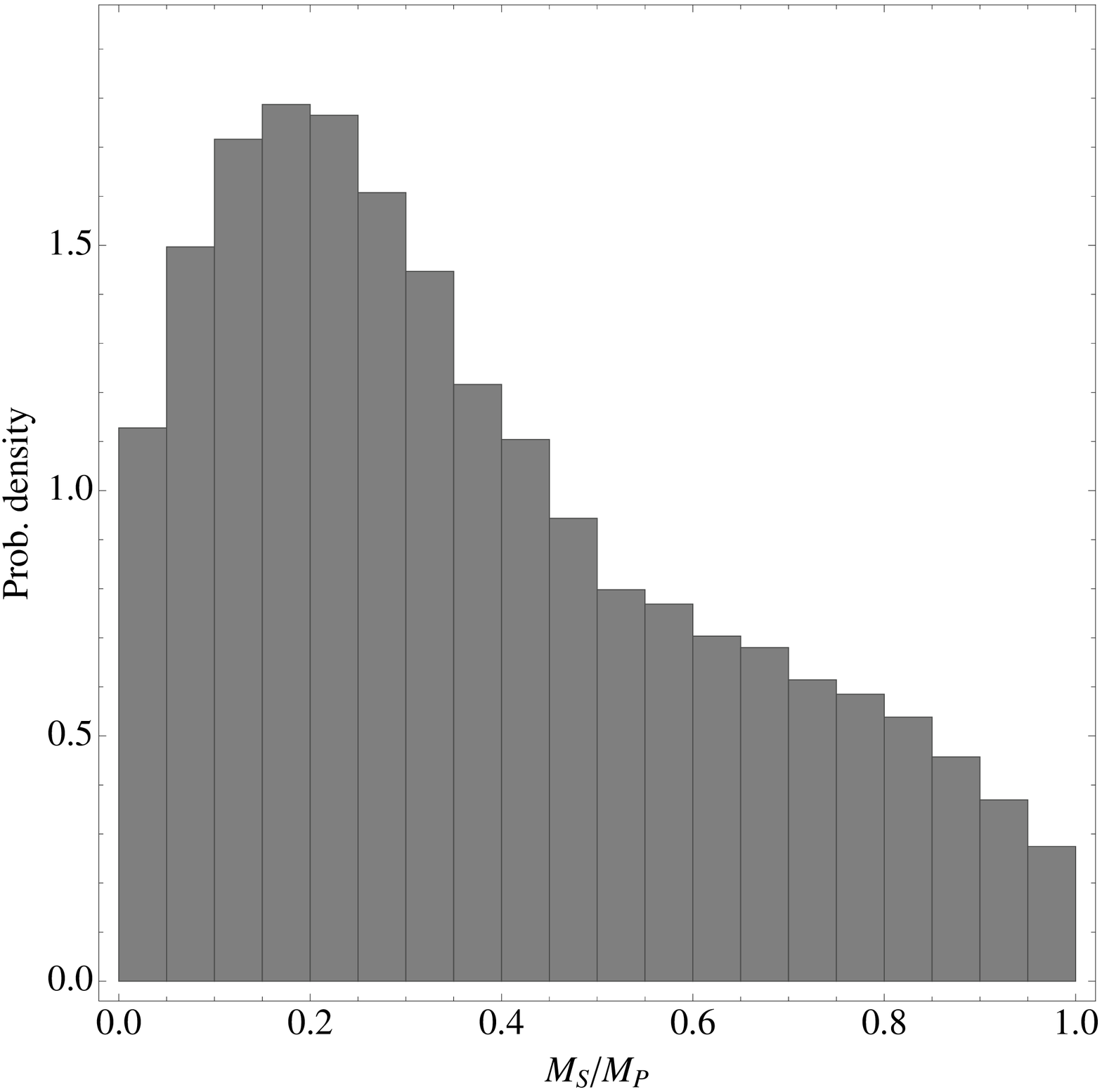,width=5.5 cm}}\\
\subfigure[\textbf{KOI-1472.01} ($M_S/M_P<0.037$)
\label{fig:KOI1472_Msp}]
{\epsfig{file=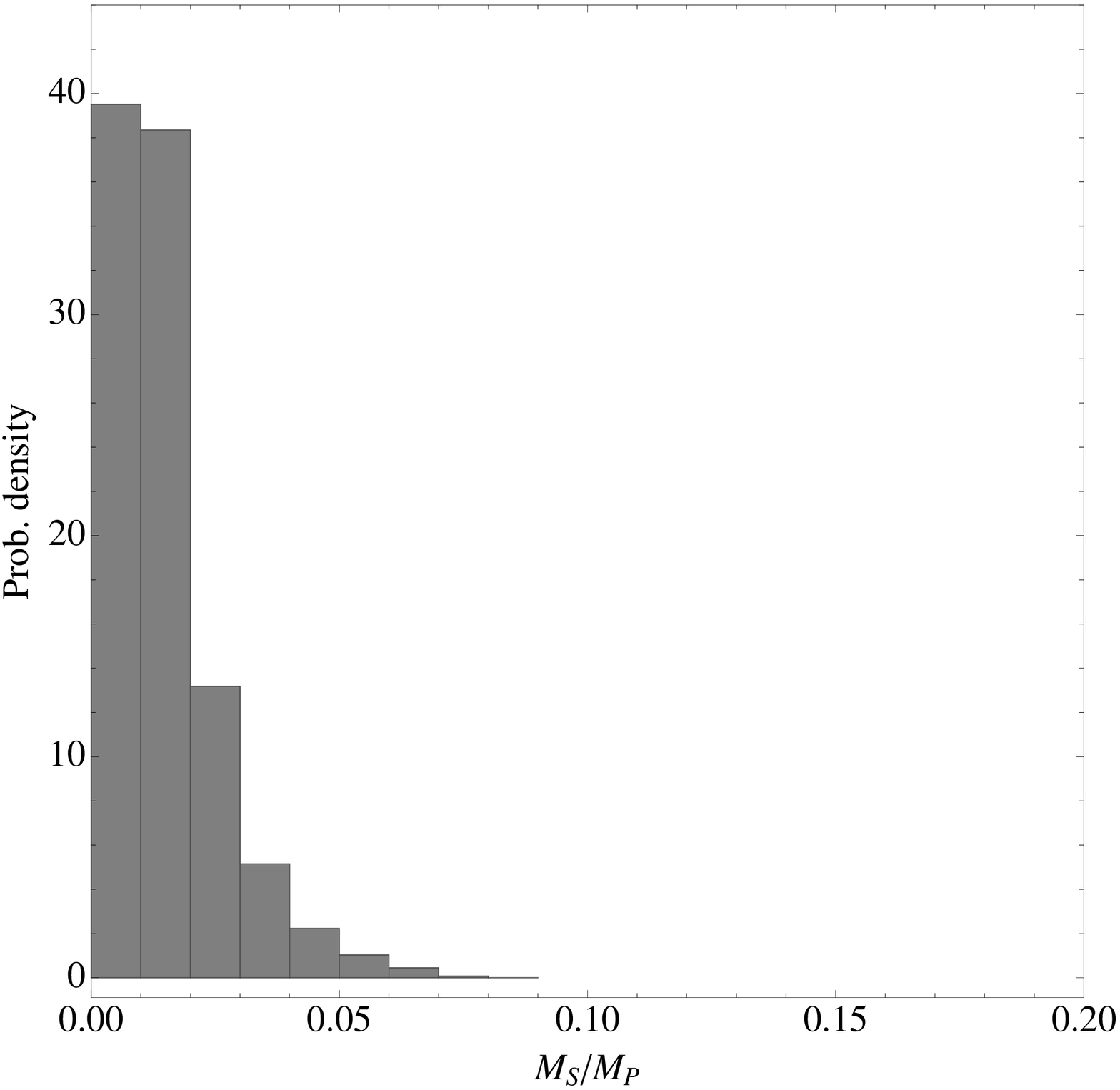,width=5.5 cm}}
\subfigure[\textbf{KOI-1857.01} ($M_S/M_P<0.028$)
\label{fig:KOI1857_Msp}]
{\epsfig{file=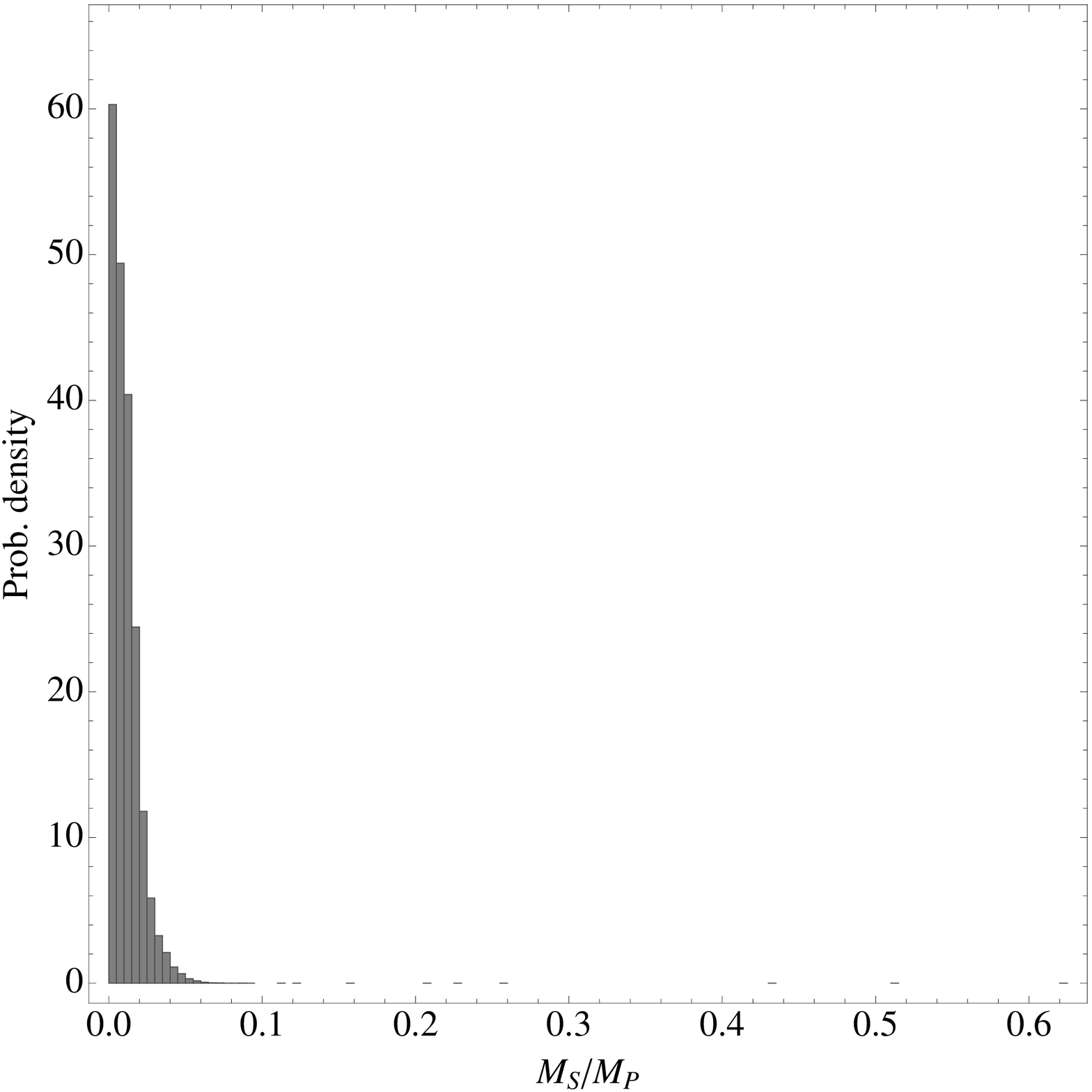,width=5.5 cm}}
\subfigure[\textbf{KOI-303.01} ($M_S/M_P<0.21$)
\label{fig:KOI303_Msp}]
{\epsfig{file=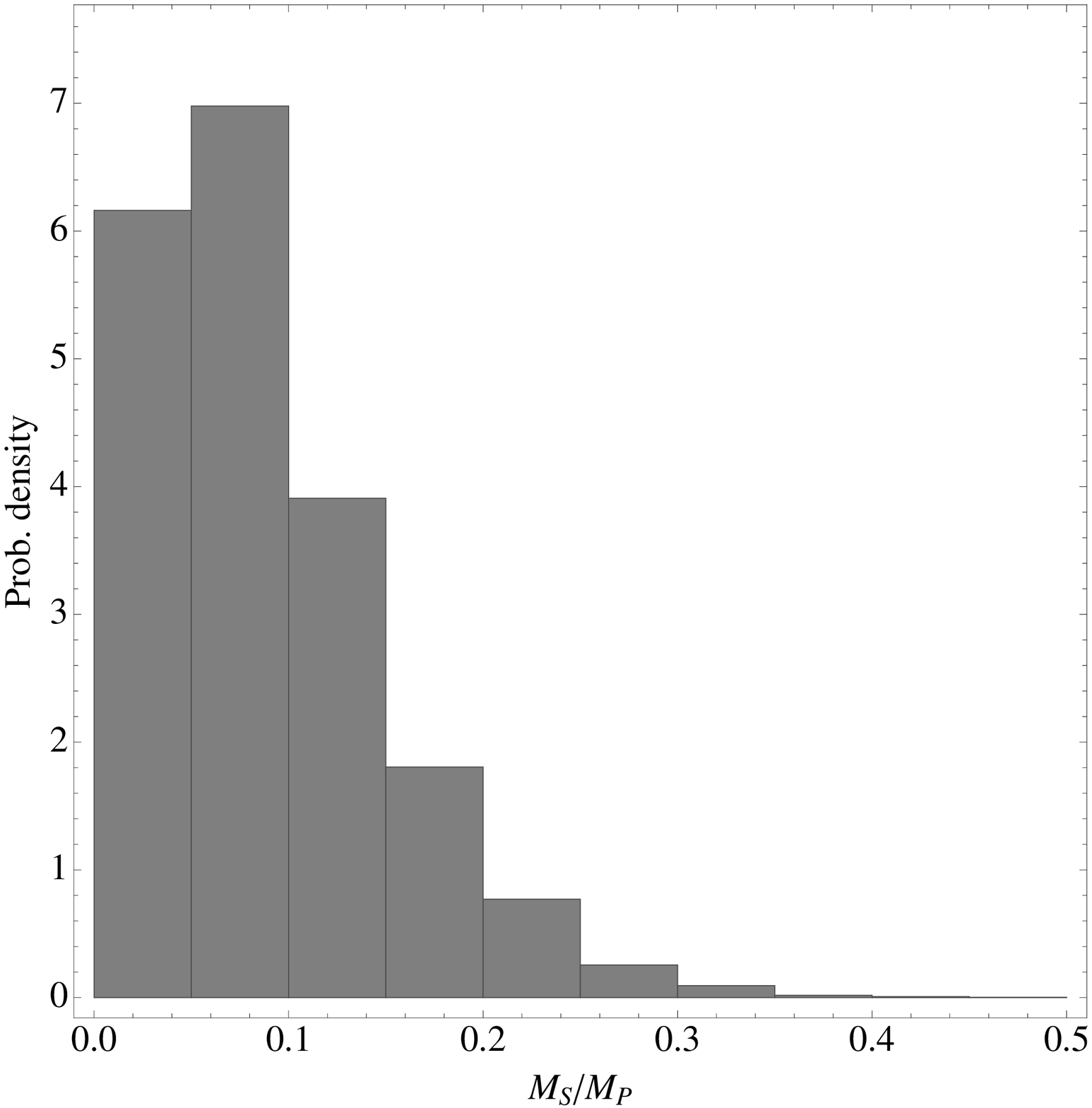,width=5.5 cm}} \\
\subfigure[\textbf{KOI-1876.01} ($M_S/M_P<0.012$)
\label{fig:KOI1876_Msp}]
{\epsfig{file=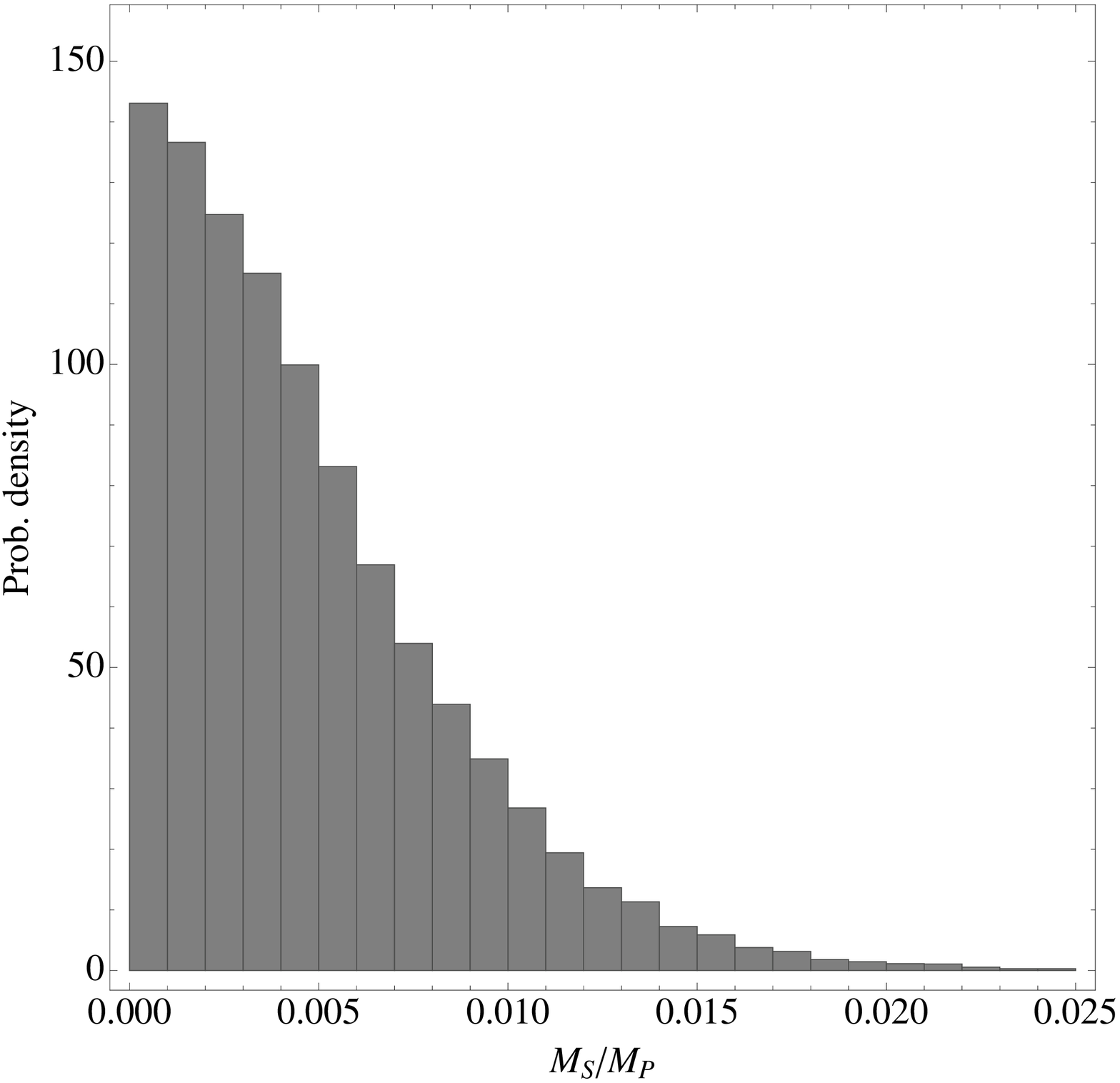,width=5.5 cm}}
\caption{
Marginalized posterior distributions of $M_S/M_P$ from model 
$\mathcal{F}_{\mathrm{S}}$ for the seven KOIs investigated here, except 
KOI-303.01 and KOI-1472.01 for which we use models $\mathcal{F}_{\mathrm{S,R0}}$ 
and $\mathcal{F}_{\mathrm{S},13}$ respectively, for reasons discussed in 
\S\ref{sub:koi303}\&\ref{sub:koi1472}. After the title of each sub-figure, we 
provide the 95\% quantile on $M_S/M_P$ in parentheses for each KOI.
\label{fig:Msp}}
\end{figure*}

The origin of the false-positive is unclear as the maximum a-posteriori fit
from $\mathcal{F}_{\mathrm{S}}$ reveals auxiliary transits (see 
Fig.~\ref{fig:KOI722_Fs}), which cannot be caused by starspot crossings. In 
this case, we consider time-correlated noise to be the most likely explanation.
The $M_S/M_P$ posterior converges on zero and we find the 95\% quantile to be
$M_S/M_P<0.016$ to the 3\,$\sigma$ quantile is $M_S/M_P<0.031$ (see 
Fig.~\ref{fig:KOI722_Msp}). Our final system parameters are provided in 
Table~\ref{tab:KOI722_parameters}.

%%% KOI-722 Fs fit

%%% KOI-722 Evidences
\begin{table}
\caption{\emph{Bayesian evidences of various fits for KOI-722.01.
A description of the different models can be found in 
\S\ref{sub:fitsoverview}.}} %title of Table
\centering % used for centering table
\begin{tabular}{l c l} % centered columns (3 columns)
\hline
Model, $\mathcal{M}$ & $\mathrm{log}\mathcal{Z}(\mathcal{M})$ & $\tilde{\mathcal{M}}_1 - \tilde{\mathcal{M}}_2$ \\ [0.5ex] % inserts table
 &  & $= \mathrm{log}\mathcal{Z}(\mathcal{M}_1) - \mathrm{log}\mathcal{Z}(\mathcal{M}_2)$ \\ [0.5ex] % inserts table
%heading
\hline
\multicolumn{3}{l}{\emph{Planet only fits...}}\\ 
%\hdashline
$\mathcal{V}_{\mathrm{P}}$  	& $38458.54 \pm 0.18$ 	& -	\\
$\mathcal{V}_{\mathrm{P,LD}}$	& $38458.60 \pm 0.19$	& $\tilde{\mathcal{V}}_{\mathrm{P,LD}}-\tilde{\mathcal{V}}_{\mathrm{P}} = (+0.05\pm0.26)$ \\
$\mathcal{F}_{\mathrm{P}}$	& $38547.22 \pm 0.11$	& $\tilde{\mathcal{F}}_{\mathrm{P}}-\tilde{\mathcal{V}}_{\mathrm{P}} = (+88.68\pm0.21)$ \\
\hline
\multicolumn{3}{l}{\emph{Planet with timing variations fits...}}\\ 
$\mathcal{F}_{\mathrm{TTV}}$	& $38519.40 \pm 0.14$	& $\tilde{\mathcal{F}}_{\mathrm{TTV}}-\tilde{\mathcal{F}}_{\mathrm{P}} = (-27.82\pm0.18)$ \\
$\mathcal{V}_{\mathrm{V}}$	& $38283.44 \pm 0.29$	& $\tilde{\mathcal{V}}_{\mathrm{V}}-\tilde{\mathcal{V}}_{\mathrm{P}} = (-175.10\pm0.34)$ \\
\hline
\multicolumn{3}{l}{\emph{Planet with moon fits...}}\\ 
$\mathcal{F}_{\mathrm{S}}$	& $38577.91 \pm 0.12$	& $\tilde{\mathcal{F}}_{\mathrm{S}}-\tilde{\mathcal{F}}_{\mathrm{P}} = (+30.68\pm0.16)$ \\ % 
$\mathcal{F}_{\mathrm{S,M0}}$	& $38583.07 \pm 0.11$	& $\tilde{\mathcal{F}}_{\mathrm{S,M0}}-\tilde{\mathcal{F}}_{\mathrm{S}} = (+5.16\pm0.17)$ \\ %  
$\mathcal{F}_{\mathrm{S,R0}}$	& $38549.42 \pm 0.11$	& $\tilde{\mathcal{F}}_{\mathrm{S,R0}}-\tilde{\mathcal{F}}_{\mathrm{S}} = (-28.49\pm0.16)$ \\ [1ex]
\hline\hline %inserts single line
\end{tabular}
\label{tab:KOI722_evidences} % is used to refer this table in the text
\end{table} % title of Table

%%% M_SP FIGURES

%%% KOI-722 System parameters
\begin{table}
\caption{\emph{System parameters for KOI-722.01 from model 
$\mathcal{V}_{\mathrm{P,LD}}$, except for $M_S/M_P$ which is derived from model
$\mathcal{F}_{\mathrm{S}}$. $^{*}$ indicates that a parameter was fixed.}}
%title of Table
\centering % used for centering table
\begin{tabular}{l l} % centered columns (2 columns)
\hline
Parameter & Value\\ [0.5ex] % inserts table
%heading
\hline
\multicolumn{2}{l}{\emph{Derived parameters...}}\\ 
$P_P$\,[days] & $46.40630_{-0.00022}^{+0.00023}$ \\
$\tau_0$\,[BJD$_{\mathrm{UTC}}$] & $2455350.8380_{-0.0012}^{+0.0012}$ \\
$R_P/R_{\star}$ & $0.02182_{-0.00043}^{+0.00063}$ \\
$b$ & $0.47_{-0.21}^{+0.24}$ \\
$(a/R_{\star})$ & $46.1_{-9.0}^{+4.4}$ \\
$i$\,[deg] & $89.41_{-0.51}^{+0.29}$ \\
$\rho_{\star}$\,[g\,cm$^{-3}$] & $0.86_{-0.41}^{+0.27}$ \\
$\tilde{T}$\,[hours] & $3.195_{-0.058}^{+0.066}$ \\
$u_1$ & $0.27_{-0.18}^{+0.31}$ \\
$(u_1+u_2)$ & $0.30_{-0.16}^{+0.18}$ \\ 
\hline
\multicolumn{2}{l}{\emph{Physical parameters...}}\\ 
$M_{\star}$\,[$R_{\odot}$] & $1.08^{*}$ \\
$R_{\star}$\,[$R_{\odot}$] & $0.83^{*}$ \\
$R_P$\,[$R_{\oplus}$] & $1.975_{-0.039}^{+0.057}$ \\
$M_S/M_P$ & $<0.016$ [95\% confidence] \\ %[1ex]
$\delta_{\mathrm{TTV}}$\,[mins] & $<4.0$ (95\% confidence) \\
$\delta_{\mathrm{TDV}}$\,[mins] & $<3.4$ (95\% confidence) \\ [1ex]
\hline\hline %inserts single line
\end{tabular}
\label{tab:KOI722_parameters} % is used to refer this table in the text
\end{table} % title of Table

\subsubsection{Summary}

We find no compelling evidence for an exomoon around KOI-722.01 and estimate
that $M_S/M_P<0.016$ to 95\% confidence. This assessment is based on the fact
the system fails the basic detection criteria B2, B3 and B4 (see 
\S\ref{sub:criteria}).

%%%%%%%%%%%%%%%%%%%%%%%%%%%%%%%%%%%%%%%%%%%%%%%%%%%%%%%%%%%%%%%%%%%%%%%%%%%%%%%%

\subsection{KOI-365.01}
\label{sub:koi365}

%% KOI-365
%%

\subsubsection{Data selection}

After detrending with \cofiam, the PA and PDC-MAP data were found to have a 
2.2\,$\sigma$ and 3.1\,$\sigma$ confidence of autocorrelation on a 30\,minute 
timescale respectively. Since only the PA data is considered acceptable 
($<3$\,$\sigma$), it will be utilized throughout the analysis that follows. 
Short-cadence data is available for quarters 3, 7, 8 and 9 and this data 
displaced the corresponding long-cadence quarters in our analysis.

\subsubsection{Planet-only fits}

% Limb darkening
When queried from MAST, the KIC effective temperature and surface gravity
were reported as $T_{\mathrm{eff}} = 5389$\,K and $\log g = 4.570$ 
\citep{brown:2011}. Using these values, we estimated quadratic limb darkening 
coefficients $u_1 = 0.5016$ and $(u_1+u_2) = 0.7064$. The initial two models we 
regressed were $\mathcal{V}_{\mathrm{P}}$ and $\mathcal{V}_{\mathrm{P,LD}}$, 
where the former uses the theoretical limb darkening coefficients as fixed 
values and the latter allows the two coefficients to be free parameters. We find 
that $\log\mathcal{Z}(\mathcal{V}_{\mathrm{P,LD}}) - 
\log\mathcal{Z}(\mathcal{V}_{\mathrm{P}}) = +5.34\pm0.24$ indicating that our 
limb darkening coefficients may not be optimal. The maximum a-posteriori limb 
darkening coefficients from the $\mathcal{V}_{\mathrm{P,LD}}$ model fit were 
$u_1 = 0.5234$ and $(u_1+u_2) = 0.6145$, which we chose to treat as fixed terms 
in all subsequent model fits.

% V -> P
KOI-365.01 has a period of $P_P = (81.73766 \pm 0.00014)$\,days (as determined 
by model $\mathcal{V}_{\mathrm{P,LD}}$) and exhibits 9 transits from Q1-Q9. 
As is typical for all cases, $\log\mathcal{Z}(\mathcal{F}_{\mathrm{P}}) > 
\log\mathcal{Z}(\mathcal{V}_{\mathrm{P}})$ indicating that allowing for 9 
independent baseline parameters is unnecessary relative to a single baseline 
term. %Removing the 8 excessive 
%parameters leads to a factor of 5.8 quicker computation times for the 
%planet-only fits.

% TTV
We find no evidence for TTVs in KOI-365.01, with 
$\log\mathcal{Z}(\mathcal{F}_{\mathrm{TTV}}) 
- \log\mathcal{Z}(\mathcal{F}_{\mathrm{P}}) = -36.98\pm0.17$, which is formally 
an 8.3\,$\sigma$ preference for a static model over a TTV model. The timing 
precision on the 9 transits ranged from 1.3 to 2.8 minutes and yields a flat TTV 
profile, as shown in Fig.~\ref{fig:KOI365_TTVs}. We calculate a standard
deviation of $\delta_{\mathrm{TTV}} = 1.5$\,minutes and $\chi_{\mathrm{TTV}}^2 
= 4.2$ for 9-2 degrees of freedom.

% TDV
The TTV+TDV model fit, $\mathcal{V}_{\mathrm{V}}$, finds consistent transit 
times with those derived by model $\mathcal{F}_{\mathrm{TTV}}$. No clear pattern 
or excessive scatter is visible in the data, shown in 
Fig.~\ref{fig:KOI365_TTVs}. We therefore conclude there is no evidence for TTVs 
or TDVs for KOI-365.01. The standard deviation of the TDVs is found to be 
$\delta_{\mathrm{TDV}} = 3.3$\,minutes and we determine $\chi_{\mathrm{TDV}}^2 = 
8.6$ for 9-1 degrees of freedom.

%%% KOI-365 TTV Table
\begin{table*}
\caption{\emph{Transit times and durations for KOI-365.01. The model used to 
calculate the supplied values is provided in parentheses next to each column
heading. BJD$_{\mathrm{UTC}}$ times offset by $2,400,000$ days.
}} % title of Table
\centering % used for centering table
\begin{tabular}{c c c c c c c} % centered columns (3 columns)
\hline
Epoch & $\tau$ [BJD$_{\mathrm{UTC}}$] ($\mathcal{F}_{\mathrm{TTV}}$) & TTV [mins] ($\mathcal{F}_{\mathrm{TTV}}$)
      & $\tau$ [BJD$_{\mathrm{UTC}}$] ($\mathcal{V}_{\mathrm{V}}$) & TTV [mins] ($\mathcal{V}_{\mathrm{V}}$)
      & $\tilde{T}$ [mins] ($\mathcal{V}_{\mathrm{V}}$) & TDV [mins] ($\mathcal{V}_{\mathrm{V}}$) \\ [0.5ex] % inserts table
%heading
\hline
-5 & $54962.9413_{-0.0021}^{+0.0019}$    & $+1.3 \pm 2.9$
   & $54962.9446_{-0.0041}^{+0.0060}$    & $+6.1 \pm 7.2$
   & $402_{-14}^{+19}$     & $+5.8 \pm 8.3$ \\
-4 & $55044.6779_{-0.0015}^{+0.0014}$    & $-0.2 \pm 2.1$
   & $55044.6775_{-0.0015}^{+0.0014}$    & $-0.6 \pm 2.1$
   & $382.5_{-5.3}^{+5.0}$ & $-3.8 \pm 2.6$ \\
-3 & $55126.4156_{-0.0011}^{+0.0013}$    & $+0.0 \pm 1.7$
   & $55126.4160_{-0.0015}^{+0.0019}$    & $+0.7 \pm 2.4$
   & $386.4_{-6.3}^{+4.5}$ & $-1.8 \pm 2.7$ \\
-2 & $55208.1519_{-0.0014}^{+0.0014}$    & $-1.9 \pm 2.0$
   & $55208.1521_{-0.0015}^{+0.0015}$    & $-1.4 \pm 2.1$
   & $391.6_{-5.0}^{+4.7}$ & $+0.8 \pm 2.4$ \\
-1 & $55289.8926_{-0.0015}^{+0.0017}$    & $+2.4 \pm 2.3$
   & $55289.8926_{-0.0015}^{+0.0015}$    & $+2.8 \pm 2.1$
   & $398.7_{-4.9}^{+4.8}$ & $+4.3 \pm 2.4$ \\
+1 & $55453.3657_{-0.0018}^{+0.0019}$    & $-0.7 \pm 2.7$
   & $55453.3660_{-0.0016}^{+0.0016}$    & $+0.3 \pm 2.3$
   & $397.1_{-5.9}^{+5.1}$ & $+3.5 \pm 2.7$ \\
+2 & $55535.1034_{-0.0012}^{+0.0012}$    & $-0.5 \pm 1.7$
   & $55535.1038_{-0.0012}^{+0.0016}$    & $+0.5 \pm 2.0$
   & $387.3_{-4.8}^{+4.0}$ & $-1.4 \pm 2.2$ \\
+3 & $55616.84083_{-0.00095}^{+0.00095}$ & $-0.8 \pm 1.4$
   & $55616.84088_{-0.00096}^{+0.00093}$ & $-0.1 \pm 1.4$
   & $390.6_{-2.9}^{+2.9}$ & $+0.3 \pm 1.5$ \\
+4 & $55698.5805_{-0.0012}^{+0.0012}$    & $+2.1 \pm 1.7$
   & $55698.5806_{-0.0013}^{+0.0014}$    & $+3.0 \pm 1.9$
   & $391.0_{-4.0}^{+4.4}$ & $+0.5 \pm 2.1$ \\ [1ex]
\hline\hline %inserts single line
\end{tabular}
\label{tab:KOI365_TTVs} % is used to refer this table in the text
\end{table*} % title of Table

\subsubsection{Moon fits}

A planet-with-moon fit, $\mathcal{F}_S$, is disfavored relative to a 
planet-only fit at $\Delta(\log\mathcal{Z}) = -2.7\pm0.2$ (see
Table~\ref{tab:KOI365_evidences}) and so detection criterion B1 is not 
satisfied. The system also fails detection criterion B2 since a zero-mass 
moon yields an improved Bayesian evidence relative to a moon with finite mass.

We may combine the posteriors from $\mathcal{F}_{\mathrm{S}}$ with the 
stellar parameters derived by B12 ($M_{\star} = 
0.99$\,$M_{\odot}$ and $R_{\star} = 0.86$\,$R_{\oplus}$) to obtain physical 
parameters for the planet-moon candidate system. We find that the planet has an 
unusually low density with $M_P=0.53_{-0.33}^{+2.71}$\,$M_{\oplus}$ and 
$R_P=2.11_{-0.31}^{+0.09}$\,$R_{\oplus}$. The moon also has a
low density with parameters $M_S=0.06_{-0.04}^{+0.45}$\,$M_{\oplus}$ and
$R_S = 0.65_{-0.21}^{+0.58}$\,$R_{\oplus}$. Whilst these parameters are
somewhat extreme they are not implausible, but at best detection criterion
B3 can be considered unclear. Finally, the $M_S/M_P$ posterior fails to
converge away from zero, failing detection criterion B4, and consistent
with the fact that model $\mathcal{F}_{\mathrm{S,M0}}$ is favored over
model $\mathcal{F}_{\mathrm{S}}$. We therefore conclude the model fits of 
$\mathcal{F}_S$ and $\mathcal{F}_{\mathrm{S,eP}}$
represent an exomoon false-positive and no convincing evidence for a satellite
around KOI-365.01 exists in Q1-9.

\begin{figure*}
\begin{center}
\includegraphics[width=18.0 cm]{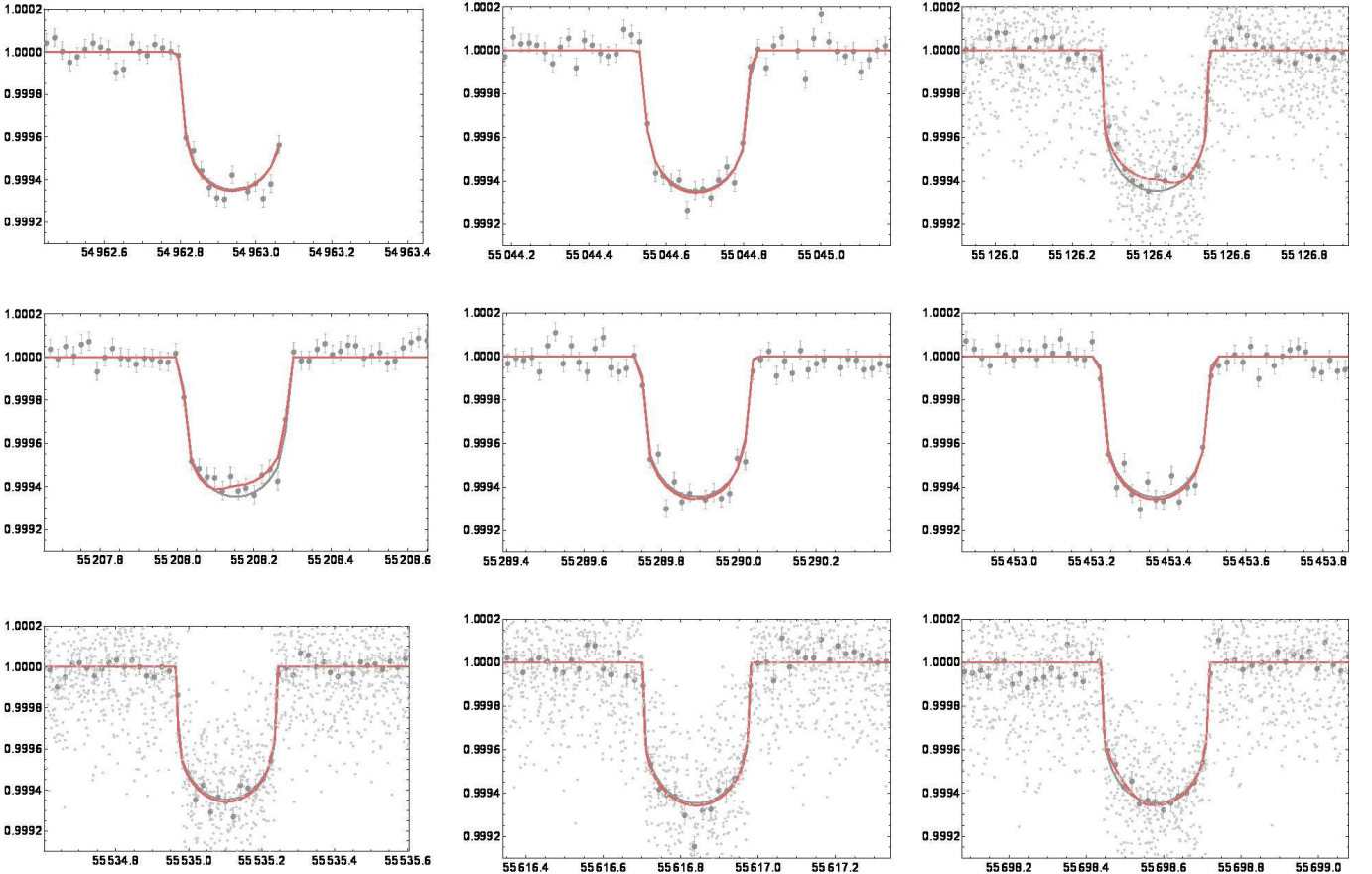}
\caption{\emph{From left-to-right then top-to-bottom we show
the chronological sequence of transits of KOI-365.01. The
first 8 panels show the Q1-9 data and the maximum a-posteriori
light curve fit of a planet-only model (gray line) and a moon model
(red line). No clear distortions due to the moon are visible in this 
close-binary type solution.}} 
\label{fig:KOI365_Fs}
\end{center}
\end{figure*}

The maximum a-posteriori model fit of $\mathcal{F}_{\mathrm{S}}$ (shown
in Fig~\ref{fig:KOI365_Fs}) does not exhibit any clear auxiliary or mutual
events, despite $R_S/R_P$ converging away from zero. The reason for this
becomes clear when one notes that the semi-major axis of the moon's orbit 
around the planet converges to $(a_{SP}/R_P) = 3.7_{-0.9}^{+1.9}$, which
shows that the moon is in very close proximity to the planet. In such a case,
the planet and moon appear almost on-top of one another and thus virtually
no light curve distortion is visible. This one way in which a fitting
algorithm can essentially ``hide a moon'' in the fits and such fits are
always suspicious. The close proximity of the moon leads to an absence of
TTVs too since TTVs scale as $M_S a_S$ \citep{kipping:2009a} and so the
solution also masks the exomoon mass.

Unlike TTVs, TDVs scale as $M_S/\sqrt{a_S}$ and so one might expect that
the exomoon mass could not be hidden from TDVs too. However, the period of 
the moon solution is also short at $P_S = 1.9_{-1.3}^{+1.5}$\,d and this causes
a problem for TDV inference. As noted in \citet{thesis:2011}, traditional
TDV theory breaks down if the moon accelerates/decelerates significantly
during a transit duration and thus the theory only holds for $P_S\gg T_{14}$.
In this case, $P_S \sim 2$\,days and $T_{14} \sim 0.3$\,days meaning that
that TDVs will not necessarily be present for even massive exomoons. Due
to these points, close-binary moon solutions tend to yield less
useful constraints on excluded mass and radius ratios. In this case,
the moon fit yields a worse Bayesian evidence than a planet-fit and
thus we know it is not the favored model, yet the upper limits are not
as reliable as with KOI-722.01 for example. The $M_S/M_P$ posterior converges on 
zero and we find the 95\% quantile to be $M_S/M_P<0.69$ and the 3\,$\sigma$ 
quantile is $M_S/M_P<0.97$ (see Fig.~\ref{fig:KOI365_Msp}). Our final system 
parameters are provided in Table~\ref{tab:KOI365_parameters}.

%%% KOI-365 Fs fit

%%% KOI-365 Evidences
\begin{table}
\caption{\emph{Bayesian evidences of various fits for KOI-365.01.
A description of the different models can be found in 
\S\ref{sub:fitsoverview}.}} %title of Table
\centering % used for centering table
\begin{tabular}{l c l} % centered columns (3 columns)
\hline
Model, $\mathcal{M}$ & $\mathrm{log}\mathcal{Z}(\mathcal{M})$ & $\tilde{\mathcal{M}}_1 - \tilde{\mathcal{M}}_2$ \\ [0.5ex] % inserts table
 &  & $= \mathrm{log}\mathcal{Z}(\mathcal{M}_1) - \mathrm{log}\mathcal{Z}(\mathcal{M}_2)$ \\ [0.5ex] % inserts table
%heading
\hline
\multicolumn{3}{l}{\emph{Planet only fits...}}\\ 
$\mathcal{V}_{\mathrm{P}}$  	& $141526.62 \pm 0.17$ 	& -	\\
$\mathcal{V}_{\mathrm{P,LD}}$	& $141531.95 \pm 0.17$	& $\tilde{\mathcal{V}}_{\mathrm{P,LD}}-\tilde{\mathcal{V}}_{\mathrm{P}} = (+5.34\pm0.24)$ \\
$\mathcal{V}_{\mathrm{P,MAP}}$	& $141535.04 \pm 0.17$	& $\tilde{\mathcal{V}}_{\mathrm{P,MAP}}-\tilde{\mathcal{V}}_{\mathrm{P}} = (+8.43\pm0.24)$ \\
$\mathcal{F}_{\mathrm{P}}$	& $141606.76 \pm 0.10$	& $\tilde{\mathcal{F}}_{\mathrm{P}}-\tilde{\mathcal{V}}_{\mathrm{P,MAP}} = (+71.72\pm0.20)$ \\
\hline
\multicolumn{3}{l}{\emph{Planet with timing variations fits...}}\\
$\mathcal{F}_{\mathrm{TTV}}$	& $141569.78 \pm 0.14$	& $\tilde{\mathcal{F}}_{\mathrm{TTV}}-\tilde{\mathcal{F}}_{\mathrm{P}} = (-36.98\pm0.17)$ \\
$\mathcal{V}_{\mathrm{V}}$	& $141394.37 \pm 0.26$	& $\tilde{\mathcal{V}}_{\mathrm{V}}-\tilde{\mathcal{V}}_{\mathrm{P,MAP}} = (-140.67\pm0.30)$ \\
\hline
\multicolumn{3}{l}{\emph{Planet with moon fits...}}\\ 
$\mathcal{F}_{\mathrm{S}}$	& $141610.16 \pm 0.12$	& $\tilde{\mathcal{F}}_{\mathrm{S}}-\tilde{\mathcal{F}}_{\mathrm{P}} = (+3.40\pm0.16)$ \\ % 
$\mathcal{F}_{\mathrm{S,M0}}$	& $141611.69 \pm 0.12$	& $\tilde{\mathcal{F}}_{\mathrm{S,M0}}-\tilde{\mathcal{F}}_{\mathrm{S}} = (+1.53\pm0.17)$ \\ %  
$\mathcal{F}_{\mathrm{S,R0}}$	& $141602.99 \pm 0.11$	& $\tilde{\mathcal{F}}_{\mathrm{S,R0}}-\tilde{\mathcal{F}}_{\mathrm{S}} = (-7.17\pm0.16)$ \\ [1ex] %
\hline\hline %inserts single line
\end{tabular}
\label{tab:KOI365_evidences} % is used to refer this table in the text
\end{table} % title of Table

%%% KOI-365 System parameters
\begin{table}
\caption{\emph{System parameters for KOI-365.01 from model 
$\mathcal{V}_{\mathrm{P,LD}}$, except for $M_S/M_P$ which is derived from model
$\mathcal{F}_{\mathrm{S}}$. $^{*}$ indicates that a parameter was fixed.}}
%title of Table
\centering % used for centering table
\begin{tabular}{l l} % centered columns (2 columns)
\hline
Parameter & Value\\ [0.5ex] % inserts table
%heading
\hline
\multicolumn{2}{l}{\emph{Derived parameters...}}\\ 
$P_P$\,[days] & $81.73766_{-0.00014}^{+0.00014}$ \\
$\tau_0$\,[BJD$_{\mathrm{UTC}}$] & $2455371.62859_{-0.00045}^{+0.00047}$ \\
$R_P/R_{\star}$ & $0.02364_{-0.00058}^{+0.00088}$ \\
$b$ & $0.47_{-0.30}^{+0.18}$ \\
$(a/R_{\star})$ & $84.5_{-11.6}^{+9.8}$ \\
$i$\,[deg] & $89.68_{-0.19}^{+0.22}$ \\
$\rho_{\star}$\,[g\,cm$^{-3}$] & $1.71_{-0.61}^{+0.66}$ \\
$\tilde{T}$\,[hours] & $6.513_{-0.038}^{+0.044}$ \\
$u_1$ & $0.47_{-0.15}^{+0.15}$ \\
$(u_1+u_2)$ & $0.652_{-0.083}^{+0.093}$ \\ 
\hline
\multicolumn{2}{l}{\emph{Physical parameters...}}\\ 
$M_{\star}$\,[$R_{\odot}$] & $0.99^{*}$ \\
$R_{\star}$\,[$R_{\odot}$] & $0.86^{*}$ \\
$R_P$\,[$R_{\oplus}$] & $2.217_{-0.054}^{+0.083}$ \\
$M_S/M_P$ & $<0.69$ [95\% confidence] \\ %[1ex]
$\delta_{\mathrm{TTV}}$\,[mins] & $<3.3$ (95\% confidence) \\
$\delta_{\mathrm{TDV}}$\,[mins] & $<6.6$ (95\% confidence) \\ [1ex]
\hline\hline %inserts single line
\end{tabular}
\label{tab:KOI365_parameters} % is used to refer this table in the text
\end{table} % title of Table

\subsubsection{Summary}

We find no compelling evidence for an exomoon around KOI-365.01 and estimate
that $M_S/M_P<0.69$ to 95\% confidence. This assessment is based on the fact
the system fails the basic detection criteria B1, B2 and B4 and B3 is considered
marginal (see \S\ref{sub:criteria}).

%%%%%%%%%%%%%%%%%%%%%%%%%%%%%%%%%%%%%%%%%%%%%%%%%%%%%%%%%%%%%%%%%%%%%%%%%%%%%%%%

\subsection{KOI-174.01}
\label{sub:koi174}

%% KOI-174
%%

\subsubsection{Data selection}

After detrending with \cofiam, the PA and PDC-MAP data were found to have a 
1.5\,$\sigma$ and 0.6\,$\sigma$ confidence of autocorrelation on a 30\,minute 
timescale respectively and therefore both were acceptable ($<3$\,$\sigma$). 
Since the PA data detrending shows no strong evidence of autocorrelation, we 
opted to use this less-manipulated data in what follows. Short-cadence data is 
available for quarters 3, 4, 5 and 6 and this data displaced the corresponding 
long-cadence quarters in our analysis.

\subsubsection{Planet-only fits}

% Limb darkening
When queried from MAST, the KIC effective temperature and surface gravity
were reported as $T_{\mathrm{eff}} = 4654$\,K and $\log g = 4.538$ 
\citep{brown:2011}. Using these values, we estimated quadratic limb darkening 
coefficients $u_1 = 0.6531$ and $(u_1+u_2) = 0.7415$. The initial two models we 
regressed were $\mathcal{V}_{\mathrm{P}}$ and $\mathcal{V}_{\mathrm{P,LD}}$, 
where the former uses the theoretical limb darkening coefficients as fixed 
values and the latter allows the two coefficients to be free parameters. We find 
that $\log\mathcal{Z}(\mathcal{V}_{\mathrm{P,LD}}) - 
\log\mathcal{Z}(\mathcal{V}_{\mathrm{P}}) = +8.22\pm0.22$ suggesting that our 
limb darkening coefficients could be improved. Strangely though, when we fix
the limb darkening coefficients to the maximum a-posteriori values from the 
$\mathcal{V}_{\mathrm{P,LD}}$ model fit, we actually obtain a worse Bayesian
evidence than $\mathcal{V}_{\mathrm{P}}$. In light of this, we decided to
continue with the theoretical coefficients for this particular system. 

% V -> P
KOI-174.01 has a period of $P_P = (56.35439 \pm 0.00019)$\,days (as determined 
by model $\mathcal{V}_{\mathrm{P,LD}}$) and exhibits full 9 transits from Q1-Q9. 
As is typical for all cases, $\log\mathcal{Z}(\mathcal{F}_{\mathrm{P}}) > 
\log\mathcal{Z}(\mathcal{V}_{\mathrm{P}})$ indicating that allowing for 9 
independent baseline parameters is unnecessary relative to a single baseline 
term. %Removing the 8 excessive 
%parameters leads to a factor of 5.8 quicker computation times for the 
%planet-only fits.

% TTV
We find no evidence for TTVs in KOI-174.01, with 
$\log\mathcal{Z}(\mathcal{F}_{\mathrm{TTV}}) 
- \log\mathcal{Z}(\mathcal{F}_{\mathrm{P}}) = -22.65\pm0.16$, which is formally 
an 6.4\,$\sigma$ preference for a static model over a TTV model. The timing 
precision on the 9 transits ranged from 2.9 to 5.9 minutes and yields a flat TTV 
profile, as shown in Fig.~\ref{fig:KOI174_TTVs}. We calculate a standard
deviation of $\delta_{\mathrm{TTV}} = 5.1$\,minutes and $\chi_{\mathrm{TTV}}^2 
= 13.0$ for 9-2 degrees of freedom.

% TDV
The TTV+TDV model fit, $\mathcal{V}_{\mathrm{V}}$, finds consistent transit 
times with those derived by model $\mathcal{F}_{\mathrm{TTV}}$. The data show no 
clear pattern or excessive scatter, visible in Fig.~\ref{fig:KOI174_TTVs}. We 
therefore conclude there is no evidence for TTVs or TDVs for KOI-174.01.
The standard deviation of the TDVs is found to be $\delta_{\mathrm{TDV}} = 
6.7$\,minutes and we determine $\chi_{\mathrm{TDV}}^2 = 10.2$ for 9-1 
degrees of freedom.

%%% KOI-174 TTV Table
\begin{table*}
\caption{\emph{Transit times and durations for KOI-174.01. The model used to 
calculate the supplied values is provided in parentheses next to each column
heading. BJD$_{\mathrm{UTC}}$ times offset by $2,400,000$ days.
}} % title of Table
\centering % used for centering table
\begin{tabular}{c c c c c c c} % centered columns (3 columns)
\hline
Epoch & $\tau$ [BJD$_{\mathrm{UTC}}$] ($\mathcal{F}_{\mathrm{TTV}}$) & TTV [mins] ($\mathcal{F}_{\mathrm{TTV}}$)
      & $\tau$ [BJD$_{\mathrm{UTC}}$] ($\mathcal{V}_{\mathrm{V}}$) & TTV [mins] ($\mathcal{V}_{\mathrm{V}}$)
      & $\tilde{T}$ [mins] ($\mathcal{V}_{\mathrm{V}}$) & TDV [mins] ($\mathcal{V}_{\mathrm{V}}$) \\ [0.5ex] % inserts table
%heading
\hline
-6 & $54977.8321_{-0.0041}^{+0.0032}$ & $-4.3 \pm 5.2$
   & $54977.8295_{-0.0044}^{+0.0046}$ & $-8.0 \pm 6.5$
   & $207_{-17}^{+17}$      & $+8.5 \pm 8.5$ \\
-5 & $55034.1934_{-0.0028}^{+0.0028}$ & $+5.8 \pm 4.0$
   & $55034.1933_{-0.0029}^{+0.0029}$ & $+5.7 \pm 4.2$
   & $190.3_{-9.8}^{+10.4}$ & $+0.0 \pm 5.0$ \\
-4 & $55090.5427_{-0.0026}^{+0.0026}$ & $-1.1 \pm 3.8$
   & $55090.5397_{-0.0030}^{+0.0036}$ & $-5.5 \pm 4.7$
   & $165_{-11}^{+13}$      & $-12.6 \pm 6.0$ \\
-3 & $55146.8945_{-0.0021}^{+0.0020}$ & $-4.6 \pm 2.9$
   & $55146.8948_{-0.0035}^{+0.0033}$ & $-4.3 \pm 4.9$
   & $187.9_{-8.6}^{+9.9}$  & $-1.2 \pm 4.6$ \\
-2 & $55203.2535_{-0.0039}^{+0.0043}$ & $+2.2 \pm 5.9$
   & $55203.2560_{-0.0045}^{+0.0032}$ & $+5.8 \pm 5.5$
   & $203_{-15}^{+12}$      & $+6.3 \pm 6.6$ \\
+0 & $55315.9626_{-0.0024}^{+0.0025}$ & $+3.1 \pm 3.5$
   & $55315.9621_{-0.0023}^{+0.0025}$ & $+2.4 \pm 3.4$
   & $198.7_{-8.0}^{+7.5}$  & $+4.2 \pm 3.9$ \\
+2 & $55428.6705_{-0.0021}^{+0.0017}$ & $+2.2 \pm 2.7$
   & $55428.6705_{-0.0021}^{+0.0020}$ & $+2.3 \pm 3.0$
   & $193.6_{-6.6}^{+6.7}$  & $+1.6 \pm 3.3$ \\
+5 & $55597.7259_{-0.0030}^{+0.0030}$ & $-8.3 \pm 4.3$
   & $55597.7263_{-0.0032}^{+0.0034}$ & $-7.8 \pm 4.8$
   & $179_{-12}^{+12}$      & $-5.9 \pm 6.0$ \\
+6 & $55654.0904_{-0.0030}^{+0.0030}$ & $+6.5 \pm 4.4$
   & $55654.0901_{-0.0028}^{+0.0028}$ & $+6.0 \pm 4.0$
   & $178.6_{-9.5}^{+10.0}$ & $-5.9 \pm 4.9$ \\ [1ex]
\hline\hline %inserts single line
\end{tabular}
\label{tab:KOI174_TTVs} % is used to refer this table in the text
\end{table*} % title of Table

\subsubsection{Moon fits}

A planet-with-moon fit, $\mathcal{F}_S$, is slightly favored relative to a 
planet-only fit at $\Delta(\log\mathcal{Z}) = +1.5\pm0.1$, but does not
satisfy detection criterion B1. The system also fails detection 
criterion B2 since a zero-mass moon yields an improved Bayesian evidence 
relative to a moon with finite mass. 

Inspection of the posteriors from model $\mathcal{F}_{\mathrm{S}}$
and using the stellar parameters of B12
($M_{\star} = 0.80$\,$M_{\odot}$ and $R_{\star} = 0.80$\,$M_{\odot}$) reveals
a set of broadly unphysical parameters. Most notably, the planet has
an unusually low density with $M_P=0.27_{-0.20}^{+1.03}$\,$M_{\oplus}$ and 
$R_P=2.28_{-0.27}^{+0.15}$\,$R_{\oplus}$. The satellite also has some odd
parameters with $M_S=0.08_{-0.06}^{+0.33}$\,$M_{\oplus}$ and 
$R_S=1.08_{-0.40}^{+0.44}$\,$R_{\oplus}$. In general, we find this combination
of masses and radii improbable and consider that detection criterion B3 is not
satisfied. Finally, the mass ratio $M_S/M_P$ does not converge away from
zero meaning detection criterion B4 is also not satisfied. We therefore conclude
that the model fit of $\mathcal{F}_{\mathrm{S}}$ represents an exomoon 
false-positive and no convincing evidence for a satellite around KOI-174.01 
exists in Q1-9.

\begin{figure*}
\begin{center}
\includegraphics[width=18.0 cm]{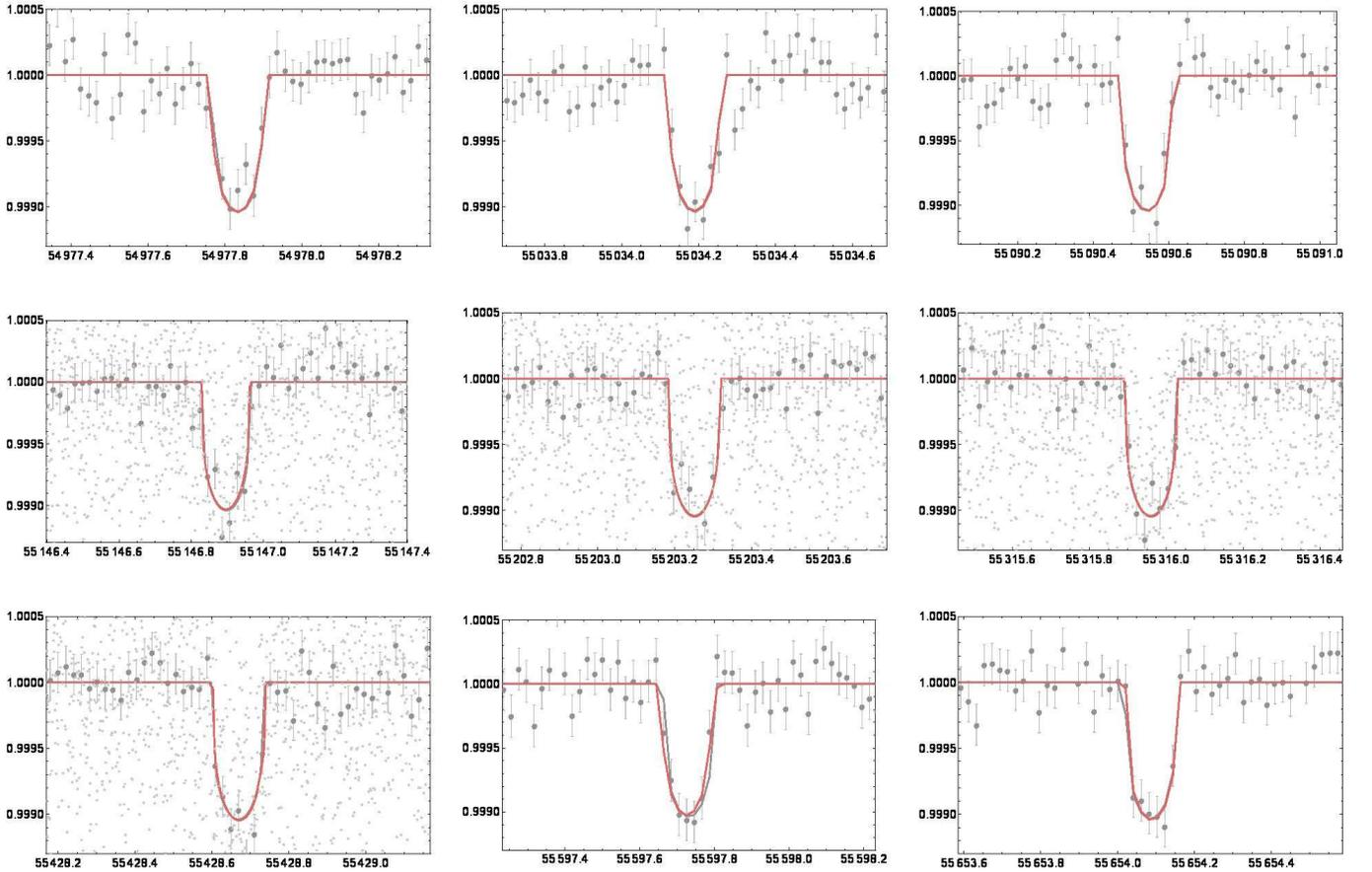}
\caption{\emph{From left-to-right then top-to-bottom we show
the chronological sequence of transits of KOI-174.01. The
first 8 panels show the Q1-9 data and the maximum a-posteriori
light curve fit of a planet-only model (gray line) and a moon model
(red line). No clear distortions due to the moon are visible in this 
close-binary type solution.}} 
\label{fig:KOI174_Fs}
\end{center}
\end{figure*}

The maximum a-posteriori model fit of $\mathcal{F}_{\mathrm{S}}$ (shown
in Fig~\ref{fig:KOI174_Fs}) does not exhibit any clear auxiliary or mutual
events, despite $R_S/R_P$ converging away from zero. The situation echoes
that of KOI-365.01 and indeed both fits can be considered close-binary
solutions. The planet-moon separation again converges to just a few
planetary radii away ($a_{SP}/R_P = 8.3_{-3.3}^{+4.9}$), close enough 
that the moon appears essentially on-top of the planet in every transit.
For the same reasons as described with KOI-365.01, this close-binary
solution results in poor constraints on the exomoon mass. As shown in
Fig.~\ref{fig:KOI174_Msp}, the $M_S/M_P$ posterior is unconverged
yielding a 95\% upper limit of $M_S/M_P<0.86$ and a 3\,$\sigma$ upper
limit of $M_S/M_P<0.99$. Our final system parameters are provided in 
Table~\ref{tab:KOI174_parameters}.

%%% KOI-174 Fs fit

%%% KOI-174 Evidences
\begin{table}
\caption{\emph{Bayesian evidences of various fits for KOI-174.01.
A description of the different models can be found in 
\S\ref{sub:fitsoverview}.}} %title of Table
\centering % used for centering table
\begin{tabular}{l c l} % centered columns (3 columns)
\hline
Model, $\mathcal{M}$ & $\mathrm{log}\mathcal{Z}(\mathcal{M})$ & $\tilde{\mathcal{M}}_1 - \tilde{\mathcal{M}}_2$ \\ [0.5ex] % inserts table
 &  & $= \mathrm{log}\mathcal{Z}(\mathcal{M}_1) - \mathrm{log}\mathcal{Z}(\mathcal{M}_2)$ \\ [0.5ex] % inserts table
%heading
\hline
\multicolumn{3}{l}{\emph{Planet only fits...}}\\ 
$\mathcal{V}_{\mathrm{P}}$  	& $68914.54 \pm 0.16$ 	& -	\\
$\mathcal{V}_{\mathrm{P,LD}}$	& $68922.76 \pm 0.16$	& $\tilde{\mathcal{V}}_{\mathrm{P,LD}}-\tilde{\mathcal{V}}_{\mathrm{P}} = (+8.22\pm0.22)$ \\
$\mathcal{V}_{\mathrm{P,MAP}}$	& $68914.48 \pm 0.16$	& $\tilde{\mathcal{V}}_{\mathrm{P,MAP}}-\tilde{\mathcal{V}}_{\mathrm{P}} = (-0.06\pm0.22)$ \\
$\mathcal{F}_{\mathrm{P}}$	& $68987.21 \pm 0.10$	& $\tilde{\mathcal{F}}_{\mathrm{P}}-\tilde{\mathcal{V}}_{\mathrm{P,MAP}} = (+72.67\pm0.19)$ \\
\hline
\multicolumn{3}{l}{\emph{Planet with timing variations fits...}}\\
$\mathcal{F}_{\mathrm{TTV}}$	& $68964.57 \pm 0.13$	& $\tilde{\mathcal{F}}_{\mathrm{TTV}}-\tilde{\mathcal{F}}_{\mathrm{P}} = (-22.65\pm0.16)$ \\
$\mathcal{V}_{\mathrm{V}}$	& $68825.27 \pm 0.21$	& $\tilde{\mathcal{V}}_{\mathrm{V}}-\tilde{\mathcal{V}}_{\mathrm{P,MAP}} = (-89.27\pm0.26)$ \\
\hline
\multicolumn{3}{l}{\emph{Planet with moon fits...}}\\ 
$\mathcal{F}_{\mathrm{S}}$	& $68988.75 \pm 0.10$	& $\tilde{\mathcal{F}}_{\mathrm{S}}-\tilde{\mathcal{F}}_{\mathrm{P}} = (+1.54\pm0.14)$ \\ % 
$\mathcal{F}_{\mathrm{S,M0}}$	& $68991.09 \pm 0.10$	& $\tilde{\mathcal{F}}_{\mathrm{S,M0}}-\tilde{\mathcal{F}}_{\mathrm{S}} = (+2.34\pm0.14)$ \\ %  
$\mathcal{F}_{\mathrm{S,R0}}$	& $68979.64 \pm 0.11$	& $\tilde{\mathcal{F}}_{\mathrm{S,R0}}-\tilde{\mathcal{F}}_{\mathrm{S}} = (-9.12\pm0.15)$ \\ [1ex] %
\hline\hline %inserts single line
\end{tabular}
\label{tab:KOI174_evidences} % is used to refer this table in the text
\end{table} % title of Table

%%% KOI-174 System parameters
\begin{table}
\caption{\emph{System parameters for KOI-174.01 from model 
$\mathcal{V}_{\mathrm{P,LD}}$, except for $M_S/M_P$ which is derived from model
$\mathcal{F}_{\mathrm{S}}$. $^{*}$ indicates that a parameter was fixed.}}
%title of Table
\centering % used for centering table
\begin{tabular}{l l} % centered columns (2 columns)
\hline
Parameter & Value\\ [0.5ex] % inserts table
%heading
\hline
\multicolumn{2}{l}{\emph{Derived parameters...}}\\ 
$P_P$\,[days] & $56.35439_{-0.00019}^{+0.00018}$ \\
$\tau_0$\,[BJD$_{\mathrm{UTC}}$] & $2455315.96099_{-0.00075}^{+0.00076}$ \\
$R_P/R_{\star}$ & $0.02908_{-0.00077}^{+0.00096}$ \\
$b$ & $0.46_{-0.12}^{+0.12}$ \\
$(a/R_{\star})$ & $119.5_{-9.9}^{+7.6}$ \\
$i$\,[deg] & $89.781_{-0.082}^{+0.066}$ \\
$\rho_{\star}$\,[g\,cm$^{-3}$] & $10.2_{-2.3}^{+2.1}$ \\
$\tilde{T}$\,[hours] & $3.195_{-0.058}^{+0.066}$ \\
$u_1$ & $0.90_{-0.26}^{+0.23}$ \\
$(u_1+u_2)$ & $0.80_{-0.14}^{+0.11}$ \\ 
\hline
\multicolumn{2}{l}{\emph{Physical parameters...}}\\ 
$M_{\star}$\,[$R_{\odot}$] & $0.80^{*}$ \\
$R_{\star}$\,[$R_{\odot}$] & $0.80^{*}$ \\
$R_P$\,[$R_{\oplus}$] & $2.537_{-0.067}^{+0.083}$ \\
$M_S/M_P$ & $<0.86$ [95\% confidence] \\ %[1ex]
$\delta_{\mathrm{TTV}}$\,[mins] & $<6.5$ (95\% confidence) \\
$\delta_{\mathrm{TDV}}$\,[mins] & $<5.8$ (95\% confidence) \\ [1ex]
\hline\hline %inserts single line
\end{tabular}
\label{tab:KOI174_parameters} % is used to refer this table in the text
\end{table} % title of Table

\subsubsection{Summary}

We find no compelling evidence for an exomoon around KOI-174.01 and estimate
that $M_S/M_P<0.86$ to 95\% confidence. This assessment is based on the fact
the system fails the basic detection criteria B1, B2, B3 and B4 (see 
\S\ref{sub:criteria}).

%%%%%%%%%%%%%%%%%%%%%%%%%%%%%%%%%%%%%%%%%%%%%%%%%%%%%%%%%%%%%%%%%%%%%%%%%%%%%%%%

\subsection{KOI-1472.01}
\label{sub:koi1472}

%% KOI-1472
%%

\subsubsection{Data selection}

After detrending with \cofiam, the PA and PDC-MAP data were both found to have a 
1.4\,$\sigma$ confidence of autocorrelation on a 30\,minute timescale and 
therefore both were acceptable ($<3$\,$\sigma$). 
As before, we opt to use the PA data in what follows. No short-cadence data is 
available for this system and so long-cadence data only was used in what 
follows.

\subsubsection{Planet-only fits}

% Limb darkening
When queried from MAST, the KIC effective temperature and surface gravity
were reported as $T_{\mathrm{eff}} = 5455$\,K and $\log g = 4.916$ 
\citep{brown:2011}. Using these values, we estimated quadratic limb darkening 
coefficients $u_1 = 0.4898$ and $(u_1+u_2) = 0.7037$. The initial two models we 
regressed were $\mathcal{V}_{\mathrm{P}}$ and $\mathcal{V}_{\mathrm{P,LD}}$, 
where the former uses the theoretical limb darkening coefficients as fixed 
values and the latter allows the two coefficients to be free parameters. We find 
that $\log\mathcal{Z}(\mathcal{V}_{\mathrm{P,LD}}) - 
\log\mathcal{Z}(\mathcal{V}_{\mathrm{P}}) = -1.51\pm0.22$ suggesting that our 
limb darkening coefficients are satisfactory.

% V -> P
KOI-1472.01 has a period of $P_P = (85.35174 \pm 0.00020)$\,days (as determined 
by model $\mathcal{V}_{\mathrm{P,LD}}$) and exhibits 9 transits from Q1-Q9, with no
transits absent from epoch -4 to +4. As is typical for all cases, 
$\log\mathcal{Z}(\mathcal{F}_{\mathrm{P}}) > 
\log\mathcal{Z}(\mathcal{V}_{\mathrm{P}})$ indicating that allowing for 9 
independent baseline parameters is unnecessary relative to a single baseline 
term.

% TTV
We find no evidence for TTVs in KOI-1472.01, with 
$\log\mathcal{Z}(\mathcal{F}_{\mathrm{TTV}}) 
- \log\mathcal{Z}(\mathcal{F}_{\mathrm{P}}) = -19.85\pm0.17$, which is formally 
an 6.0\,$\sigma$ preference for a static model over a TTV model. The timing 
precision on the 9 transits ranged from 2.0 to 3.6 minutes (from model 
$\mathcal{F}_{\mathrm{TTV}}$) and yields a flat TTV 
profile, as shown in Fig.~\ref{fig:KOI1472_TTVs}. We calculate a standard
deviation of $\delta_{\mathrm{TTV}} = 4.6$\,minutes and $\chi_{\mathrm{TTV}}^2 
= 34.9$ for 9-2 degrees of freedom. Although the $\chi^2$ is somewhat excessive,
the excess is not significant in light of the increased parameter volume
required to produce these results i.e. the Bayesian evidence is lower than a 
static model.

% TDV
The TTV+TDV model fit, $\mathcal{V}_{\mathrm{V}}$, finds consistent transit 
times with those derived by model $\mathcal{F}_{\mathrm{TTV}}$. As before, the
data show no clear pattern or excessive scatter, visible in 
Fig.~\ref{fig:KOI1472_TTVs}. We therefore conclude there is presently no evidence 
for TTVs or TDVs for KOI-1472.01. The standard deviation of the TDVs is found to 
be $\delta_{\mathrm{TDV}} = 6.0$\,minutes and we determine 
$\chi_{\mathrm{TDV}}^2 = 34.2$ for 9-1 degrees of freedom.

%%% KOI-1472 TTV Table
\begin{table*}
\caption{\emph{Transit times and durations for KOI-1472.01. The model used to 
calculate the supplied values is provided in parentheses next to each column
heading. BJD$_{\mathrm{UTC}}$ times offset by $2,400,000$ days.
}} % title of Table
\centering % used for centering table
\begin{tabular}{c c c c c c c} % centered columns (3 columns)
\hline
Epoch & $\tau$ [BJD$_{\mathrm{UTC}}$] ($\mathcal{F}_{\mathrm{TTV}}$) & TTV [mins] ($\mathcal{F}_{\mathrm{TTV}}$)
      & $\tau$ [BJD$_{\mathrm{UTC}}$] ($\mathcal{V}_{\mathrm{V}}$) & TTV [mins] ($\mathcal{V}_{\mathrm{V}}$)
      & $\tilde{T}$ [mins] ($\mathcal{V}_{\mathrm{V}}$) & TDV [mins] ($\mathcal{V}_{\mathrm{V}}$) \\ [0.5ex] % inserts table
%heading
\hline
-4 & $54994.0093_{-0.0014}^{+0.0014}$ & $+5.1 \pm 2.0$
   & $54994.0093_{-0.0014}^{+0.0014}$ & $5.81 \pm 2.0$
   & $387.9_{-4.5}^{+4.6}$ & $2.9 \pm 2.3$ \\
-3 & $55079.3573_{-0.0026}^{+0.0024}$ & $-0.3 \pm 3.6$
   & $55079.3530_{-0.0120}^{+0.0080}$ & $-6 \pm 14$
   & $394_{-26}^{+38}$ & $6 \pm 16$ \\
-2 & $55164.7086_{-0.0014}^{+0.0015}$ & $-0.9 \pm 2.1$
   & $55164.7086_{-0.0015}^{+0.0014}$ & $-0.4 \pm 2.1$
   & $383.8_{-4.9}^{+4.9}$ & $0.9 \pm 2.5$ \\
-1 & $55250.0537_{-0.0016}^{+0.0015}$ & $-10.5 \pm 2.2$
   & $55250.0554_{-0.0018}^{+0.0020}$ & $-7.5 \pm 2.8$
   & $362.6_{-6.3}^{+6.0}$ & $-9.7 \pm 3.1$ \\
+0 & $55335.4153_{-0.0015}^{+0.0015}$ & $+3.7 \pm 2.2$
   & $55335.4153_{-0.0015}^{+0.0016}$ & $4.0 \pm 2.2$
   & $386.8_{-5.0}^{+5.1}$ & $2.4 \pm 2.5$ \\
+1 & $55420.7624_{-0.0016}^{+0.0016}$ & $-3.1 \pm 2.3$
   & $55420.7626_{-0.0016}^{+0.0016}$ & $-2.5 \pm 2.3$
   & $401.4_{-5.1}^{+5.3}$ & $9.7 \pm 2.6$ \\
+2 & $55506.1166_{-0.0014}^{+0.0014}$ & $+0.5 \pm 2.0$
   & $55506.1169_{-0.0015}^{+0.0015}$ & $1.2 \pm 2.2$
   & $373.1_{-5.0}^{+5.1}$ & $-4.5 \pm 2.5$ \\
+3 & $55591.4683_{-0.0019}^{+0.0020}$ & $+0.4 \pm 2.8$
   & $55591.4683_{-0.0019}^{+0.0021}$ & $0.5 \pm 2.9$
   & $387.7_{-6.4}^{+6.4}$ & $2.8 \pm 3.2$ \\
+4 & $55676.8215_{-0.0014}^{+0.0015}$ & $+2.6 \pm 2.1$
   & $55676.8216_{-0.0014}^{+0.0014}$ & $2.8 \pm 2.0$
   & $373.5_{-4.6}^{+4.6}$ & $-4.3 \pm 2.3$ \\ [1ex]
\hline\hline %inserts single line
\end{tabular}
\label{tab:KOI1472_TTVs} % is used to refer this table in the text
\end{table*} % title of Table

\subsubsection{Moon fits}

A planet-with-moon fit, $\mathcal{F}_S$, is favored relative to a 
planet-only fit at $\Delta(\log\mathcal{Z}) = +9.5\pm0.2$ (see
Table~\ref{tab:KOI1472_evidences}). The formal significance equates to 
3.96\,$\sigma$ and can therefore be considered to lie on the margin of 
satisfying detection criterion B1.

The posteriors reveal a solution hitting against the lower boundary 
condition on the planet-moon separation, yielding $a_{SP}/R_P = 
3.2_{-0.9}^{+11.0}$. The proximity between the two objects is particularly 
extreme when one notes that the fit also finds a substantial satellite size of
$R_S/R_P = 0.40_{-0.14}^{+0.16}$. The solution is therefore consistent
with a close-binary, as we also found for KOI-365.01 and KOI-174.01.
Due to the low density of the planet solution relative to that of the 
satellite, the posteriors indicate a Roche limit well-inside the planet and 
thus we cannot exclude the satellite with a tidal disruption argument.

\begin{figure*}
\begin{center}
\includegraphics[width=18.0 cm]{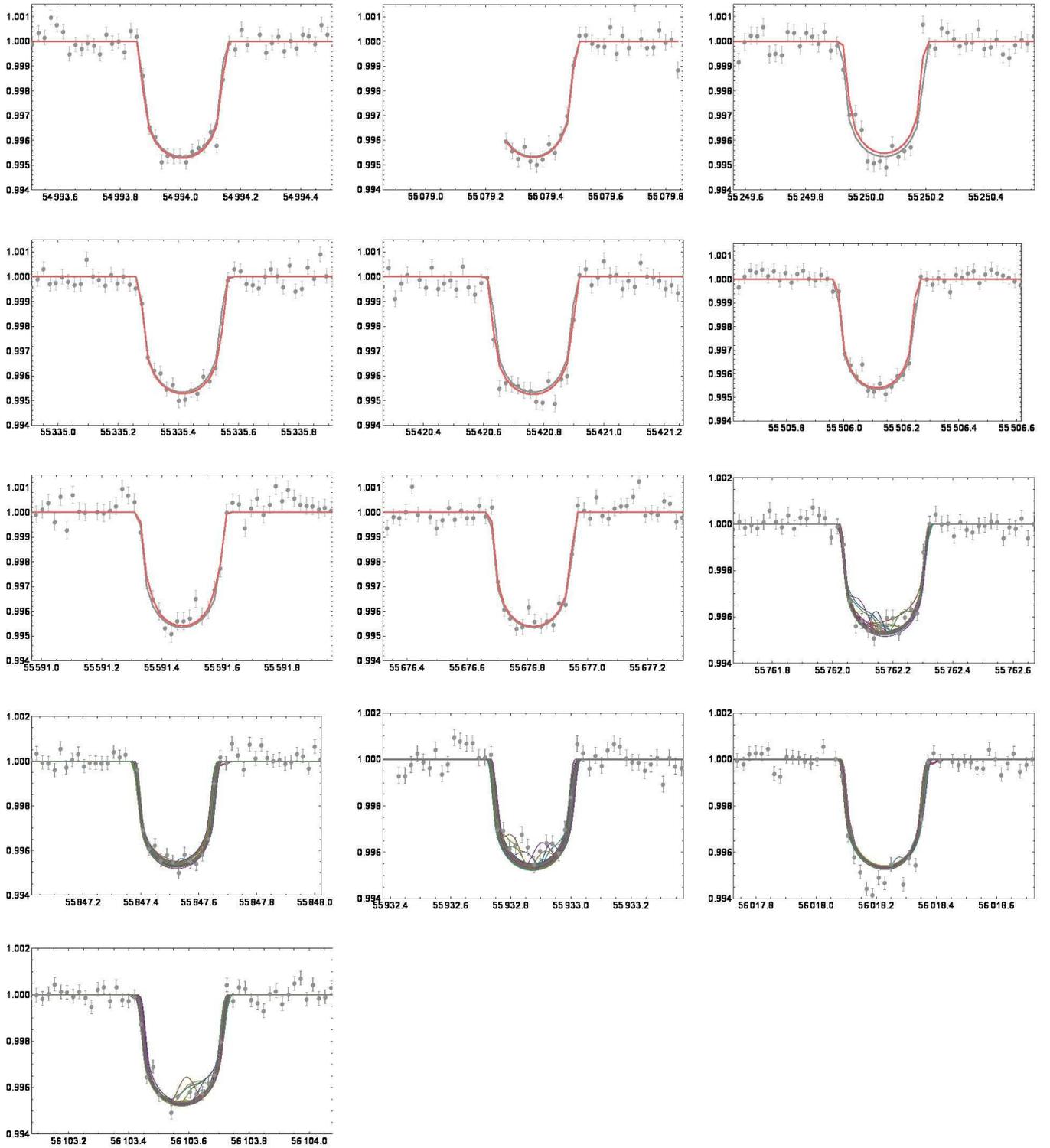}
\caption{\emph{From left-to-right then top-to-bottom we show
the chronological sequence of transits of KOI-1472.01. The
first 8 panels show the Q1-9 data and the maximum a-posteriori
light curve fit of a planet-only model (gray line) and a moon model
(red line). The last 5 panels show the Q10-13 data with 50
extrapolations (50 different shadings used) of the moon model overlaid 
(parameters randomly drawn from the joint posteriors), which exhibit poor 
predictive power.}} 
\label{fig:KOI1472_Fs}
\end{center}
\end{figure*}

The stellar parameters for the host star are estimated in B12
as $M_{\star} = 0.93$\,$M_{\odot}$ and $R_{\star} = 0.56$\,$R_{\odot}$, from 
which we can estimate physical parameters for the candidate planet-moon system. 
From model $\mathcal{F}_{\mathrm{S}}$, we determine planetary
parameters of $M_P = 18.6_{-11.5}^{+29.0}$\,$M_{\oplus}$ and
$R_P = 3.50_{-0.21}^{+0.16}$\,$R_{\oplus}$ indicating a low density
of $\rho_P = 0.87_{-0.59}^{+1.50}$\,g\,cm$^{-3}$, making the planet
a small Neptune-type planet. The satellite returns 
$M_S = 9.3_{-7.0}^{+21.2}$\,$M_{\oplus}$ and 
$R_S = 1.40_{-0.46}^{+0.44}$\,$R_{\oplus}$ suggesting a Super-Earth
type moon. The satellite's period is $P_S = 0.89_{-0.44}^{+3.46}$\,days
and thus low enough (relative to the transit duration of 0.26\,days)
to conveniently mask TTV and TDV effects. As with the previous two
close-binary solutions, radius-effects are also difficult to identify
in the maximum a-posteriori fit shown in Fig.~\ref{fig:KOI1472_Fs}.
Although the satellite candidate is just a few planetary radii away from
its host, the posteriors are broadly physical and we therefore conclude 
that detection criterion B3 is satisfied.

Out of all of the moon fits attempted, the highest Bayesian evidence comes
from $\mathcal{F}_{\mathrm{S,R0}}$ (a zero-radius moon model) as shown
in Table~\ref{tab:KOI1472_evidences}, meaning that detection criterion
B2 is failed. We note that the $M_S/M_P$ ratio is not well-converged in 
the $\mathcal{F}_{\mathrm{S}}$ model, but does appear divergent
from zero, broadly satisfying criterion B4. 

In conclusion, KOI-1472.01 passes B1, B3 and B4 but fails criterion B2.
We therefore consider that more data may resolve whether the moon
hypothesis is plausible or not.

%%% KOI-1472 Fs fit

\subsubsection{Predictive power of the moon model}

From a $\chi^2$ perspective, the maximum a-posteriori realization from
model $\mathcal{F}_{\mathrm{S}}$ is naturally lower than that of
$\mathcal{F}_{\mathrm{P}}$ where we find 2181.01 versus 2248.48
respectively, for 1495 data points spanning Q1-9. At the time of
writing, Q10-13 had recently become available and this data may be
used as a test between the planet and planet-with-moon hypotheses.
If our moon model is genuine, then extrapolating the model into
Q10-13 should yield a better prediction (in a $\chi^2$ sense) than
the simple planet-only model. We downloaded and detrended this
data accordingly using \cofiam\ and the PA time series, which covers
5 new transits and 815 new LC data points. Given that we got a $\chi^2$
improvement of 67.5 over Q1-9 (1495 points), we might expect the
moon model to yield an improvement of $\sim30$ in Q10-13.

We extrapolated the maximum a-posteriori realization of model 
$\mathcal{F}_{\mathrm{S}}$ into Q10-13 and find $\chi^2 = 1222.85$ for 815 data
points. Repeating the process for $\mathcal{F}_{\mathrm{P}}$ yields $\chi^2 = 
1196.54$. We therefore find that the moon model has substantially \emph{worse} 
predictive power than a simple planet-only model. Since the maximum 
a-posteriori realization is just a single realization, we decided to 
extrapolate 50 realizations randomly drawn from the joint posteriors of 
$\mathcal{F}_{\mathrm{S}}$ in order to account for the parameter uncertainties 
(shown in Fig.~\ref{fig:KOI1472_Fs}). From these 50 realizations, we compute
50 $\chi^2$ values from which we take the mean to be $1233\pm22$. Repeating
the process for $\mathcal{F}_{\mathrm{P}}$ yields $\chi^2=1201.7\pm8.5$,
again supporting the planet-only model over the planet-with-moon model.

These results clearly show that the planet-only model has superior predictive
power to that of the moon model, thus the moon hypothesis fails detection
criterion F2. 

\subsubsection{Refitting the updated data set}

The $M_S/M_P$ posterior derived from Q1-9 is converged off-zero due to a 
spurious detection. In order to place constraints on this this parameter, we
require a posterior derived from a null detection instead. To this end,
we decided to re-fit the Q1-13 data with an updated moon model, 
$\mathcal{F}_{\mathrm{S},13}$. A secondary goal was to check whether detection 
criterion F3 could be satisfied by refitting i.e. the same signal is retrieved as 
before but with higher significance.

The regression yielded a distinct solution with both the semi-major axis
and period of the moon moving outwards by many sigma. Specifically, we find
$a_{SP}/R_P = 69_{-11}^{+18}$ and $P_S = 29.9_{-7.7}^{+16.1}$\,days. The
mass ratio posterior, $M_S/M_P$, also shows a distinct profile and now
converges close to zero with $M_S/M_P = 0.0117_{-0.0052}^{+0.0118}$, which
is consistent with a null-detection. The fact the $M_S/M_P$ posterior converges
close to zero means that criterion F1 is not satisfied and the fact that the 
solution is distinct from that obtained by the Q1-9 data means that criterion F3 
is also not satisfied. In summary then, all three follow-up criteria are not 
satisfied and on this basis we conclude that there is no evidence for an exomoon 
around KOI-1472.01.

In Table~\ref{tab:KOI1472_parameters} we display our final system parameters for
KOI-1472.01. Given that the Q1-9 fits yielded a spurious moon detection but the
Q1-13 fits are consistent with a null-detection, we use the latter for our
constraints on $M_S/M_P$. The posterior of $M_S/M_P$ from 
$\mathcal{F}_{\mathrm{S},13}$, shown in Fig.~\ref{fig:KOI1472_Msp}, yields a
95\% quantile of $M_S/M_P<0.037$ and a 3\,$\sigma$ quantile of $M_S/M_P<0.063$.

%%% KOI-1472 Evidences
\begin{table}
\caption{\emph{Bayesian evidences of various fits for KOI-1472.01.
A description of the different models can be found in \S\ref{sub:fitsoverview}.
The ``13'' subscript denotes that Q1-13 data was used rather than Q1-9.}} %title of Table
\centering % used for centering table
\begin{tabular}{l c l} % centered columns (3 columns)
\hline
Model, $\mathcal{M}$ & $\mathrm{log}\mathcal{Z}(\mathcal{M})$ & $\tilde{\mathcal{M}}_1 - \tilde{\mathcal{M}}_2$ \\ [0.5ex] % inserts table
 &  & $= \mathrm{log}\mathcal{Z}(\mathcal{M}_1) - \mathrm{log}\mathcal{Z}(\mathcal{M}_2)$ \\ [0.5ex] % inserts table
%heading
\hline
\multicolumn{3}{l}{\emph{Planet only fits...}}\\ 
$\mathcal{V}_{\mathrm{P}}$  	& $9485.63 \pm 0.15$ 	& -	\\
$\mathcal{V}_{\mathrm{P,LD}}$	& $9484.12 \pm 0.15$	& $\tilde{\mathcal{V}}_{\mathrm{P,LD}}-\tilde{\mathcal{V}}_{\mathrm{P}} = (-1.51\pm0.22)$ \\
$\mathcal{F}_{\mathrm{P}}$	& $9541.89 \pm 0.10$	& $\tilde{\mathcal{F}}_{\mathrm{P}}-\tilde{\mathcal{V}}_{\mathrm{P}} = (+56.26\pm0.18)$ \\
\hline
\multicolumn{3}{l}{\emph{Planet with timing variations fits...}}\\ 
$\mathcal{F}_{\mathrm{TTV}}$	& $9522.04 \pm 0.13$	& $\tilde{\mathcal{F}}_{\mathrm{TTV}}-\tilde{\mathcal{F}}_{\mathrm{P}} = (-19.85\pm0.17)$ \\
$\mathcal{V}_{\mathrm{V}}$	& $9385.63 \pm 0.21$	& $\tilde{\mathcal{V}}_{\mathrm{V}}-\tilde{\mathcal{V}}_{\mathrm{P}} = (-100.00\pm0.26)$ \\
\hline
\multicolumn{3}{l}{\emph{Planet with moon fits...}}\\ 
$\mathcal{F}_{\mathrm{S}}$	& $9551.38 \pm 0.11$	& $\tilde{\mathcal{F}}_{\mathrm{S}}-\tilde{\mathcal{F}}_{\mathrm{P}} = (+9.49\pm0.15)$ \\ % 
$\mathcal{F}_{\mathrm{S,M0}}$	& $9550.19 \pm 0.11$	& $\tilde{\mathcal{F}}_{\mathrm{S,M0}}-\tilde{\mathcal{F}}_{\mathrm{S}} = (-1.18\pm0.16)$ \\ %  
$\mathcal{F}_{\mathrm{S,R0}}$	& $9552.32 \pm 0.11$	& $\tilde{\mathcal{F}}_{\mathrm{S,R0}}-\tilde{\mathcal{F}}_{\mathrm{S}} = (+1.03\pm0.16)$ \\ [1ex] %
\hline
$\mathcal{F}_{\mathrm{P},13}$	& $14782.31 \pm 0.10$	& - \\
$\mathcal{F}_{\mathrm{S},13}$	& $14810.85 \pm 0.13$	& $\tilde{\mathcal{F}}_{\mathrm{S},13}-\tilde{\mathcal{F}}_{\mathrm{P},13} = (+28.54\pm0.16)$ \\
\hline\hline %inserts single line
\end{tabular}
\label{tab:KOI1472_evidences} % is used to refer this table in the text
\end{table} % title of Table

%%% KOI-1472 System parameters
\begin{table}
\caption{\emph{System parameters for KOI-1472.01 from model 
$\mathcal{V}_{\mathrm{P,LD}}$, except for $M_S/M_P$ which is derived from model
$\mathcal{F}_{\mathrm{S},13}$. $^{*}$ indicates that a parameter was fixed.}}
%title of Table
\centering % used for centering table
\begin{tabular}{l l} % centered columns (2 columns)
\hline
Parameter & Value\\ [0.5ex] % inserts table
%heading
\hline
\multicolumn{2}{l}{\emph{Derived parameters...}}\\ 
$P_P$\,[days] & $85.35174_{-0.00020}^{+0.00020}$ \\
$\tau_0$\,[BJD$_{\mathrm{UTC}}$] & $2455335.41282_{-0.00053}^{+0.00054}$ \\
$R_P/R_{\star}$ & $0.06422_{-0.0023}^{+0.0028}$ \\
$b$ & $0.52_{-0.33}^{+0.14}$ \\
$(a/R_{\star})$ & $88.5_{-9.7}^{+12.5}$ \\
$i$\,[deg] & $89.66_{-0.14}^{+0.23}$ \\
$\rho_{\star}$\,[g\,cm$^{-3}$] & $1.80_{-0.53}^{+0.88}$ \\
$\tilde{T}$\,[hours] & $6.276_{-0.080}^{+0.081}$ \\
$u_1$ & $0.63_{-0.22}^{+0.31}$ \\
$(u_1+u_2)$ & $0.53_{-0.13}^{+0.17}$ \\ 
\hline
\multicolumn{2}{l}{\emph{Physical parameters...}}\\ 
$M_{\star}$\,[$R_{\odot}$] & $0.93^{*}$ \\
$R_{\star}$\,[$R_{\odot}$] & $0.56^{*}$ \\
$R_P$\,[$R_{\oplus}$] & $3.92_{-0.14}^{+0.17}$ \\
$M_S/M_P$ & $<0.037$ (95\% confidence) \\ %[1ex]
$\delta_{\mathrm{TTV}}$\,[mins] & $<2.7$ (95\% confidence) \\
$\delta_{\mathrm{TDV}}$\,[mins] & $<2.3$ (95\% confidence) \\ [1ex]
\hline\hline %inserts single line
\end{tabular}
\label{tab:KOI1472_parameters} % is used to refer this table in the text
\end{table} % title of Table

\subsubsection{Summary}

We find no compelling evidence for an exomoon around KOI-1472.01 and estimate
that $M_S/M_P<0.037$ to 95\% confidence. This assessment is based on the fact
the system fails the basic detection criterion B2 as well as the follow-up
criteria F1, F2 and F3 (see \S\ref{sub:criteria}).

%%%%%%%%%%%%%%%%%%%%%%%%%%%%%%%%%%%%%%%%%%%%%%%%%%%%%%%%%%%%%%%%%%%%%%%%%%%%%%%%

\subsection{KOI-1857.01}
\label{sub:koi1857}

%% KOI-1857
%%

\subsubsection{Data selection}

After detrending with \cofiam, the PA and PDC-MAP data were found to have a 
2.3\,$\sigma$ and 2.3\,$\sigma$ confidence of autocorrelation on a 30\,minute 
timescale respectively and therefore both were acceptable ($<3$\,$\sigma$). As
with previous systems, we choose to use the PA data over the PDC-MAP data as
both are acceptable. No short-cadence data is available for this system and so 
long-cadence data only was used in what follows.

\subsubsection{Planet-only fits}

% Limb darkening
When queried from MAST, the KIC effective temperature and surface gravity
were reported as $T_{\mathrm{eff}} = 5619$\,K and $\log g = 4.527$ 
\citep{brown:2011}. Using these values, we estimated quadratic limb darkening 
coefficients $u_1 = 0.4567$ and $(u_1+u_2) = 0.6910$. The initial two models we 
regressed were $\mathcal{V}_{\mathrm{P}}$ and $\mathcal{V}_{\mathrm{P,LD}}$, 
where the former uses the theoretical limb darkening coefficients as fixed 
values and the latter allows the two coefficients to be free parameters. We find 
that $\log\mathcal{Z}(\mathcal{V}_{\mathrm{P,LD}}) - 
\log\mathcal{Z}(\mathcal{V}_{\mathrm{P}}) = +0.58\pm0.21$ suggesting that our 
limb darkening coefficients may not be optimal. We thus chose to set the
coefficients to the maximum a-posteriori values of $u_1 = 0.5409$ and 
$(u_1+u_2) = 0.8872$.

% V -> P
KOI-1857.01 has a period of $P_P = (88.64486 \pm 0.00075)$\,days (as determined 
by model $\mathcal{V}_{\mathrm{P,LD}}$) and exhibits 8 transits from Q1-Q9 from 
epoch -4 to +4, except epoch -2 which is absent. As is typical for all cases, 
$\log\mathcal{Z}(\mathcal{F}_{\mathrm{P}}) > 
\log\mathcal{Z}(\mathcal{V}_{\mathrm{P}})$ indicating that allowing for 8 
independent baseline parameters is unnecessary relative 
to a single baseline term.

% TTV
We find no evidence for TTVs in KOI-1857.01, with 
$\log\mathcal{Z}(\mathcal{F}_{\mathrm{TTV}}) 
- \log\mathcal{Z}(\mathcal{F}_{\mathrm{P}}) = -18.59\pm0.15$, which is formally 
an 5.8\,$\sigma$ preference for a static model over a TTV model. The timing 
precision on the 8 transits ranged from 6.8 to 9.5 minutes (from model 
$\mathcal{F}_{\mathrm{TTV}}$) and yields a flat TTV 
profile, as shown in Fig.~\ref{fig:KOI1857_TTVs}. We calculate a standard
deviation of $\delta_{\mathrm{TTV}} = 11.0$\,minutes and $\chi_{\mathrm{TTV}}^2 
= 13.4$ for 8-2 degrees of freedom.

% TDV
The TTV+TDV model fit, $\mathcal{V}_{\mathrm{V}}$, finds consistent transit 
times with those derived by model $\mathcal{F}_{\mathrm{TTV}}$. We detect no 
clear pattern or excessive scatter in the data, visible in 
Fig.~\ref{fig:KOI1857_TTVs}. We therefore conclude there is presently no 
evidence for TTVs or TDVs for KOI-1857.01. The standard deviation of the TDVs is 
found to be $\delta_{\mathrm{TDV}} = 18.6$\,minutes and we determine 
$\chi_{\mathrm{TDV}}^2 = 7.4$ for 8-1 degrees of freedom.

%%% KOI-1857 TTV Table
\begin{table*}
\caption{\emph{Transit times and durations for KOI-1857.01. The model used to 
calculate the supplied values is provided in parentheses next to each column
heading. BJD$_{\mathrm{UTC}}$ times offset by $2,400,000$ days.
}} % title of Table
\centering % used for centering table
\begin{tabular}{c c c c c c c} % centered columns (3 columns)
\hline
Epoch & $\tau$ [BJD$_{\mathrm{UTC}}$] ($\mathcal{F}_{\mathrm{TTV}}$) & TTV [mins] ($\mathcal{F}_{\mathrm{TTV}}$)
      & $\tau$ [BJD$_{\mathrm{UTC}}$] ($\mathcal{V}_{\mathrm{V}}$) & TTV [mins] ($\mathcal{V}_{\mathrm{V}}$)
      & $\tilde{T}$ [mins] ($\mathcal{V}_{\mathrm{V}}$) & TDV [mins] ($\mathcal{V}_{\mathrm{V}}$) \\ [0.5ex] % inserts table
%heading
\hline
-4 & $54978.4743_{-0.0047}^{+0.0048}$ & $+11.6 \pm 6.9$
   & $54978.4744_{-0.0047}^{+0.0050}$ & $+7.1 \pm 7.0$
   & $471_{-17}^{+21}$ & $+8 \pm 9.7$ \\
-3 & $55067.1001_{-0.0056}^{+0.0056}$ & $-16.2 \pm 8.1$
   & $55067.1017_{-0.0071}^{+0.0075}$ & $-18 \pm 11$
   & $461_{-27}^{+27}$ & $+3 \pm 14$ \\
-1 & $55244.4050_{-0.0047}^{+0.0047}$ & $+4.8 \pm 6.8$
   & $55244.4029_{-0.0078}^{+0.0056}$ & $-0.2 \pm 9.7$
   & $493_{-23}^{+34}$ & $+18 \pm 14$ \\
+0 & $55333.0409_{-0.0051}^{+0.0051}$ & $-8.5 \pm 7.3$
   & $55333.0430_{-0.0059}^{+0.0063}$ & $-6.5 \pm 8.8$
   & $450_{-22}^{+23}$ & $-3 \pm 11$ \\
+1 & $55421.6891_{-0.0050}^{+0.0054}$ & $-4.1 \pm 7.5$
   & $55421.6896_{-0.0056}^{+0.0058}$ & $-3.6 \pm 8.3$
   & $471_{-19}^{+21}$ & $+7 \pm 10$ \\
+2 & $55510.3429_{-0.0058}^{+0.0061}$ & $+8.4 \pm 8.6$
   & $55510.3446_{-0.0070}^{+0.0065}$ & $+11.7 \pm 9.7$
   & $507_{-24}^{+29}$ & $+25 \pm 13$ \\
+3 & $55598.9909_{-0.0059}^{+0.0057}$ & $+12.5 \pm 8.4$
   & $55598.9899_{-0.0052}^{+0.0054}$ & $+12.8 \pm 7.6$
   & $450_{-19}^{+21}$ & $-3 \pm 10$ \\
+4 & $55687.6201_{-0.0068}^{+0.0064}$ & $-10.4 \pm 9.5$
   & $55687.6241_{-0.0080}^{+0.0088}$ & $-2 \pm 12$
   & $528_{-53}^{+128}$ & $+36 \pm 45$ \\ [1ex]
\hline\hline %inserts single line
\end{tabular}
\label{tab:KOI1857_TTVs} % is used to refer this table in the text
\end{table*} % title of Table

\subsubsection{Moon fits}

A planet-with-moon fit, $\mathcal{F}_S$, is slightly favoured relative to a 
planet-only fit at $\Delta(\log\mathcal{Z}) = +2.16\pm0.14$, or 1.57\,$\sigma$, 
meaning detection criterion B1 is not satisfied. The zero-mass moon model is 
also preferred meaning criterion B2 is not satisfied either.

\begin{figure*}
\begin{center}
\includegraphics[width=18.0 cm]{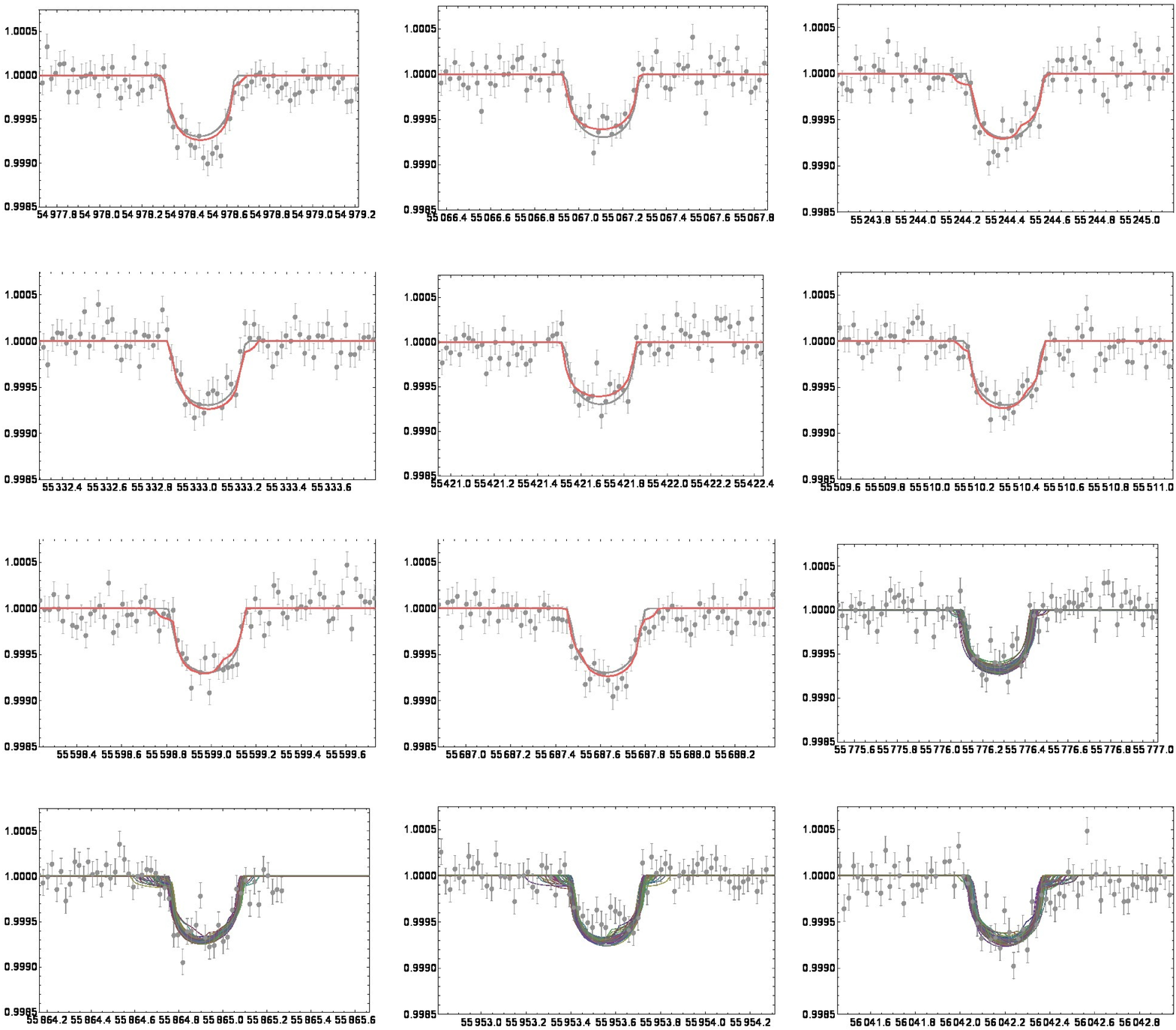}
\caption{\emph{From left-to-right then top-to-bottom we show
the chronological sequence of transits of KOI-1857.01. The
first 8 panels show the Q1-9 data and the maximum a-posteriori
light curve fit of a planet-only model (gray line) and a moon model
(red line). The last 4 panels show the Q10-13 data with 50
extrapolations (50 different shadings used) of the moon model overlaid 
(parameters randomly drawn from the joint posteriors), which do not exhibit 
significant predictive power.}} 
\label{fig:KOI1857_Fs}
\end{center}
\end{figure*}

Despite failing B1 and B2, the fits yield broadly physical parameters and
the low confidence could be indicative of a low signal-to-noise moon
embedded in the data. Using $M_{\star}=0.94$\,$M_{\odot}$ and 
$R_{\star}=0.85$\,$R_{\odot}$ from B12, we determine 
$M_P = 6.1_{-4.6}^{+19.3}$\,$M_{\oplus}$ for 
$R_P = 2.067_{-0.069}^{+0.077}$\,$R_{\oplus}$ and 
$M_S = 0.54_{-0.39}^{+1.55}$\,$M_{\oplus}$ for 
$R_S = 0.88_{-0.16}^{+0.14}$\,$R_{\oplus}$. The solution is also clearly
not a close-binary with $a_{SP}/R_P = 42_{-25}^{+12}$ and auxiliary
transits driving the fit, as evident from Fig.~\ref{fig:KOI1857_Fs}.
Finally, the mass and radius ratio posteriors show convergence away
from zero meaning criteria B3 and B4 are satisfied.

%%% KOI-1857 Fs fit

\subsubsection{Predictive power of the moon model}

At this stage, we considered KOI-1857.01 to be a potential candidate and
further data may confirm/reject the signal. We therefore detrended the
PA data for KOI-1857 from Q10-13 covering 4 new transits, which had 
recently become available at the time of writing. Before we attempted to 
re-fit the updated data set, we extrapolated the light curve model fit 
$\mathcal{F}_{\mathrm{S}}$ into Q10-13 for both the maximum 
a-posteriori solution and 50 randomly sampled solutions from the joint 
posteriors, which are shown in Fig.~\ref{fig:KOI1857_Fs}. The
maximum a-posteriori moon solution can be compared to the detrended
data, where we compute $\chi^2 = 744.8$ for 621 data points. For
comparison, we repeated the process for the maximum a-posteriori
planet-only model and compute $\chi^2 = 743.6$ i.e. nearly
identical but slightly worse. 

This comparison suggests that the moon model has no significant predictive 
power, which is a major concern for accepting the moon hypothesis. We 
should expect a moon model to give a better $\chi^2$ by around $\gtrsim10$ 
based upon the improvement of the best fits on the Q1-9 data. For Q1-9, we 
found $\Delta\chi^2 = 30.3$ between the two best fits for 1437 data points. 
The 50 realizations of light curve predictions, shown in 
Fig.~\ref{fig:KOI1857_Fs}, do not seem to show a convincing agreement with the 
data either. Indeed, the distribution of the $\chi^2$ values for these fits
yields $\chi^2 = 724\pm12$ whereas repeating the process for
50 predictions from $\mathcal{F}_{\mathrm{P}}$ yields $719\pm10$, which again
suggests that the planetary model actually has slightly better predictive 
power.

\subsubsection{Refitting the updated data set}

The moon model $\mathcal{F}_{\mathrm{S}}$ exhibits approximately
the same predictive power as a simple planet-only model (although technically
slightly worse). As the difference in predictive power is not substantial
(unlike with what we had with KOI-1472.01), we considered that the moon 
hypothesis could not yet be excluded. To fully exploit the Q10-13 data, we 
stitched the detrended data onto the previous Q1-9 time series and re-fitted 
the updated data with a planet-only model ($\mathcal{F}_{\mathrm{P}13}$) and a 
planet-with-moon model ($\mathcal{F}_{\mathrm{S}13}$).

We find that the significance of the moon solution has indeed been enhanced
moving from 1.57\,$\sigma$ to 3.81\,$\sigma$, although still below the
threshold of detection criterion B1.

The new solution finds a much more massive planet than before with 
$M_P = 82_{-53}^{+210}$\,$M_{\oplus}$ for
$R_P = 2.31_{-0.14}^{+0.29}$\,$R_{\oplus}$. The satellite appears
broadly unchanged in terms of mass and radius with
$M_S = 0.59_{-0.38}^{+1.57}$\,$M_{\oplus}$ for 
$R_S = 0.971_{-0.090}^{+0.095}$\,$R_{\oplus}$. Despite this, the orbital
solution is distinct with $a_{SP}/R_P = 78_{-42}^{+11}$ i.e. the moon
is at about twice the orbital separation than before. One significant
difference is that with just Q1-9 data, both the planet-only fit
and the planet-with-moon fit obtained a $\rho_{\star}$ value which appeared
consistent with the KIC spectral classification (late G). Specifically, 
we found $\rho_{\star}=1.4_{-0.4}^{+0.2}$\,g\,cm$^{-3}$ from the moon fit and 
from $1.3_{-0.7}^{+0.3}$\,g\,cm$^{-3}$ from the planet fit. However, the
$\mathcal{F}_{\mathrm{S},13}$ model pushes down to 
$0.53_{-0.33}^{+0.39}$\,g\,cm$^{-3}$ meaning that the star is now more
consistent with an F-dwarf. The KIC catalogue is known to be inaccurate
for K/M dwarfs but G/F dwarfs are quite reliably identified and thus this
discrepancy is suspicious. It is possible the planet is eccentric but given
the wide orbit of the moon solution, even a moderate eccentricity would
mean that the moon was unstable. We consider that detection criterion B3
is marginally satisfied. However, in these new fits the $M_S/M_P$
posterior now appears well-converged on zero, failing criterion B4.

With Q1-9, we found that the zero-mass moon model was preferred over the
physical moon model with $\Delta\log\mathcal{Z} = 7.77\pm0.15$, thus
failing detection criterion B2. This result is further supported by
the Q1-13 analysis where the same comparison yields 
$\Delta\log\mathcal{Z} = 10.09\pm0.16$. We therefore find that follow-up
detection criteria F1, F2 and F3 are all not satisfied and the candidate
can be dismissed. There is therefore no evidence for an exomoon around
KOI-1857.01 from Q1-13. From the Q1-9 analysis only, the $M_S/M_P$
posterior converges off zero and indicates $M_S/M_P<0.49$ to 95\% 
confidence and $M_S/M_P<0.95$ to 3\,$\sigma$ confidence. Including the 
Q10-13 data eliminates the mass convergent solution and leads to much
tighter constraints of $M_S/M_P<0.028$ to 95\% confidence and 
$M_S/M_P<0.052$ to 3\,$\sigma$ confidence (see Fig.~\ref{fig:KOI1857_Msp}).

%%% KOI-1857 Evidences
\begin{table}
\caption{\emph{Bayesian evidences of various fits for KOI-1857.01.
A description of the different models can be found in \S\ref{sub:fitsoverview}.
The ``13'' subscript denotes that Q1-13 data was used rather than Q1-9.}} 
%title of Table
\centering % used for centering table
\begin{tabular}{l c l} % centered columns (3 columns)
\hline
Model, $\mathcal{M}$ & $\mathrm{log}\mathcal{Z}(\mathcal{M})$ & $\tilde{\mathcal{M}}_1 - \tilde{\mathcal{M}}_2$ \\ [0.5ex] % inserts table
 &  & $= \mathrm{log}\mathcal{Z}(\mathcal{M}_1) - \mathrm{log}\mathcal{Z}(\mathcal{M}_2)$ \\ [0.5ex] % inserts table
%heading
\hline
\multicolumn{3}{l}{\emph{Planet only fits...}}\\ 
$\mathcal{V}_{\mathrm{P}}$  	& $10535.69 \pm 0.15$ 	& -	\\
$\mathcal{V}_{\mathrm{P,LD}}$	& $10536.27 \pm 0.15$	& $\tilde{\mathcal{V}}_{\mathrm{P,LD}}-\tilde{\mathcal{V}}_{\mathrm{P}} = (+0.58\pm0.21)$ \\
$\mathcal{V}_{\mathrm{P,MAP}}$  	& $10537.26 \pm 0.15$   & $\tilde{\mathcal{V}}_{\mathrm{P,MAP}}-\tilde{\mathcal{V}}_{\mathrm{P}} = (+1.57\pm0.21)$ \\
$\mathcal{F}_{\mathrm{P}}$	& $10592.10 \pm 0.09$	& $\tilde{\mathcal{F}}_{\mathrm{P}}-\tilde{\mathcal{V}}_{\mathrm{P,MAP}} = (+54.84\pm0.18)$ \\
\hline
\multicolumn{3}{l}{\emph{Planet with timing variations fits...}}\\
$\mathcal{F}_{\mathrm{TTV}}$	& $10573.51 \pm 0.12$	& $\tilde{\mathcal{F}}_{\mathrm{TTV}}-\tilde{\mathcal{F}}_{\mathrm{P}} = (-18.59\pm0.15)$ \\
$\mathcal{V}_{\mathrm{V}}$	& $10445.32 \pm 0.22$	& $\tilde{\mathcal{V}}_{\mathrm{V}}-\tilde{\mathcal{V}}_{\mathrm{P,MAP}} = (-91.94\pm0.27)$ \\
\hline
\multicolumn{3}{l}{\emph{Planet with moon fits...}}\\ 
$\mathcal{F}_{\mathrm{S}}$	& $10594.26 \pm 0.10$	& $\tilde{\mathcal{F}}_{\mathrm{S}}-\tilde{\mathcal{F}}_{\mathrm{P}} = (+2.16\pm0.14)$ \\ % 
$\mathcal{F}_{\mathrm{S,M0}}$	& $10602.03 \pm 0.11$	& $\tilde{\mathcal{F}}_{\mathrm{S,M0}}-\tilde{\mathcal{F}}_{\mathrm{S}} = (+7.77\pm0.15)$ \\ %  
$\mathcal{F}_{\mathrm{S,R0}}$	& $10592.41 \pm 0.10$	& $\tilde{\mathcal{F}}_{\mathrm{S,R0}}-\tilde{\mathcal{F}}_{\mathrm{S}} = (-1.85\pm0.15)$ \\ 
\hline
$\mathcal{F}_{\mathrm{P},13}$ & $15158.41 \pm 0.10$	& -  \\
$\mathcal{F}_{\mathrm{S},13}$ & $15167.30 \pm 0.10$	& $\tilde{\mathcal{F}}_{\mathrm{S},13}-\tilde{\mathcal{F}}_{\mathrm{P},13} = (+8.89\pm0.15)$ \\
$\mathcal{F}_{\mathrm{S,M0},13}$ & $15177.36 \pm 0.12$	& $\tilde{\mathcal{F}}_{\mathrm{S,M0},13}-\tilde{\mathcal{F}}_{\mathrm{S},13} = (+10.06\pm0.16)$ \\
$\mathcal{F}_{\mathrm{S,R0},13}$ & $15156.57 \pm 0.10$	& $\tilde{\mathcal{F}}_{\mathrm{S,R0},13}-\tilde{\mathcal{F}}_{\mathrm{S},13} = (-10.73\pm0.15)$ \\ [1ex]
\hline\hline %inserts single line
\end{tabular}
\label{tab:KOI1857_evidences} % is used to refer this table in the text
\end{table} % title of Table

%%% KOI-1857 System parameters
\begin{table}
\caption{\emph{System parameters for KOI-1857.01 from model 
$\mathcal{V}_{\mathrm{P,LD}}$, except for $M_S/M_P$ which is derived from model
$\mathcal{F}_{\mathrm{S},13}$. $^{*}$ indicates that a parameter was fixed.}}
%title of Table
\centering % used for centering table
\begin{tabular}{l l} % centered columns (2 columns)
\hline
Parameter & Value\\ [0.5ex] % inserts table
%heading
\hline
\multicolumn{2}{l}{\emph{Derived parameters...}}\\ 
$P_P$\,[days] & $88.64486_{-0.00074}^{+0.00076}$ \\
$\tau_0$\,[BJD$_{\mathrm{UTC}}$] & $2455333.0474_{-0.0020}^{+0.0020}$ \\
$R_P/R_{\star}$ & $0.02384_{-0.00094}^{+0.00255}$ \\
$b$ & $0.43_{-0.30}^{+0.33}$ \\
$(a/R_{\star})$ & $80.7_{-21.7}^{+7.8}$ \\
$i$\,[deg] & $89.70_{-0.43}^{+0.22}$ \\
$\rho_{\star}$\,[g\,cm$^{-3}$] & $1.27_{-0.77}^{+0.40}$ \\
$\tilde{T}$\,[hours] & $7.55_{-0.17}^{+0.18}$ \\
$u_1$ & $0.72_{-0.32}^{+0.32}$ \\
$(u_1+u_2)$ & $0.81_{-0.18}^{+0.13}$ \\ 
\hline
\multicolumn{2}{l}{\emph{Physical parameters...}}\\ 
$M_{\star}$\,[$R_{\odot}$] & $0.94^{*}$ \\
$R_{\star}$\,[$R_{\odot}$] & $0.85^{*}$ \\
$R_P$\,[$R_{\oplus}$] & $2.210_{-0.087}^{+0.236}$ \\
$M_S/M_P$ & $<0.028$ (95\% confidence) \\ %[1ex]
$\delta_{\mathrm{TTV}}$\,[mins] & $<5.9$ (95\% confidence) \\
$\delta_{\mathrm{TDV}}$\,[mins] & $<7.7$ (95\% confidence) \\ [1ex]
\hline\hline %inserts single line
\end{tabular}
\label{tab:KOI1857_parameters} % is used to refer this table in the text
\end{table} % title of Table

\subsubsection{Summary}

We find no compelling evidence for an exomoon around KOI-1857.01 and estimate
that $M_S/M_P<0.028$ to 95\% confidence. This assessment is based on the fact
the system fails the basic detection criteria B1 and B2 as well as the follow-up
criteria F1, F2 and F3 (see \S\ref{sub:criteria}).

%%%%%%%%%%%%%%%%%%%%%%%%%%%%%%%%%%%%%%%%%%%%%%%%%%%%%%%%%%%%%%%%%%%%%%%%%%%%%%%%

\subsection{KOI-303.01}
\label{sub:koi303}

%% KOI-303
%%

\subsubsection{Data selection}

After detrending with \cofiam, the PA and PDC-MAP data were found to have a 
2.0\,$\sigma$ and 2.6\,$\sigma$ confidence of autocorrelation on a 30\,minute 
timescale respectively and therefore both were acceptable ($<3$\,$\sigma$). 
The less manipulated PA data is selected over the PDC-MAP data throughout.
Short-cadence data is available for this system for quarters 6, 7 and 8 and 
this data displaced the corresponding long-cadence time series in what follows.

\subsubsection{Improved stellar parameters}

During the analysis of this candidate, we were able to acquire two high 
resolution spectra of KOI-303.01 using the ARCES spectrograph on the 3.5\,m 
Astrophysical Research Consortium Telescope at the Apache Point Observatory on 
31 August 2012. We used a $1\farcs 6 \times 3\farcs 2$ slit with an exposure 
time of 20\,minutes yielding a signal-to-noise of 25 per pixel with a resolution 
of 31,500. The echelle spectra were extracted using standard IRAF tools.

We used the Stellar Parameter Classification (SPC) method \citep{buchhave:2012} 
to derive the stellar atmosphere parameters. SPC cross-correlates the observed 
spectrum against a grid of synthetic spectra drawn from a library calculated by 
John Laird using Kurucz models \citep{kurucz:1992}. The synthetic spectra cover 
a window of 300\,\AA\ centered near the gravity-sensitive Mgb features and have a 
spacing of 250\,K in effective temperature, 0.5\,dex in gravity, 0.5\,dex in 
metallicity and 1\,km\,s$^{-1}$ in rotational velocity. To derive the precise 
stellar parameters between the grid points, the normalized cross-correlation 
peaks were fitted with a three dimensional polynomial as a function of effective 
temperature, surface gravity and metallicity. This procedure was carried out for 
different rotational velocities and the final stellar parameters were determined 
by a weighted mean of the values from the spectral orders covered by the 
library.

We used the Yonsei-Yale (YY) isochrones \citet{yi:2001} to determine the 
physical properties of the host star using the stellar parameters derived with 
SPC. The isochrone analysis made used of the stellar effective temperature 
$T_{\mathrm{eff}} = (5576 \pm 50)$\,K, the surface gravity $\log(g) = (4.38 \pm 
0.10)$ (cgs) and the metallicity [Fe/H] $= -0.28 \pm 0.08$ from the SPC 
analysis. The surface gravity acts as a luminosity indicator for the star. 
Whilst $a/R_*$ derived from light curve fitting of the transits can also be used 
as a luminosity indicator \citep{seager:2003}, this parameter can often vary 
significantly between planet-only versus planet-with-moon models. Therefore, 
using $\log(g)$ makes no prior assumption about which model is correct. Using 
the Monte Carlo method, we generated random realizations of these three 
parameters and found matching isochrones and their corresponding physical 
parameters. The physical parameter realizations are then used to produce a 
kernel density estimator to calculate the mode along with the 68.3\% confidence 
intervals. This process leads us to estimate 
$M_* = 0.774_{-0.017}^{+0.036}$\,$M_{\odot}$ and $R_* =
0.833_{-0.039}^{+0.061}$\,$R_{\odot}$, which may be compared to the 
B12 values of $M_* = 1.03$\,$M_{\odot}$ and $R_* = 
0.95$\,$R_{\odot}$.

\subsubsection{Planet-only fits}

% Limb darkening
When queried from MAST, the KIC effective temperature and surface gravity
were reported as $T_{\mathrm{eff}} = 5497$\,K and $\log g = 4.497$ 
\citep{brown:2011}. Both of these estimates are consistent with our SPC
determination and we decided to use these values to estimate the quadratic
limb darkening coefficients, $u_1 = 0.4848$ and $(u_1+u_2) = 0.6990$. The 
initial two models we regressed were $\mathcal{V}_{\mathrm{P}}$ and 
$\mathcal{V}_{\mathrm{P,LD}}$, where the former uses the theoretical limb 
darkening coefficients as fixed values and the latter allows the two 
coefficients to be free parameters. We find that 
$\log\mathcal{Z}(\mathcal{V}_{\mathrm{P,LD}}) - 
\log\mathcal{Z}(\mathcal{V}_{\mathrm{P}}) = +237.56\pm0.25$ indicating that our 
limb darkening coefficients were not optimal. We thus chose to set the
coefficients to the maximum a-posteriori values of $u_1 = 0.1389$ and 
$(u_1+u_2) = 0.8147$.

% V -> P
KOI-303.01 has a period of $P_P = (60.92884 \pm 0.00013)$\,days (as determined 
by model $\mathcal{V}_{\mathrm{P,LD}}$) and exhibits 10 full transits from Q1-Q9 
from epoch -6 to +6, except epochs 0, +2 and +3 which are absent. As is typical 
for all cases, $\log\mathcal{Z}(\mathcal{F}_{\mathrm{P}}) > 
\log\mathcal{Z}(\mathcal{V}_{\mathrm{P}})$ indicating 
that allowing for 10 independent baseline parameters is unnecessary relative 
to a single baseline term.

% TTV
We find no evidence for TTVs in KOI-303.01, with 
$\log\mathcal{Z}(\mathcal{F}_{\mathrm{TTV}}) 
- \log\mathcal{Z}(\mathcal{F}_{\mathrm{P}}) = -24.20\pm0.18$, which is formally 
an 6.6\,$\sigma$ preference for a static model over a TTV model. The timing 
precision on the 10 transits ranged from 1.9 to 3.0 minutes (from model 
$\mathcal{F}_{\mathrm{TTV}}$) and yields a flat TTV 
profile, as shown in Fig.~\ref{fig:KOI303_TTVs}. We calculate a standard
deviation of $\delta_{\mathrm{TTV}} = 3.8$\,minutes and $\chi_{\mathrm{TTV}}^2 
= 18.2$ for 10-2 degrees of freedom.

% TDV
The TTV+TDV model fit, $\mathcal{V}_{\mathrm{V}}$, finds consistent transit 
times with those derived by model $\mathcal{F}_{\mathrm{TTV}}$. The timings show 
no clear pattern or excessive scatter, visible in Fig.~\ref{fig:KOI303_TTVs}. We 
therefore conclude that there is presently no evidence for TTVs or TDVs for 
KOI-303.01. The standard deviation of the TDVs is found to be 
$\delta_{\mathrm{TDV}} = 3.2$\,minutes and we determine 
$\chi_{\mathrm{TDV}}^2 = 10.3$ for 10-1 degrees of freedom.

%%% KOI-303 TTV Table
\begin{table*}
\caption{\emph{Transit times and durations for KOI-303.01. The model used to 
calculate the supplied values is provided in parentheses next to each column
heading. BJD$_{\mathrm{UTC}}$ times offset by $2,400,000$ days.
}} % title of Table
\centering % used for centering table
\begin{tabular}{c c c c c c c} % centered columns (3 columns)
\hline
Epoch & $\tau$ [BJD$_{\mathrm{UTC}}$] ($\mathcal{F}_{\mathrm{TTV}}$) & TTV [mins] ($\mathcal{F}_{\mathrm{TTV}}$)
      & $\tau$ [BJD$_{\mathrm{UTC}}$] ($\mathcal{V}_{\mathrm{V}}$) & TTV [mins] ($\mathcal{V}_{\mathrm{V}}$)
      & $\tilde{T}$ [mins] ($\mathcal{V}_{\mathrm{V}}$) & TDV [mins] ($\mathcal{V}_{\mathrm{V}}$) \\ [0.5ex] % inserts table
%heading
\hline
-6 & $55006.3643_{-0.0017}^{+0.0018}$ & $-4.3 \pm 2.5$
   & $55006.3646_{-0.0019}^{+0.0019}$ & $-3.6 \pm 2.7$
   & $380.5_{-6.4}^{+6.8}$ & $+4.6 \pm 3.3$ \\
-5 & $55067.2957_{-0.0015}^{+0.0015}$ & $-0.6 \pm 2.2$
   & $55067.2953_{-0.0017}^{+0.0016}$ & $-0.9 \pm 2.4$
   & $379.2_{-5.4}^{+5.7}$ & $+3.9 \pm 2.8$ \\
-4 & $55128.2232_{-0.0018}^{+0.0017}$ & $-2.6 \pm 2.5$
   & $55128.2228_{-0.0019}^{+0.0019}$ & $-2.6 \pm 2.8$
   & $380.1_{-6.5}^{+6.7}$ & $+4.3 \pm 3.3$ \\
-3 & $55189.1563_{-0.0017}^{+0.0018}$ & $+3.6 \pm 2.5$
   & $55189.1562_{-0.0016}^{+0.0016}$ & $+4.1 \pm 2.3$
   & $363.5_{-5.3}^{+5.4}$ & $-4.0 \pm 2.7$ \\
-2 & $55250.0824_{-0.0018}^{+0.0017}$ & $-0.4 \pm 2.5$
   & $55250.0824_{-0.0018}^{+0.0017}$ & $+0.3 \pm 2.5$
   & $373.1_{-5.7}^{+5.8}$ & $+0.9 \pm 2.9$ \\
-1 & $55311.0139_{-0.0019}^{+0.0019}$ & $+3.4 \pm 2.7$
   & $55311.0138_{-0.0018}^{+0.0018}$ & $+4.2 \pm 2.6$
   & $375.4_{-5.9}^{+5.8}$ & $+2.0 \pm 2.9$ \\
+1 & $55432.8720_{-0.0018}^{+0.0018}$ & $+4.1 \pm 2.6$
   & $55432.8719_{-0.0021}^{+0.0021}$ & $+5.1 \pm 3.0$
   & $370.1_{-6.1}^{+6.6}$ & $-0.6 \pm 3.2$ \\
+4 & $55615.6569_{-0.0013}^{+0.0013}$ & $+1.7 \pm 1.9$
   & $55615.6566_{-0.0016}^{+0.0016}$ & $+2.7 \pm 2.3$
   & $365.6_{-5.3}^{+4.9}$ & $-2.9 \pm 2.6$ \\
+5 & $55676.5837_{-0.0021}^{+0.0021}$ & $-1.2 \pm 3.0$
   & $55676.5829_{-0.0026}^{+0.0031}$ & $-0.9 \pm 4.1$
   & $375.6_{-9.9}^{+8.2}$ & $+2.1 \pm 4.5$ \\
+6 & $55737.5080_{-0.0021}^{+0.0020}$ & $-7.8 \pm 3.0$
   & $55737.5074_{-0.0026}^{+0.0023}$ & $-7.0 \pm 3.5$
   & $366.6_{-8.8}^{+7.6}$ & $-2.4 \pm 4.1$ \\ [1ex]
\hline\hline %inserts single line
\end{tabular}
\label{tab:KOI303_TTVs} % is used to refer this table in the text
\end{table*} % title of Table

\subsubsection{Moon fits}

A planet-with-moon fit, $\mathcal{F}_S$, is slightly favoured relative to a 
planet-only fit at $\Delta(\log\mathcal{Z}) = +12.34\pm0.16$, or 
4.59\,$\sigma$, meaning detection criterion B1 is met. The zero-mass 
moon model is slightly preferred though at $\Delta\log\mathcal{Z} = 
3.13\pm0.17$.

\begin{figure*}
\begin{center}
\includegraphics[width=18.0 cm]{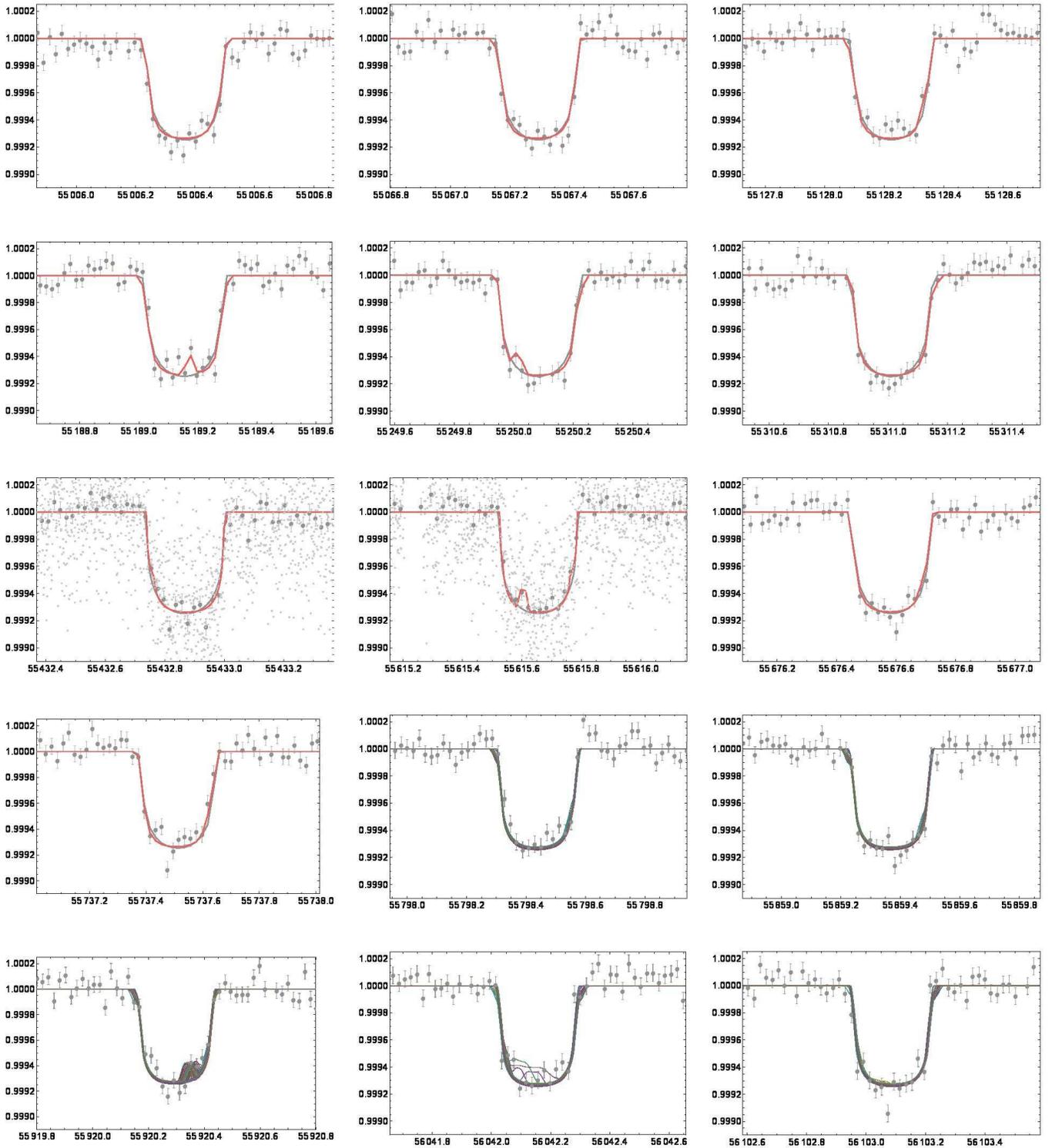}
\caption{\emph{From left-to-right then top-to-bottom we show
the chronological sequence of transits of KOI-303.01. The
first 10 panels show the Q1-9 data and the maximum a-posteriori
light curve fit of a planet-only model (gray line) and a moon model
(red line). The last 5 panels show the Q10-13 data with 50
extrapolations (50 different shadings used) of the moon model overlaid 
(parameters randomly drawn from the joint posteriors), which do not exhibit 
significant predictive power.}} 
\label{fig:KOI303_Fs}
\end{center}
\end{figure*}

Despite failing B2, the fits yield broadly physical parameters and
the zero-mass moon preference could be indicative of a low 
signal-to-noise TTV/TDV data at this stage. Using $M_{\star}=0.774$\,$M_{\odot}$ 
and $R_{\star}=0.833$\,$R_{\odot}$ from our spectroscopic observations and SPC 
analysis, we determine $M_P = 19.6_{-15,4}^{+9.2}$\,$M_{\oplus}$ for 
$R_P = 2.044_{-0.046}^{+0.039}$\,$R_{\oplus}$ and 
$M_S = 0.47_{-0.40}^{+0.64}$\,$M_{\oplus}$ for 
$R_S = 0.958_{-0.089}^{+0.094}$\,$R_{\oplus}$. The solution is also clearly
not a close-binary with $M_S/M_P = 0.030_{-0.018}^{+0.025}$ and three 
mutual transits driving the fit, as evident from Fig.~\ref{fig:KOI303_Fs}.
Finally, the mass ratio posterior very slightly peaks away from zero
suggesting a low signal-to-noise hint of a TTV/TDV mass signal. In conclusion, 
we evaluate that detection criterion B3 is satisfied and B4 is unclear.

%%% KOI-303 Fs fit

\subsubsection{Predictive power of the moon model}

At this stage, we considered KOI-303.01 to be a potential candidate and
further data may confirm/reject the signal. We therefore detrended the
PA data for KOI-303 from Q10-13 covering 5 new transits, which had 
recently become available at the time of writing. Before we attempted to 
re-fit the updated data set, we extrapolated the light curve model fit 
$\mathcal{F}_{\mathrm{S}}$ into Q10-13 for both the maximum 
a-posteriori solution and 50 randomly sampled solutions from the joint 
posteriors, which are shown in Fig.~\ref{fig:KOI303_Fs}. The
maximum a-posteriori moon solution can be compared to the detrended
data, where we compute $\chi^2 = 1295.21$ for 836 data points. For
comparison, we repeated the process for the maximum a-posteriori
planet-only model and compute $\chi^2 = 1276.85$. To provide some
reference, the two best fits yield a $\chi^2$ of $11688.64$ and
$11746.60$ over Q1-9 for $\mathcal{F}_{\mathrm{S}}$ and 
$\mathcal{F}_{\mathrm{P}}$ respectively over 10344 data points.

This simple test suggests the moon model lacks predictive power, which is a 
major concern for accepting the moon hypothesis. In order to investigate
the uncertainty in the best-fit models, we extrapolated the transit light 
curve for 50 realizations where the system parameters are randomly drawn
from the joint posteriors. The 50 realizations from the model 
$\mathcal{F}_{\mathrm{S}}$ are shown in Fig.~\ref{fig:KOI303_Fs}. 
The distribution of the $\chi^2$ values for these fits
yields $\chi^2 = 1300\pm21$ whereas repeating the process for
50 predictions from $\mathcal{F}_{\mathrm{P}}$ yields $1280\pm25$, which again
suggests that the planetary model has superior predictive power. We
therefore conclude that KOI-303.01 does not satisfy detection criterion F2. On
this basis, we reject KOI-303.01 as a possible candidate and consider
that there is no evidence to support the presence of an exomoon around
this target with present data.

\begin{figure}
\begin{center}
\includegraphics[width=8.4 cm]{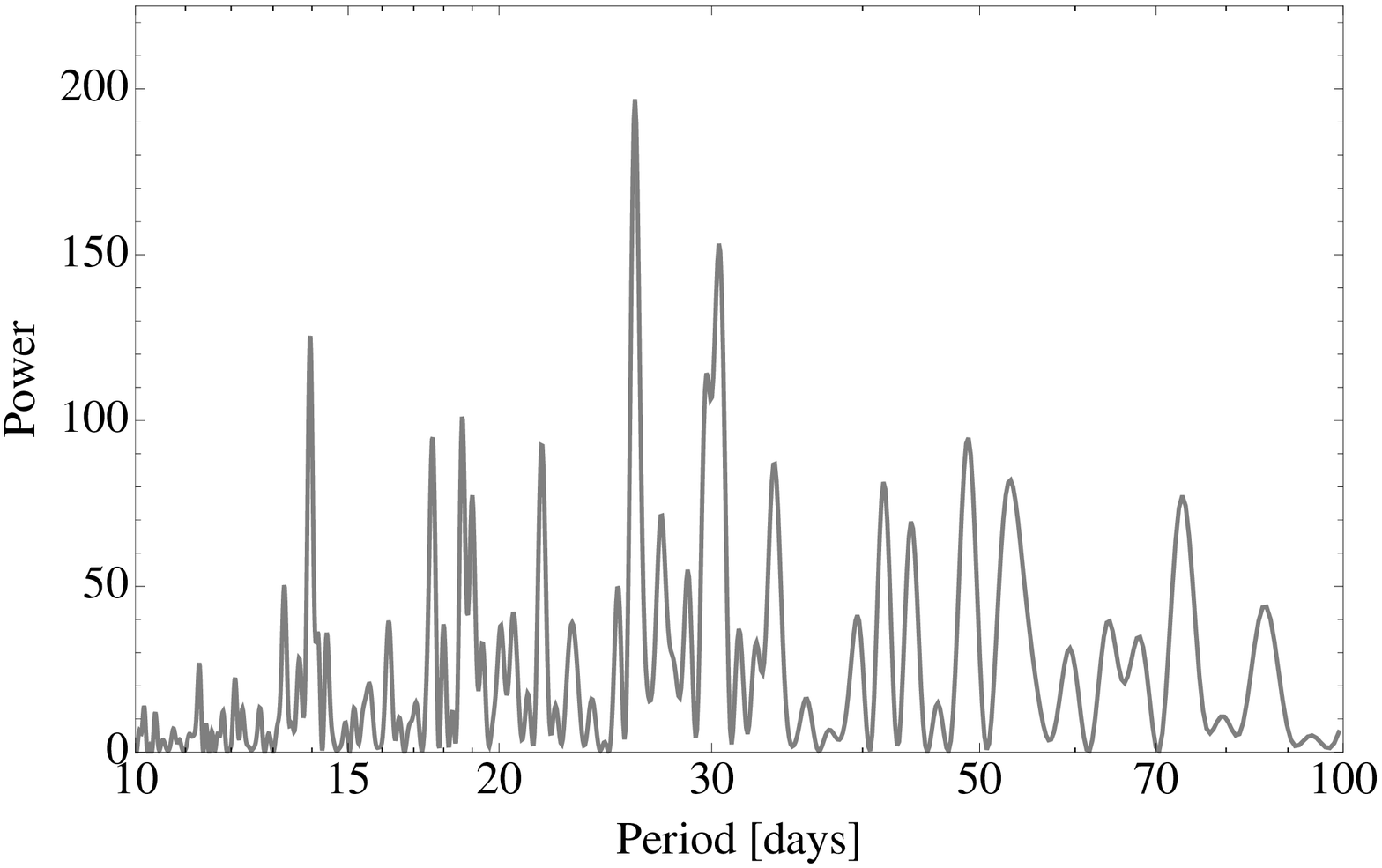}
\caption{\emph{Lomb-Scargle periodogram of the out-of-transit normalized
flux of KOI-303 from quarters 1-11 and quarter 13. We zoom in on 10-100\,days
where significant power is identified, exhibitting a complex set of 
periodicities indicative of differential rotation. The peak at 30.41\,days
is consistent with the timings of the transit anomalies.}} 
\label{fig:periodogram}
\end{center}
\end{figure}

The three mutual events fitted by the moon model morphologically resemble
starspot crossing events (e.g. see \citealt{rabus:2009}). We consider that
this is the most likely source of false-positive for KOI-303.01 in light of the 
unmet detection criteria and the presence of flux variations
consistent with rotational modulations. These modulations reveal a complex
set of periodicities when processed with a Lomb-Scargle periodogram (see
Fig.~\ref{fig:periodogram}), indicative of significant differential rotation. 
Analyzing Q1-13 (except Q12 which is incoherent with the other quarters), the 
highest power occurs at 25.92\,days and the second highest at 30.41\,days.
We estimated spot anomaly timings of BJD$_{\mathrm{UTC}}$\,2455189.173, 
2455250.013 and 2455615.604. These timings are inconsistent with a single spot
rotating every 25.92\,days but do give excellent agreement with a single
spot rotating every 30.41\,days. We therefore consider that KOI-303 exhibits
differential rotation and the transit crosses a spot which rotates every 
30.41\,days.

%%% KOI-303 periodogram

One major effect of spots are such crossing events yielding a false $R_S/R_P$ 
signal. To a lesser degree, spot crossings can perturb the transit profile and
cause small false timing variations, giving a false $M_S/M_P$ signal.
However, in general this latter distortion will be small and already
our timing analysis suggests a lack of TTVs. In light of this, we use
the $\mathcal{F}_{\mathrm{S,R0}}$ model for our final $M_S/M_P$ posterior,
shown in Fig.~\ref{fig:KOI303_Msp}. The 95\% quantile from this
posterior constrains $M_S/M_P<0.21$ and the 3\,$\sigma$ quantile
constrains $M_S/M_P<0.33$.

%%% KOI-303 Evidences
\begin{table}
\caption{\emph{Bayesian evidences of various fits for KOI-303.01.
A description of the different models can be found in \S\ref{sub:fitsoverview}.
}} 
%title of Table
\centering % used for centering table
\begin{tabular}{l c l} % centered columns (3 columns)
\hline
Model, $\mathcal{M}$ & $\mathrm{log}\mathcal{Z}(\mathcal{M})$ & $\tilde{\mathcal{M}}_1 - \tilde{\mathcal{M}}_2$ \\ [0.5ex] % inserts table
 &  & $= \mathrm{log}\mathcal{Z}(\mathcal{M}_1) - \mathrm{log}\mathcal{Z}(\mathcal{M}_2)$ \\ [0.5ex] % inserts table
%heading
\hline
\multicolumn{3}{l}{\emph{Planet only fits...}}\\ 
$\mathcal{V}_{\mathrm{P}}$  	& $69125.46 \pm 0.18$ 	& -	\\
$\mathcal{V}_{\mathrm{P,LD}}$	& $69363.02 \pm 0.18$	& $\tilde{\mathcal{V}}_{\mathrm{P,LD}}-\tilde{\mathcal{V}}_{\mathrm{P}} = (+237.56\pm0.25)$ \\
$\mathcal{V}_{\mathrm{P,MAP}}$  	& $69365.13 \pm 0.18$   & $\tilde{\mathcal{V}}_{\mathrm{P,MAP}}-\tilde{\mathcal{V}}_{\mathrm{P}} = (+239.67\pm0.25)$ \\
$\mathcal{F}_{\mathrm{P}}$	& $69439.45 \pm 0.11$	& $\tilde{\mathcal{F}}_{\mathrm{P}}-\tilde{\mathcal{V}}_{\mathrm{P,MAP}} = (+74.32\pm0.21)$ \\
\hline
\multicolumn{3}{l}{\emph{Planet with timing variations fits...}}\\ 
$\mathcal{F}_{\mathrm{TTV}}$	& $69415.25 \pm 0.14$	& $\tilde{\mathcal{F}}_{\mathrm{TTV}}-\tilde{\mathcal{F}}_{\mathrm{P}} = (-24.20\pm0.18)$ \\
$\mathcal{V}_{\mathrm{V}}$	& $69218.80 \pm 0.27$	& $\tilde{\mathcal{V}}_{\mathrm{V}}-\tilde{\mathcal{V}}_{\mathrm{P,MAP}} = (-146.33\pm0.32)$ \\
\hline
\multicolumn{3}{l}{\emph{Planet with moon fits...}}\\ 
$\mathcal{F}_{\mathrm{S}}$	& $69451.79 \pm 0.12$	& $\tilde{\mathcal{F}}_{\mathrm{S}}-\tilde{\mathcal{F}}_{\mathrm{P}} = (+12.34\pm0.16)$ \\ % 
$\mathcal{F}_{\mathrm{S,M0}}$	& $69454.92 \pm 0.12$	& $\tilde{\mathcal{F}}_{\mathrm{S,M0}}-\tilde{\mathcal{F}}_{\mathrm{S}} = (+3.13\pm0.17)$ \\ %  
$\mathcal{F}_{\mathrm{S,R0}}$	& $69447.02 \pm 0.16$	& $\tilde{\mathcal{F}}_{\mathrm{S,R0}}-\tilde{\mathcal{F}}_{\mathrm{S}} = (-4.77\pm0.20)$ \\ [1ex]
\hline\hline %inserts single line
\end{tabular}
\label{tab:KOI303_evidences} % is used to refer this table in the text
\end{table} % title of Table

%%% KOI-303 System parameters
\begin{table}
\caption{\emph{System parameters for KOI-303.01 from model 
$\mathcal{V}_{\mathrm{P,LD}}$, except for $M_S/M_P$ which is derived from model
$\mathcal{F}_{\mathrm{S,R0}}$.}}
%title of Table
\centering % used for centering table
\begin{tabular}{l l} % centered columns (2 columns)
\hline
Parameter & Value\\ [0.5ex] % inserts table
%heading
\hline
\multicolumn{2}{l}{\emph{Derived parameters...}}\\ 
$P_P$\,[days] & $60.92884_{-0.00013}^{+0.00013}$ \\
$\tau_0$\,[BJD$_{\mathrm{UTC}}$] & $2455371.94026_{-0.00053}^{+0.00052}$ \\
$R_P/R_{\star}$ & $0.02504_{-0.00044}^{+0.00095}$ \\
$b$ & $0.36_{-0.24}^{+0.27}$ \\
$(a/R_{\star})$ & $70.6_{-11.1}^{+4.2}$ \\
$i$\,[deg] & $89.71_{-0.31}^{+0.20}$ \\
$\rho_{\star}$\,[g\,cm$^{-3}$] & $1.79_{-0.72}^{+0.34}$ \\
$\tilde{T}$\,[hours] & $6.150_{-0.063}^{+0.065}$ \\
$u_1$ & $0.23_{-0.14}^{+0.17}$ \\
$(u_1+u_2)$ & $0.70_{-0.13}^{+0.15}$ \\ 
\hline
\multicolumn{2}{l}{\emph{Physical parameters...}}\\ 
$M_{\star}$\,[$R_{\odot}$] (SPC) & $0.774_{-0.017}^{+0.036}$ \\
$R_{\star}$\,[$R_{\odot}$] (SPC) & $0.833_{-0.039}^{+0.061}$ \\
$R_P$\,[$R_{\oplus}$] & $2.30_{-0.13}^{+0.17}$ \\
$M_S/M_P$ & $<0.21$ (95\% confidence) \\ %[1ex]
$\delta_{\mathrm{TTV}}$\,[mins] & $<3.6$ (95\% confidence) \\
$\delta_{\mathrm{TDV}}$\,[mins] & $<3.9$ (95\% confidence) \\ [1ex]
\hline\hline %inserts single line
\end{tabular}
\label{tab:KOI303_parameters} % is used to refer this table in the text
\end{table} % title of Table

\subsubsection{Summary}

We find no compelling evidence for an exomoon around KOI-303.01 and estimate
that $M_S/M_P<0.21$ to 95\% confidence. This assessment is based on the fact
the system fails the basic detection criterion B2 and is marginal for B4. Further
investigation reveals that the candidate also fails the follow-up criterion F2 
(see \S\ref{sub:criteria}).

%%%%%%%%%%%%%%%%%%%%%%%%%%%%%%%%%%%%%%%%%%%%%%%%%%%%%%%%%%%%%%%%%%%%%%%%%%%%%%%%

\subsection{KOI-1876}
\label{sub:koi1876}

%% KOI-1876
%%

\subsubsection{Data selection}

After detrending with \cofiam, the PA and PDC-MAP data were found to have a 
1.7\,$\sigma$ and 1.9\,$\sigma$ confidence of autocorrelation on a 30\,minute 
timescale respectively and therefore both were acceptable ($<3$\,$\sigma$). In 
general, we always prefer using the raw data and so we opted for the PA data in 
all subsequent analysis of this system. We note that only long-cadence data was 
available for this system.

\subsubsection{Planet-only fits}

% Limb darkening
When queried from MAST, the KIC effective temperature and surface gravity
were reported as $T_{\mathrm{eff}} = 4230$\,K and $\log g = 4.387$ 
\citep{brown:2011}. Using these values, we estimated quadratic limb darkening 
coefficients $u_1 = 0.6747$ and $(u_1+u_2) = 0.7511$. The initial two models we 
regressed were $\mathcal{V}_{\mathrm{P}}$ and $\mathcal{V}_{\mathrm{P,LD}}$ 
where the former uses the aforementioned limb darkening coefficients as fixed 
values and the latter allows the two coefficients to be free parameters. We find 
that $\log\mathcal{Z}(\mathcal{V}_{\mathrm{P,LD}}) - 
\log\mathcal{Z}(\mathcal{V}_{\mathrm{P}}) = -0.24\pm0.20$ indicating that the 
theoretical limb darkening coefficients are adequate and will be adopted in
all subsequent fits.

% V -> P
KOI-1876.01 has a period of $P_P = 82.53238\pm0.00071$\,days (as 
determined by model $\mathcal{V}_{\mathrm{P,LD}}$) and exhibits 8 transits from 
Q1-Q9. As is typical for all cases, $\log\mathcal{Z}(\mathcal{F}_{\mathrm{P}}) > 
\log\mathcal{Z}(\mathcal{V}_{\mathrm{P}})$ indicating that allowing for 8 
independent baseline parameters is unnecessary relative to a single baseline 
term.

% TTV
We find no evidence for TTVs in KOI-1876.01, with 
$\log\mathcal{Z}(\mathcal{F}_{\mathrm{TTV}}) 
- \log\mathcal{Z}(\mathcal{F}_{\mathrm{P}}) = -17.27\pm0.15$, which is formally 
an 5.5\,$\sigma$ preference for a static model over a TTV model. The timing 
precision on the 8 transits ranged from 6.0 to 7.9 minutes. The TTVs, shown in 
Fig.~\ref{fig:KOI1876_TTVs}, show no clear pattern and exhibit a standard 
deviation of $\delta_{\mathrm{TTV}} = 10.8$\,minutes and $\chi_{\mathrm{TTV}}^2 
= 16.3$ for 8-2 degrees of freedom.

% TDV
The TTV+TDV model fit, $\mathcal{V}_{\mathrm{V}}$, finds consistent transit 
times with those derived by model $\mathcal{F}_{\mathrm{TTV}}$. No clear pattern 
or excessive scatter is visible in the data, shown in 
Fig.~\ref{fig:KOI1876_TTVs}. We therefore conclude there is no evidence for TTVs 
or TDVs for KOI-1876.01. The standard deviation of the TDVs is found to be 
$\delta_{\mathrm{TDV}} = 14.2$\,minutes and we determine $\chi_{\mathrm{TDV}}^2 
= 4.6$ for 8-1 degrees of freedom.

\subsubsection{Moon fits}

A planet-with-moon with, $\mathcal{F}_{\mathrm{S}}$, is preferable to a 
planet-only fit at a formally moderate significance level of 3.4\,$\sigma$,
close to our 4\,$\sigma$ threshold of criterion B1. Detection criterion B2
is certainly satisfied with $\mathcal{F}_{\mathrm{S}}$ favored over the
zero-mass moon model at 6.0\,$\sigma$ and over the zero-radius moon model at
13.6\,$\sigma$ (see Table~\ref{tab:KOI1876_evidences}).

We may combine the posteriors from $\mathcal{F}_{\mathrm{S}}$ with the 
stellar parameters derived by B12 ($M_{\star} = 
0.51$\,$M_{\odot}$ and $R_{\star} = 0.49$\,$R_{\oplus}$) to obtain physical 
parameters for the planet-moon candidate system. The planet's parameters are 
consistent with a dense Super-Earth/mini-Neptune with 
$M_P = 34.2_{-5.9}^{+5.5}$\,$M_{\oplus}$ for 
$R_P = 2.081_{-0.044}^{+0.057}$\,$R_{\oplus}$. In contrast, the moon appears to
have an unphysically low-density with $M_S = 0.13_{-0.09}^{+0.16}$\,$M_{\oplus}$
for $R_P = 0.995_{-0.091}^{+0.090}$\,$R_{\oplus}$. We conclude that this moon
fit fails detection criterion B3. The $M_S/M_P$ posterior, shown in
Figure~\ref{fig:KOI1876_Msp}, also fails to converge
away from zero, failing criterion B4. The 95\% and 3\,$\sigma$ quantiles of
this posterior are $M_S/M_P<0.012$ and $M_S/M_P<0.022$.

\begin{figure*}
\begin{center}
\includegraphics[width=18.0 cm]{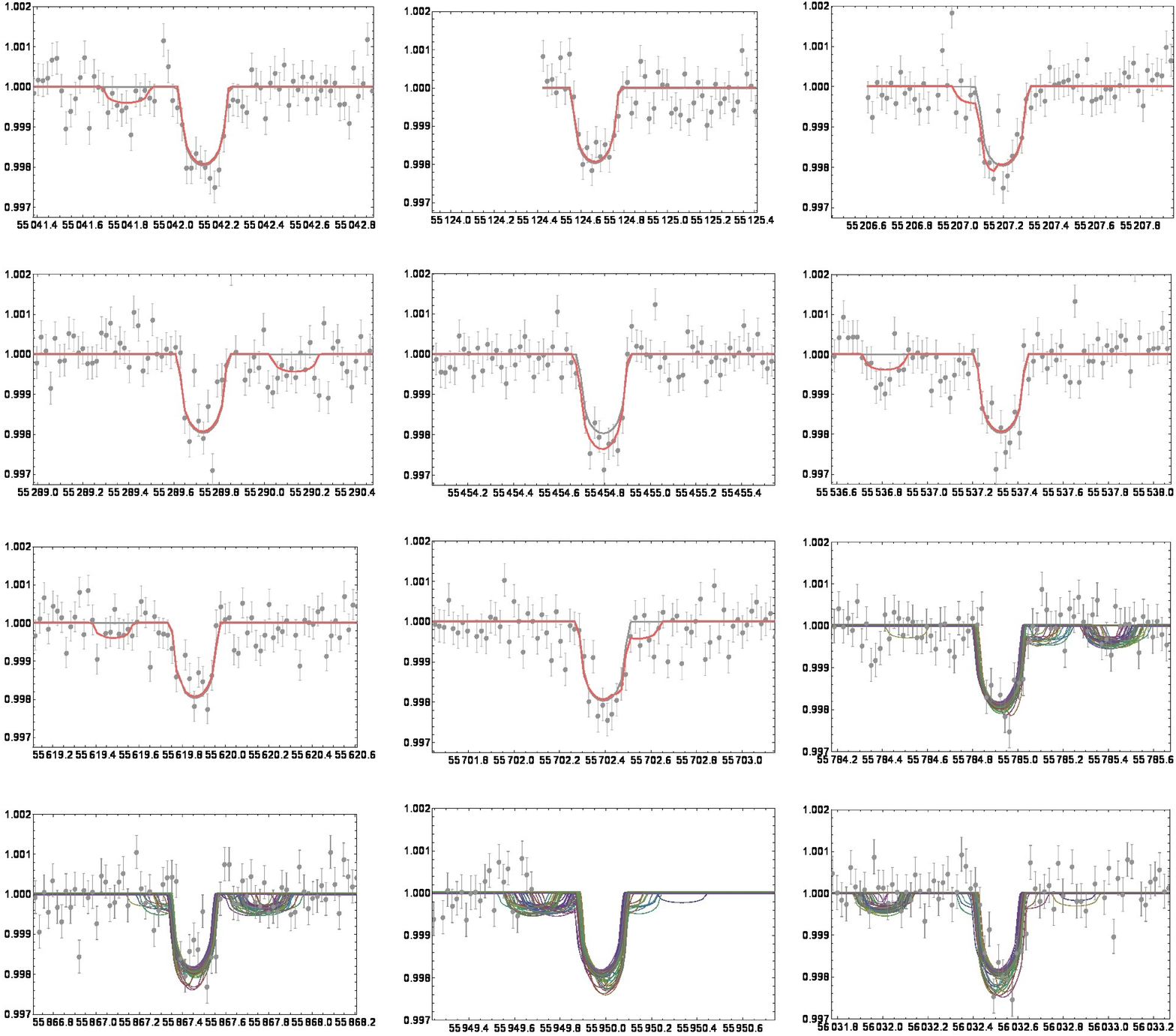}
\caption{\emph{From left-to-right then top-to-bottom we show
the chronological sequence of transits of KOI-1876.01. The
first 8 panels show the Q1-9 data and the maximum a-posteriori
light curve fit of a planet-only model (gray line) and a moon model
(red line). The last 4 panels show the Q10-13 data with 50
extrapolations (50 different shadings used) of the moon model overlaid 
(parameters randomly drawn from the joint posteriors), which do not exhibit 
significant predictive 
power.}} 
\label{fig:KOI1876_Fs}
\end{center}
\end{figure*}

The zero-peaked moon mass posterior suggests that the spurious signal is
driven by the spurious moon's radius. Inspection of the maximum a-posteriori 
realization of model $\mathcal{F}_{\mathrm{S}}$ (Fig.~\ref{fig:KOI1876_Fs})
reveals auxiliary transits, which cannot be due to starspot crossings. For this
reason, similar to KOI-1857.01, we decided to investigate the predictive power
of the moon model, to ensure we are not overlooking a physical signal.

%%% KOI-1876 Fs fit

\subsubsection{Predictive power of the moon model}

We detrended the PA data for KOI-1876 from Q10-13 covering 4 new transits, which 
had recently become available at the time of writing. We extrapolated the light 
curve model fit $\mathcal{F}_{\mathrm{S}}$ into Q10-13 for both the maximum 
a-posteriori solution and 50 randomly sampled solutions from the joint 
posteriors, which are shown in Fig.~\ref{fig:KOI1876_Fs}. The
maximum a-posteriori moon solution can be compared to the detrended
data, where we compute $\chi^2 = 621.9$ for 516 data points. For
comparison, we repeated the process for the maximum a-posteriori
planet-only model and compute $\chi^2 = 577.7$ i.e. substantially worse.

This comparison suggests that the moon model has no significant predictive 
power, which is a major concern for accepting the moon hypothesis. We 
should expect a moon model to give a better $\chi^2$ by around $\gtrsim25$ 
based upon the improvement of the best fits on the Q1-9 data. For Q1-9, we 
found $\Delta\chi^2 = 52.9$ between the two best fits for 1129 data points. 
The 50 realizations of light curve predictions, shown in 
Fig.~\ref{fig:KOI1876_Fs}, do not seem to show a convincing agreement with the 
data either. Indeed, the distribution of the $\chi^2$ values for these fits
yields $\chi^2 = 631\pm25$ whereas repeating the process for
50 predictions from $\mathcal{F}_{\mathrm{P}}$ yields $593\pm20$, which again
suggests that the planetary model actually has slightly better predictive 
power.

The lack of predictive power of model $\mathcal{F}_{\mathrm{S}}$ (detection 
criterion F2) confirms that the moon model is spurious and we conclude the most
probable origin of the spurious signal to be time-correlated noise, similar 
to KOI-1857.01. There is therefore no evidence for an exomoon around KOI-1876.01
based on the presently available data. Our final system parameters for this
system are provided in Table~\ref{tab:KOI1876_parameters}.

%%% KOI-1876 Evidences
\begin{table}
\caption{\emph{Bayesian evidences of various fits for KOI-1876.01.
A description of the different models can be found in 
\S\ref{sub:fitsoverview}.}} %title of Table
\centering % used for centering table
\begin{tabular}{l c l} % centered columns (3 columns)
\hline
Model, $\mathcal{M}$ & $\mathrm{log}\mathcal{Z}(\mathcal{M})$ & $\tilde{\mathcal{M}}_1 - \tilde{\mathcal{M}}_2$ \\ [0.5ex] % inserts table
 &  & $= \mathrm{log}\mathcal{Z}(\mathcal{M}_1) - \mathrm{log}\mathcal{Z}(\mathcal{M}_2)$ \\ [0.5ex] % inserts table
%heading
\hline
\multicolumn{3}{l}{\emph{Planet only fits...}}\\ 
$\mathcal{V}_{\mathrm{P}}$  	& $6956.16 \pm 0.14$ 	& -	\\
$\mathcal{V}_{\mathrm{P,LD}}$	& $6955.92 \pm 0.14$	& $\tilde{\mathcal{V}}_{\mathrm{P,LD}}-\tilde{\mathcal{V}}_{\mathrm{P}} = (-0.24\pm0.20)$ \\
$\mathcal{F}_{\mathrm{P}}$	& $7003.43 \pm 0.09$	& $\tilde{\mathcal{F}}_{\mathrm{P}}-\tilde{\mathcal{V}}_{\mathrm{P}} = (+47.27\pm0.17)$ \\
\hline
\multicolumn{3}{l}{\emph{Planet with timing variations fits...}}\\
$\mathcal{F}_{\mathrm{TTV}}$	& $6986.16 \pm 0.12$	& $\tilde{\mathcal{F}}_{\mathrm{TTV}}-\tilde{\mathcal{F}}_{\mathrm{P}} = (-17.27\pm0.15)$ \\
$\mathcal{V}_{\mathrm{V}}$	& $6874.39 \pm 0.21$	& $\tilde{\mathcal{V}}_{\mathrm{V}}-\tilde{\mathcal{V}}_{\mathrm{P}} = (-81.77\pm0.25)$ \\
\hline
\multicolumn{3}{l}{\emph{Planet with moon fits...}}\\ 
$\mathcal{F}_{\mathrm{S}}$	& $7010.71 \pm 0.11$	& $\tilde{\mathcal{F}}_{\mathrm{S}}-\tilde{\mathcal{F}}_{\mathrm{P}} = (+7.27\pm0.14)$ \\ % 
$\mathcal{F}_{\mathrm{S,M0}}$	& $6916.01 \pm 0.11$	& $\tilde{\mathcal{F}}_{\mathrm{S,M0}}-\tilde{\mathcal{F}}_{\mathrm{S}} = (-94.70\pm0.15)$ \\ %  
$\mathcal{F}_{\mathrm{S,R0}}$	& $6990.39 \pm 0.12$	& $\tilde{\mathcal{F}}_{\mathrm{S,R0}}-\tilde{\mathcal{F}}_{\mathrm{S}} = (-20.32\pm0.16)$ \\ [1ex] %
\hline\hline %inserts single line
\end{tabular}
\label{tab:KOI1876_evidences} % is used to refer this table in the text
\end{table} % title of Table

%%% KOI-1876 final system parameters
\begin{table}
\caption{\emph{System parameters for KOI-1876.01 from model 
$\mathcal{V}_{\mathrm{P,LD}}$, except for $M_S/M_P$ which is derived from model
$\mathcal{F}_{\mathrm{S}}$. $^{*}$ indicates that a parameter was fixed.}}
%title of Table
\centering % used for centering table
\begin{tabular}{l l} % centered columns (2 columns)
\hline
Parameter & Value\\ [0.5ex] % inserts table
%heading
\hline
\multicolumn{2}{l}{\emph{Derived parameters...}}\\ 
$P_P$\,[days] & $82.53238_{-0.00071}^{+0.00073}$ \\
$\tau_0$\,[BJD$_{\mathrm{UTC}}$] & $2455372.2615_{-0.0017}^{+0.0017}$ \\
$R_P/R_{\star}$ & $0.02384_{-0.00094}^{+0.00255}$ \\
$b$ & $0.46_{-0.32}^{+0.34}$ \\
$(a/R_{\star})$ & $118_{-35}^{+13}$ \\
$i$\,[deg] & $89.78_{-0.33}^{+0.16}$ \\
$\rho_{\star}$\,[g\,cm$^{-3}$] & $4.5_{-3.0}^{+1.7}$ \\
$\tilde{T}$\,[hours] & $4.72_{-0.18}^{+0.17}$ \\
$u_1$ & $0.61_{-0.37}^{+0.47}$ \\
$(u_1+u_2)$ & $0.64_{-0.27}^{+0.24}$ \\ 
\hline
\multicolumn{2}{l}{\emph{Physical parameters...}}\\
$M_{\star}$\,[$R_{\odot}$] & $0.51^{*}$ \\
$R_{\star}$\,[$R_{\odot}$] & $0.49^{*}$ \\
$R_P$\,[$R_{\oplus}$] & $2.202_{-0.096}^{+0.229}$ \\
$M_S/M_P$ & $<0.012$ (95\% confidence) \\ %[1ex]
$\delta_{\mathrm{TTV}}$\,[mins] & $<5.4$ (95\% confidence) \\
$\delta_{\mathrm{TDV}}$\,[mins] & $<30.8$ (95\% confidence) \\ [1ex]
\hline\hline %inserts single line
\end{tabular}
\label{tab:KOI1876_parameters} % is used to refer this table in the text
\end{table} % title of Table

\subsubsection{Summary}

We find no compelling evidence for an exomoon around KOI-1876.01 and estimate
that $M_S/M_P<0.012$ to 95\% confidence. This assessment is based on the fact
the system fails the basic detection criteria B3 and B4 as well as the follow-up
criterion F2 (see \S\ref{sub:criteria}).

\section{DISCUSSION \& CONCLUSIONS}
\label{sec:discussion}

%% DISCUSSION
%%

%%% PART ONE - SUMMARIZE

We have presented the first systematic search for the moons of extrasolar
planets. In this work, we have focussed on a sub-sample of seven transiting
planet candidates which our automatic target selection (TSA) algorithm
identified as being not only dynamically viable satellite hosts, but also
exhibiting sufficiently high signal-to-noise photometry that Earth-sized moons 
should be detectable. We also point out that this sub-sample was selected with
the additional filters that the planetary candidates have radii 
$<6$\,$R_{\oplus}$ and that no other planetary candidates are known to exist in 
the system. The former filter is applied since Jovian-sized candidates have a 
higher false-positive rate \citep{santerne:2012} and the latter since the
presence of perturbing planets complicates our analysis.

We find no compelling evidence for an extrasolar moon in any of the seven
systems. This determination is aided by the introduction of exomoon detection
criteria, which we have proposed in this paper (see \S\ref{sub:criteria}). 
Although some perverse configurations of satellite orbits could still hide large 
moons in the systems analyzed, we are able to marginalize over the parameter 
volume to place estimated upper limits on the satellite-to-planet mass ratio,
$M_S/M_P$ (see Fig.~\ref{fig:Msp}).

In the cases of KOI-722.01, KOI-1472.01, KOI-1857.01 and KOI-1876.01, we estimate 
tight constraints with a 95\% upper quantiles of $M_S/M_P<0.04$. Since all four 
objects are relatively low-radii (2.0\,$R_{\oplus}$, 3.9\,$R_{\oplus}$, 
2.2\,$R_{\oplus}$ \& 2.2\,$R_{\oplus}$ respectively), these limits likely probe 
down to sub-Earth masses, although an exact determination cannot be made without 
an estimate of the planetary masses. For KOI-365.01, KOI-174.01 and KOI-303.01
we are unable to derive tight upper limits due to the fits favoring close-binary 
type solutions. A satellite in close proximity shows negligible transit timing
effects since the orbital period of the moon is comparable to that of the
transit duration. For this reason, almost any $M_S/M_P$ value is allowed
and our upper limits on this term for these three candidates are less useful.
In the case of KOI-303.01, this spurious signal is probably due to the presence
of three starspot crossing events driving a spurious fit. In the cases of
KOI-365.01 and KOI-174.01, the origin is less clear but we suggest here the
possibility of erroneous limb darkening parameters, time-correlated noise or
stellar activity inducing slight transit profile distortions.

In all seven cases, we are able to derive upper limits on the presence of
moon-induced transit timing variations (TTV) and transit duration variations
(TDV), which are typically of the order of a few minutes. We find that all of 
the candidates favor a static model over a model including TTVs. Transit times 
and durations are also made available and are visible in Fig.~\ref{fig:TTVs}. 
Revised transit parameters are presented for all planetary candidates, including 
revised stellar parameters for KOI-303.01 based upon new observations using the 
ARCES spectrograph on the 3.5\,m Astrophysical Research Consortium Telescope at 
the Apache Point Observatory. The light curve derived stellar densities from our 
fits are consistent with the KIC stellar classifications, suggesting no obvious 
evidence for a blend or a wildly eccentric orbit in any case.

%%% PART TWO - FINAL THOUGHTS

% Few moons around mini-Neptunes?
With only seven candidates analyzed and four yielding strong upper limits, we
emphasize that our HEK survey is only just beginning and it is too early to draw
any meaningful statistical conclusions at this stage. The four candidates
with strong limits have radii between 2\,$R_{\oplus}$ and 4\,$R_{\oplus}$
and may be classed as Super-Earths/Mini-Neptunes. Our results therefore
cautiously suggest the preliminary conclusion that such objects do not acquire 
large moons with a high frequency. This may be because the objects never
acquired a large moon or alternatively because such large moons are lost
quicker than the timescales estimated in \citet{barnes:2002}, which we used for
our target selection procedure. The latter argument may be investigated by
inspecting larger radii planetary candidates since the hosts will be presumably
more massive and thus capable of maintaining a large satellite for a greater
duration. The relative paucity of Jovian-sized objects in the 
\emph{Kepler}-sample (B12) means that they are generally
less favorable for a HEK analysis (e.g. fewer bright host stars), but 
nevertheless we will survey these candidates in future work.

% Few moons around single KOIs?
The other unique characteristic of this survey was that we focussed on planetary
candidates in systems where no other planetary candidates had been identified. 
This choice simplifies our TTV analysis but also represents a sample bias. 
Closely-packed systems with multiple planets may translate to a greater 
possibility of large moons, perhaps due to a larger initial reservoir of 
planetessimals from which to form such bodies. In contrast, the system Kepler-36 
\citep{carter:2012} offers a speculative counter-example since the only two 
known transiting bodies reside in 6:7 resonance and may have obtained such a 
configuration through the stripping of one of them as a primordial moon. Since 
only two transiting bodies are known, these objects would have appeared as a 
single transiting binary in a previous epoch, similar to the candidates focused 
on in this survey. It is clear that we are only just beginning to unearth 
evidence for or against large moons and future HEK surveys will test these 
hypotheses by sampling different systems.

% #####################################################################
%% Acknowledgements
\acknowledgements
\section*{Acknowledgements}

% Reviewer thanks
We would like to thank the anonymous reviewer for their thoughtful
comments which improved the quality of our manuscript.
% Dodds thanks
This work made use of the Michael Dodds Computing Facility, for which
we are grateful to Michael Dodds, 
% $250 thanks
Carl Allegretti, David Van Buren, Anthony Grange, Cameron Lehman, Ivan Longland, 
Dell Lunceford, Gregor Rothfuss, Matt Salzberg, Richard Sundvall, Graham 
Symmonds, Kenneth Takigawa,
% $100 thanks
Marion Adam, Dour High Arch, Mike Barrett, Greg Cole, Sheena Dean, Steven 
Delong, Robert Goodman, Mark Greene, Stephen Kitt, Robert Leyland, Matthias 
Meier, Roy Mitsuoka, David Nicholson, Nicole Papas, Steven Purcell, Austen 
Redman, Michael Sheldon, Ronald Sonenthal, Nicholas Steinbrecher, Corbin Sydney, 
John Vajgrt, Louise Valmoria, Hunter Williams, Troy Winarski and Nigel Wright.
% Personal funding
DMK is funded by the NASA Carl Sagan Fellowships. JH and GB acknowledge partial 
support from NSF grant AST-1108686 and NASA grant NNX12AH91H. DN acknowledges 
support from NSF AST-1008890.
% APO acknowledgement
Based on observations with the Apache Point Observatory 3.5\,meter telescope,
which is owned and operated by the Astrophysical Research Consortium.
% Kepler acknowledgement
We offer our thanks and praise to the extraordinary scientists, engineers
and individuals who have made the \emph{Kepler Mission} possible. Without
their continued efforts and contribution, our project would not be possible.
%
%% EOF Acknowledgements

% #####################################################################
%% Bibliography

%\section*{Appendix}
%
%\clearpage

\end{document}